\documentclass[%
 reprint,
 amsmath,amssymb,
 aps,
floatfix,
]{revtex4-2}
\pdfoutput=1
\usepackage[utf8]{inputenc}
\usepackage[english]{babel}
\usepackage[T1]{fontenc}
\usepackage{amsmath}
\usepackage{hyperref}
\usepackage{amsfonts}
\usepackage{dcolumn}
\usepackage{bm}
\usepackage{tikz}
\usepackage{graphicx}
\usepackage{physics}
\usepackage{soul}
\usepackage{xcolor}
\usepackage{graphicx}
\usepackage{float}
\usepackage{subfigure}

\newcommand{\highlight}[1]{
  \colorbox{red!50}{$\displaystyle#1$}}

\usepackage{amsmath,amsfonts,amssymb,times,natbib}
\usepackage[standard]{ntheorem}

\begin{document}

\title{Quantum cooling activated by coherently-controlled thermalisation}

\author{Hanlin Nie}
\affiliation{Centre for Quantum Technologies, National University of Singapore, Block S15, 3 Science Drive 2, 117543, Singapore}

\author{Tianfeng Feng}

\affiliation{State Key Laboratory of Optoelectronic Materials and Technologies and School of Physics, Sun Yat-sen University, Guangzhou, People's Republic of China}

\author{Samuel Longden}
\affiliation{Department of Physics, Imperial College London, Prince Consort Road, London, SW7 2AZ, UK}

\author{Vlatko Vedral}
\affiliation{Centre for Quantum Technologies, National University of Singapore, Block S15, 3 Science Drive 2, 117543, Singapore}
\affiliation{Clarendon Laboratory, Department of Physics, University of Oxford, Oxford, OX1 3PU, England}
\affiliation{Department of Physics, National University of Singapore, 2 Science Drive 3, Singapore 117542}

\date{3 Feb 2022}
\begin{abstract}
It has been shown in a recent letter\cite {felce2020quantum} that thermalising a quantum system with two identical baths of temperature $T$ in an indefinite causal order (ICO) can output a system with temperature different to $T$, allowing for novel non-classical thermodynamic cooling cycle. However, it is in general hard to extract heat from extremely cold system, and for the existing ICO fridge with qubit working system and two thermal reservoirs, the amount of heat the working system can extract decreases rapidly for very low reservoirs temperatures.

In this paper, we show that it is possible to significantly boost the heat extraction ability of the ICO fridge by applying $N$ identical thermalising channels in a superposition of $N$ cyclic causal orders\cite{chiribella2021quantum}, and that this can be further boosted in the ultracold regime by replacing the working qubit with a quDit working substance. Moreover, we show that for the alternative controlled-SWAPs scheme presented in \cite{felce2020quantum} where one additionally has access to the reservoir qubits which are quantum correlated with the control-target system, the performance can be greatly enhanced in general (tripled for all $N$'s and temperatures). Then inspired by \cite{abbott2020communication,wood2021operational}, we show that quantum coherent control between thermalising a working system with one of $N$ identical thermalising channels (where causal indefiniteness plays no role) yield same advantages in controlled-SWAPs scheme compared to the generalised $N$-SWITCH protocol for the thermodynamic task described in \cite{felce2020quantum}. We also provide an experimental simulatable quantum cooling protocol with coherently-controlled thermalising channels and notice that it can outperform ICO refrigerator with some specific implementations of the thermalising channel in the case when we only have access to the control-target system. These 2 quantum cooling protocols bear much lower circuit complexity compared to the one with indefinite causal order which makes it more accessible for implementation of this type of nonclassical refrigerator with cutting edge quantum technologies. 
\end{abstract}

\maketitle

\section{Introduction}
Quantum SWITCH, a quantum supermap\cite{chiribella2008transforming} that transforms a set of quantum channels into the new quantum channel, enables the causal ordering of quantum operations performed to be in a quantum superposition when controlled on a quantum coherent controlled system.

Recent works have shown advantages in some quantum information and thermodyamic tasks which utilise these indefinite causal orders, for instance in quantum communication\cite{ebler2018enhanced,guo2020experimental,chiribella2021indefinite,wei2019experimental,procopio2019communication}, quantum computation\cite{chiribella2013quantum,PRXQuantum.2.010320,araujo2014computational,renner2021experimentally}, metrology\cite{zhao2020quantum,chapeau2021noisy,mukhopadhyay2018superposition} and algorithmic cooling\cite{goldberg2021breaking}. In \cite{felce2020quantum}, author claims that applying identical thermalising channels enables heat transfer from a cold to a hot reservoir if the order of applying the channels is in a superposition and measurement of the control system is allowed.

A more recent letter\cite{chiribella2021quantum} shows ICO can provide even more striking advantages in quantum communication---by placing $N$ completely depolarising channels in a superposition of N cyclic orders, one can achieve a heralded, nearly perfect transmission of quantum information when $N$ is large enough which differs greatly from $N=2$ case, where quantum information can't be effectively transmitted. An important question is then---does the more complex patterns of correlations induced by more than 2 alternative causal orders yield even greater advantages in thermodyamic task described in \cite{felce2020quantum}? Our results answer this question in the affirmative. We then derive the fundamental limitation on this non-classical refrigeration and notice that it is independent of number of thermalising channels and close related to the ratio of the sizes of the hot and cold reservoirs.

But this is not the end of the story. Debate about whether ICO is a necessary ingredient to attain quantum communication advantages claimed in \cite{ebler2018enhanced,chiribella2021indefinite} is heating up. In \cite{guerin2019communication}, which has inspired more careful considerations about role of the quantum trajectories in enhancing communication, it is shown that utilising quantum control of superposed communication channels and quantum-controlled operations in series can also lead to similar communication advantages. While the authors of \cite{kristjansson2020resource} provided a well-structured framework to argue that ICO does provide the greatest advantages and is indispensable in the communication task described in \cite{ebler2018enhanced,chiribella2021indefinite}, a recent experiment\cite{rubino2021experimental} suggests that within the framework of quantum interferometry, the use of channels in series with quantum-controlled operations yields the greatest advantages and that all three schemes use the same experimental resources, in contrast with the strictly theoretical viewpoint in \cite{kristjansson2020resource}. 

Motivated by the enlightening exchange mentioned above, we show that in the controlled-SWAPs scheme where reservoir qubits are accessible and so also serve as resources for cooling when the cooling branch is obtained, quantum coherent control between thermalising a working system with one of N identical reservoirs can provide the same advantage as the one with thermalisation in a superposition of N cyclic causal orders. Moreover, we propose a quantum refrigerator where ICO plays no role which outperforms the generalised N-SWITCH protocol (with regard to coefficient of performance\cite{felce2020quantum} and the ability to cool down the cold reservoirs initialised at an arbitrarily high temperature to temperature which is sufficiently close to absolute zero) when only the control-target state is accessible. An important difference between this and the N-SWITCH protocol - which only required a description of the channel (for example in terms of Kraus operators) to fully describe the action of a coherent-controlled quantum channel - is that in this scheme we also need to specify the transformation matrix\cite{abbott2020communication} of the thermalising channel which depends on its implementation. Meanwhile, the experimental complexity of the cooling schemes where ICO plays no role is much lower than the generalised N-SWITCH protocol, especially when $N$ is large (even we just need $N$ instead of $N!$ alternative causal orders when it comes to cyclic orders). As \cite{felce2021refrigeration} suggests, implementing the unitary circuit used to construct the quantum SWITCH of 2 identical thermalising channels on IBMQ requires 51 gates (19 are two qubit gates), but it would be much more resource intensive to implement the generalised N-SWITCH fridge with $N>2$ in experimental platforms like in \cite{felce2021refrigeration,nie2020experimental}. 

The rest of this paper is structured as follows. We first generalise the quantum SWITCH in \cite{felce2020quantum} to one with $N$ identical thermalising channels and $N$ carefully selected causal orders, then compare this scheme to the original scheme presented in \cite{felce2020quantum} by considering: the amount of heat that can be transferred when attaining cooling branch; the average heat transfer (weighted by the probabilities of attaining the cooling/heating branch) and the coefficient of performance, in order to illustrate the advantages (especially at the ultracold temperature) and trade-off when exploiting more causal orders indefinitely. We then evaluate the lowest temperature attainable for the cold reservoirs when initialised at a fixed temperature and establish a suitable lower bound. Next, we demonstrate how quantum cooling can be achieved with $N$ coherently-controlled identical thermalising channels. We first focus on the controlled-SWAPs scheme which has more advantages for the cooling task, and show that there exists a protocol without ICO that provides the same advantages as the one assisted with the generalised N-SWITCH with cyclic orders. We then describe a simulatable quantum cooling protocol with coherently-controlled thermalising channels and provide a specific implementation (obeys general constraints on the transformation matrices\cite{abbott2020communication} of the thermalising channel) whose performance is then compared with the quantum N-SWITCH refrigerator. Finally, we outline some interesting phenomena from a Maxwell-demon-like scenario involving thermalisation of a sample of particles in superposition of quantum trajectories, before drawing our conclusions and discussing possible future directions to take this research.

\section{Generalised quantum SWITCH with N identical thermalising channels} \label{icoSW}

\subsection{Outline for the previous ICO cooling protocol }\label{ico2N}

Let's first briefly review the protocol of non-classical heat extraction with ICO\cite{felce2020quantum}. A quantum SWITCH can let a quantum state $\rho$ pass through two channels $N_{1}$ and $N_{2}$ in a superposition of causal orders. The Kraus operators of the quantum SWITCH of channels $N_{1}$ and $N_{2}$ are

\begin{figure}[H]
\centering
\includegraphics[width=0.4\textwidth]{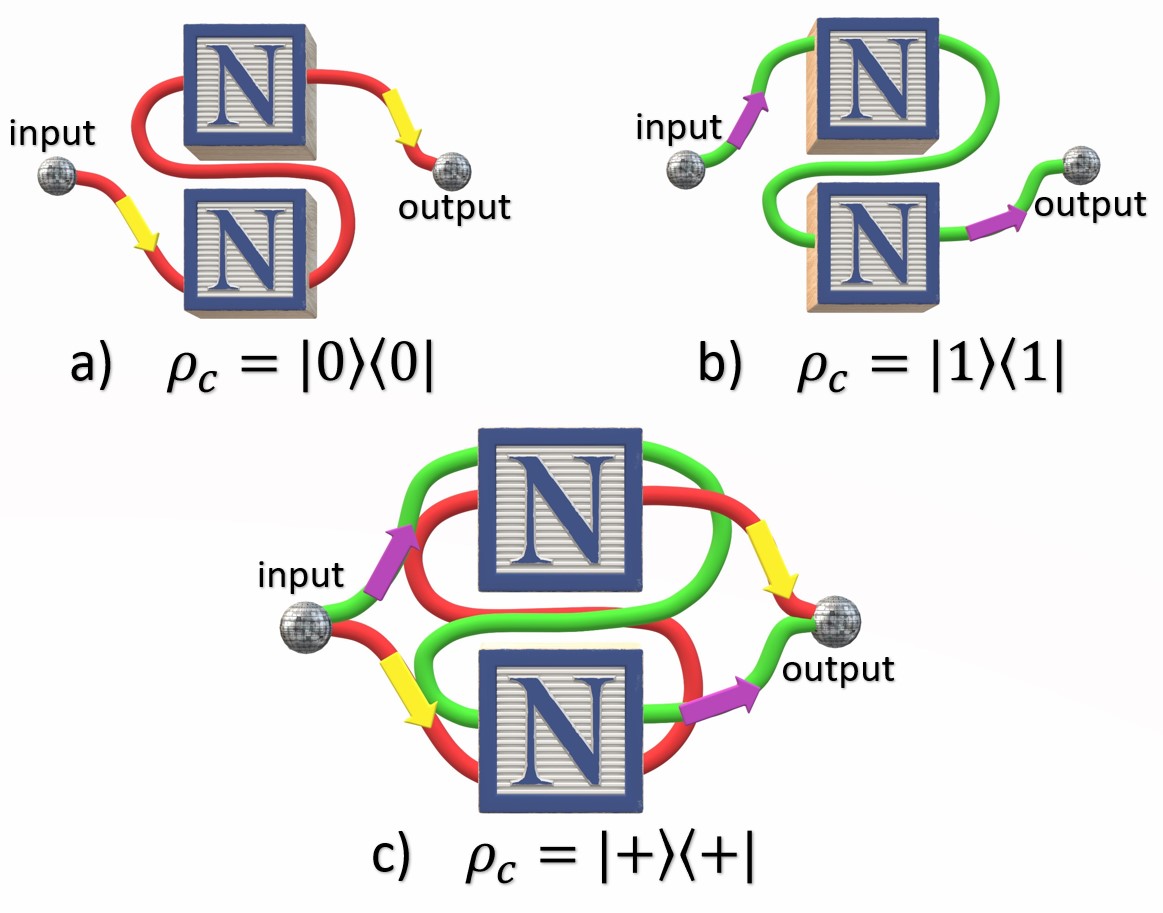}
\caption{\textbf{Two identical channels are placed in a definite causal order in (a) and (b) when the control system is in state $\dyad{0}{0}$ and $\dyad{1}{1}$ respectively. For (c) we can see that a quantum SWITCH can place these 2 channels in a superposition of alternative orders with the state of the control system being in state $\dyad{+}{+}$.}}
\label{qsw2}
\end{figure}

\begin{equation}\label{1}
W_{ij}=|0\rangle\langle 0|_c \otimes K_i^{(2)}K_j^{(1)}+|1\rangle\langle 1|_c \otimes K_j^{(1)}K_i^{(2)},
\end{equation}where the subscript c denotes the control qubit and the operators $K_i^{(2)}$
 and $K_j^{(1)}$ denote the Kraus operators for $N_{1}$ and $N_{2}$ respectively.

These Kraus operators act on a target quantum state $\rho$ and a control state $\rho_c$ so that the quantum SWITCH of the two channels gives
\begin{equation}\label{2}
S(N_{1},N_{2})(\rho_c\otimes\rho)=\sum_{i,j}W_{ij}(\rho_c\otimes\rho)W_{ij}^{\dagger}.
\end{equation}

When $N_{1}$ and $N_{2}$ are two thermalzing channels with same temperture (share same thermal state $T$). Setting $\rho=\rho_T$ and $\rho_c=|+\rangle\langle +|$, we have 

\begin{equation}\label{3}
S(N_{1},N_{2})(\rho_c\otimes\rho)=\frac{1}{2}[I\otimes T +(|0\rangle \langle 1|+|1\rangle \langle 0|)\otimes T^3].
\end{equation}
If the control qubit is projected into $|\pm\rangle=\frac{1}{\sqrt{2}}(|0\rangle\pm|1\rangle)$, the state of target system collapses into 
\begin{align}
\rho_{c}=\frac{1}{2p_{c}}(T+T^3), \label{4}\\
\rho_{h}=\frac{1}{2p_{h}}(T-T^3),\label{5}
\end{align}
where $p_{c}=\frac{1}{2}tr(T+T^3)$ and $p_{h}=\frac{1}{2}tr(T-T^3)$ are the probabilities of attaining $|+\rangle_c$ and $|-\rangle_c$ respectively. $\rho_h$ and $\rho_c$ represent the density matrix of the working system when attaining heating (cooling) branch.

\textit{Working system in ICO refrigerator}---In general, for a quantum system $S$ with finite energy levels with Hamiltonian $H_{S}$, its Gibbs (thermal) state is 
\begin{equation}\label{6}
    \rho^{Gibbs}_{S}=\sum_{i}\frac{e^{-\beta E_{i}}}{Z}\ket{i}\bra{i}= \sum_{i}p_{i}\ket{i}\bra{i}.
\end{equation}where $\beta=\frac{1}{k_{B}T}$ is the inverse temperature, $E_{i}$ denotes the \textit{i}-th energy eigenvalue with corresponding eigenstate $\ket{i}$ and $Z=\sum_{i}e^{-\beta E_{i}}$ is the partition function. A thermal state can be regarded as a statistical mixture of the energy eigenstates locally, but its decomposition can be infinitely many other ensembles. Ref.\cite{felce2020quantum} analyses the case where the working system's Hamiltonian is $H=\Delta|e\rangle\langle e|$ (we will generalised the case to a $D$ dimensions working system and evaluate how the performance of the ICO refrigerator will vary with $D$ in section \ref{Dwork}), so the thermal state with effective temperature $T$ is

\begin{equation}\label{7}
T=\frac{1}{Z_T}\begin{pmatrix} 1 & 0 \\ 0 & e^{-\beta_T\Delta}
\end{pmatrix},
\end{equation}where $\Delta$ is the energy gap between excited and ground state of the working system. So we have

\begin{equation}\label{8}
p_h=\frac{1}{2}[1-\frac{1+r^3}{(1+r)^3}]=\frac{3r}{2(1+r)^2},
\end{equation}
where we denote $r=e^{-\beta_T\Delta}$. When $r$ is small, by making use of first-order Taylor expansion, one can see that $p_h\approx r$. The weighted energy change (for heating branch) is
\begin{equation}\label{9}
 \Delta\tilde{E}_{h}=p_h[tr(\rho_hH)-tr(TH)]=\frac{r(1-r)}{2(1+r)^3}\Delta,
\end{equation}
Similarly, when $r$ is small, $ \Delta\tilde{E}\approx \frac{r\Delta}{2} \approx \frac{1}{2}\Delta p_h$. So in such extreme case, the effective heat extraction depends on probability $p_h$. One may wonder why we evaluate weighted energy change for heating branch when we are talking about a quantum cooling scheme. Actually $\Delta\tilde{E}_{h}=-\Delta\tilde{E}_{c}$ based on the fact that energy is conserved for the whole process on average. And for the heating branches, we release heat to the hot reservoir, so the larger the weighted energy change for the heating branch, the larger the energy decrease is in cold reservoirs (more heat that can be extracted) on average. We analyse a Maxwell-demon like scenario in next section to illustrate the physical meaning of weighted energy change and explain why it is also a good quantity which can help us to evaluate the performance of the ICO fridge.

\begin{widetext}

\begin{figure}
\centering
\includegraphics[width=0.7\textwidth]{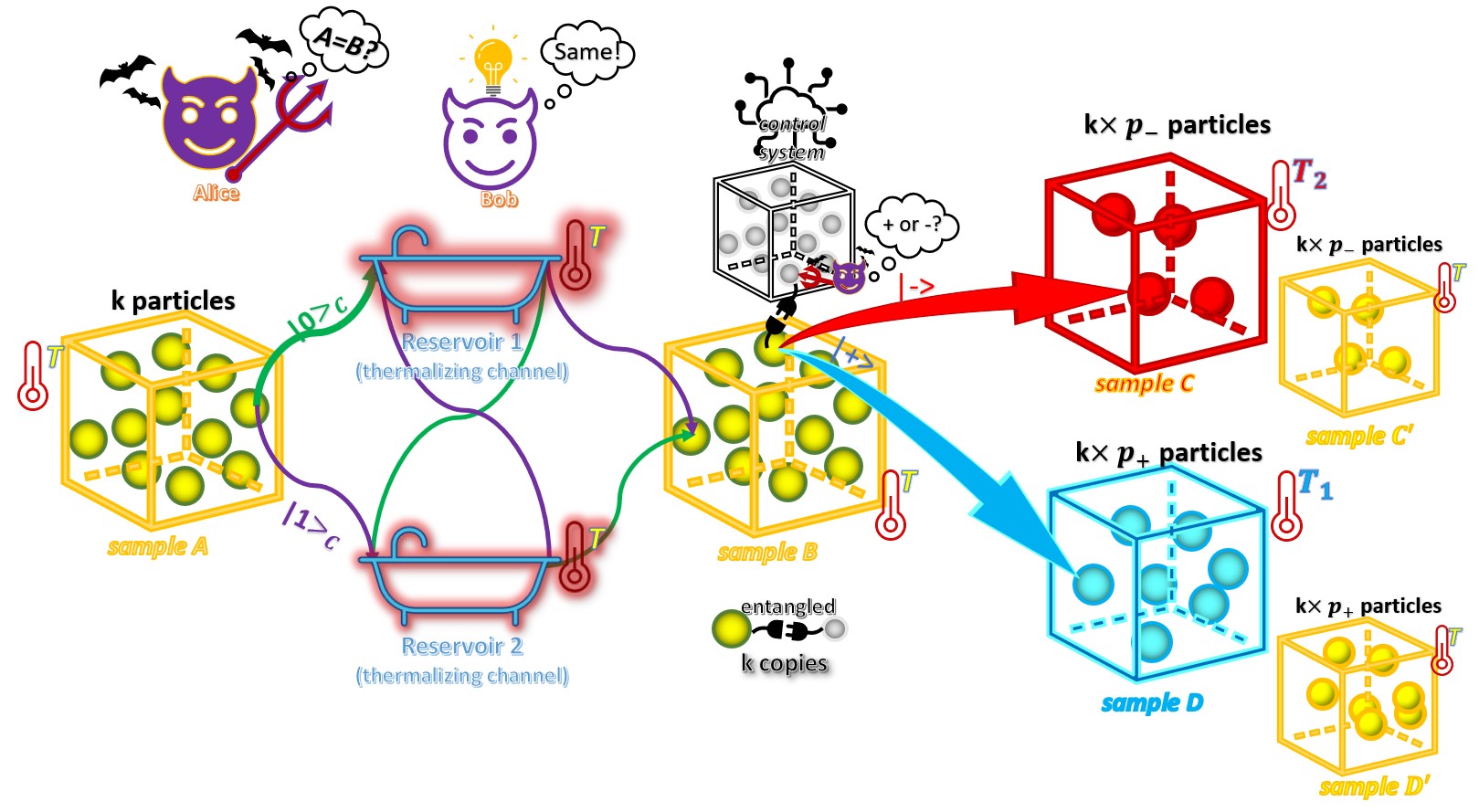}
\caption{\textbf{A Maxwell-demon-like scenario with indefinite causal order. The demon Alice possesses Sample A which contains $k$ identical particles each of thermal state $T$. Alice thermalises all of the particles in an indefinite causal order assisted by $k$ control systems, preparing $k$ identical copies of the entangled control-target systems. Bob, who only has the access to the local target systems can't tell the differences between Sample A and B. Alice now gets to work measuring the states of each of the $k$ control systems and tells Bob each result. Bob uses this information to sort the $k$ particles into two boxes, Sample C and D, of effective temperatures higher and lower than $T$, respectively.}}
\label{f1}
\end{figure}

\end{widetext}
\subsection{Maxwell-demon-like scenario with indefinite causal order} \label{Maxlike}
Besides realizing a standard cooling machine supplied by external driving force to reverse the natural heat flow, implementing a Maxwell demon can conduct the similar task. The essence of the the thermodynamic task describe in \cite{felce2020quantum} is a Maxwell-demon-like scenario, but the demon doesn't act directly on the target system (in general, this will cause disturbances in the thermodynamical situation of the working system\cite{beyer2019steering,elouard2017extracting}). We also notice that there is recent work about quantum cooling fueled by invasive quantum measurements\cite{buffoni2019quantum} on the target system directly, but this is different from the scheme we want to discuss here. 

In \cite{jennings2010entanglement}, the authors introduce a `global demon' who can access the large entangled quantum system, that makes it possible for she to confuse the `local demon' who can only conduct operations locally. For example the global demon can arrange thermodynamic processes that local demon can run `backward' (heat flows from a cold to hot reservoirs etc) fueled by the pre-existing quantum correlations in the global state. Similarly, we can also introduce Alice (global demon) and Bob in our scheme, but Alice here is a much weaker vision of the mischievous global demon in \cite{jennings2010entanglement}, the \textit{only} allowed global operation for Alice is thermalising a target system with some reservoirs in an indefinite causal order (or different superpositions of quantum trajectories we will discuss later) assisted by a control system, so she can prepare a (or many copies of) entangled control-target state(s). And as we mentioned above, Alice won't act on the target system directly, Bob who only has access to the target system wants to conduct some thermodynamical tasks with the local thermal states. Even though we focus on the cooling task here, via the global state $\rho_{CT}$ Alice can prepare (as long as long as it is not a product state\cite{morris2019assisted}) , Bob can also extract work from the target system if Alice conducts suitable measurements on the control system and has classical communication with Bob.

As Fig.\ref{f1} shown, Alice first prepares a sample A consisting of k particles which are already thermalised by a reservoir with temperature $T$. Then she thermalises all the particles from A with reservoirs in same temperature $T$ in an indefinite causal order assisted by k control qubits. So after this step, she prepares k copies of entangled states described in Eqn.(\ref{3}). But at this stage, Bob can't tell the differences between samples A and B since he doesn't have access to the global system. Then based on the measurement result of control system in \{$\ket{+},\ket{-}$\} shared by Alice, Bob sorts $k$ particles into two boxes C and D, with effective temperature higher or lower than $T$, respectively. But do notice that not like the traditional Maxwell demon scenario, the number of particles in sample C or D here are highly depending on the temperature of the reservoirs (and you will also see later by changing the number of reservoirs and dimension of the target system we can actually manipulate the number of particles in each sample). And if Alice thermalises the particles in definite causal order or she doesn't measure the control system (or measures it in computational basis), she can't help Bob to prepare samples like C and D. Then in this setup we can see that weighted energy actually quantifies \textit{on average} how much heat that can be transferred by a particle in sample C which becomes hotter in Bob's point of view such that we can create a colder sample D. So when we evaluate fridge assisted by different quantum SWITCH (superposition of different alternative orders), it will help us to determine the ability of heat extraction for the working system in each ICO fridge at certain temperature. As results from computer simulations in section \ref{sim} shown, the larger the weighted energy change, the fewer cycles needed to be performed to cool down the cold reservoirs from some fixed starting temperatures to some given attainable final temperatures. And you will also see in section\ref{coeffp}, coefficent of performance defined in \cite{felce2020quantum} is actually the absolute value of weighted energy change for cooling branch divided by the work needed to reset the register for the optimal case when the temperatures of the cold and hot reservoirs are the same.\medskip

\subsection{Interference terms in the entangled target-control state for generalised N-SWITCH} \label{offterm}
\begin{figure}
\centering
\includegraphics[width=0.4\textwidth]{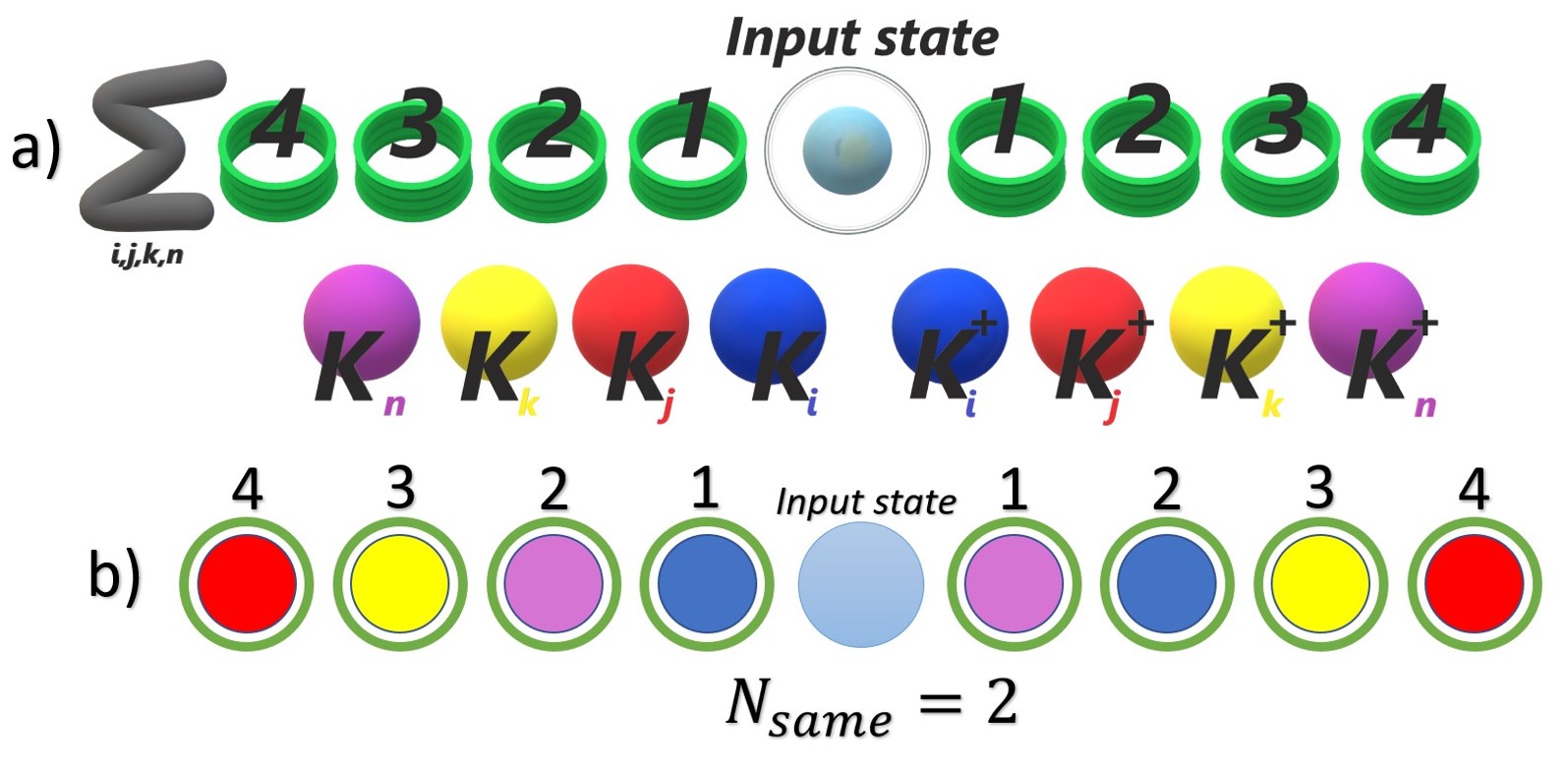}
\caption{\textbf{Ball tossing scheme helps us visualise terms in state generated by quantum SWITCH. The figure shows the case for $N=4$. The balls with different colors denote Kraus operators from different channels, and the aim is to fill in all the holes with these balls, and balls on the right-hand-side can only go to one of the holes on the right-hand-side of the input state and vice versa. Once filled, $N_{same}$ is the number of colour-number pairs on the right-hand-side that match that on the left-hand-side. (b) illustrates an example for $N_{same}=2$. When $N_{same}=N$, it denotes one of the diagonal terms.} }
\label{f2}
\end{figure}

Off-diagonal (interference) terms in final state generated by the quantum SWITCH are the origin of the advantages in those quantum information tasks assisted by indefinite causal order, so it is natural to think whether more complex patterns of correlations induced by more alternative causal orders will provide us with greater advantages. But for the generalised N-SWITCH making use of all possible causal orders, the complexity increases dramatically with $N$. And for the experimental implementation, for example, the realization of the unitary circuit used to construct quantum SWITCH of 2 identical thermalising channels in \cite{felce2020quantum} already needs 51 gates (19 are two qubit gates) for platforms other than photonic one (NMR\cite{nie2020experimental} and IBMQ\cite{felce2021refrigeration}). In \cite{wilson2020diagrammatic}, the authors use a diagrammatic approach to analyse the generalised N-SWITCH and show that why focusing on the cyclic orders is a good strategy to construct a N-SWITCH. And a recent letter\cite{chiribella2021quantum} shows that utilisation of $N$ completely depolarising channels in a superposition of N cyclic orders, one can achieve a heralded, nearly perfect transmission of quantum information when $N$ is large enough, and this is advantage can't be provided by the 2-SWITCH. Following similar spirit, we want to investigate whether a generalised N-SWITCH can provide greater advantages in the quantum cooling task in \cite{felce2020quantum}. But since we are aiming at task other than quantum communication, it is also necessary to investigate whether other choices of causal orders can provide similar or even greater advantages than the cyclic ones. 

\begin{figure}
\centering
\includegraphics[width=0.3\textwidth]{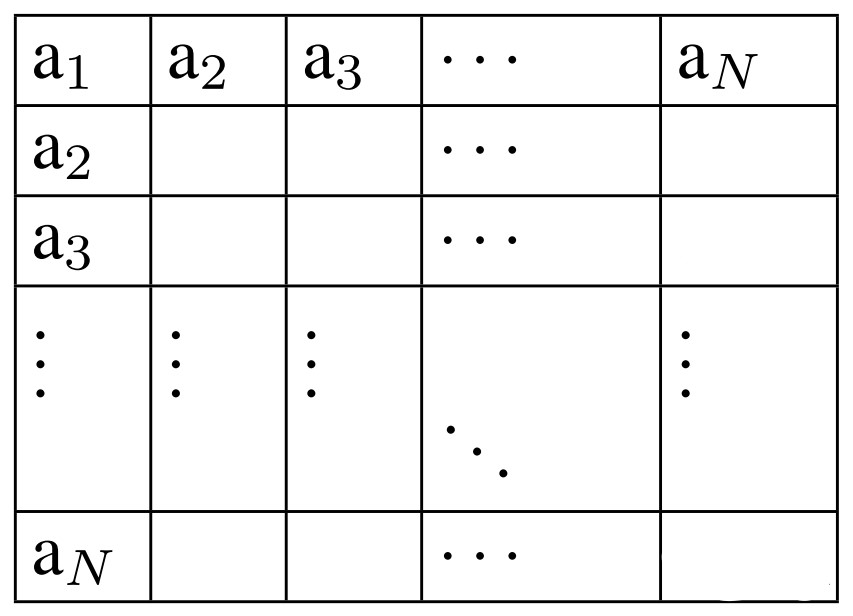}
\caption{\textbf{Latin-square-like chart which can help us to select $N$ particular casual orders with $N$ identical thermalising channels.}}
\label{f3}
\end{figure}

\begin{figure}
\centering
\includegraphics[width=0.3\textwidth]{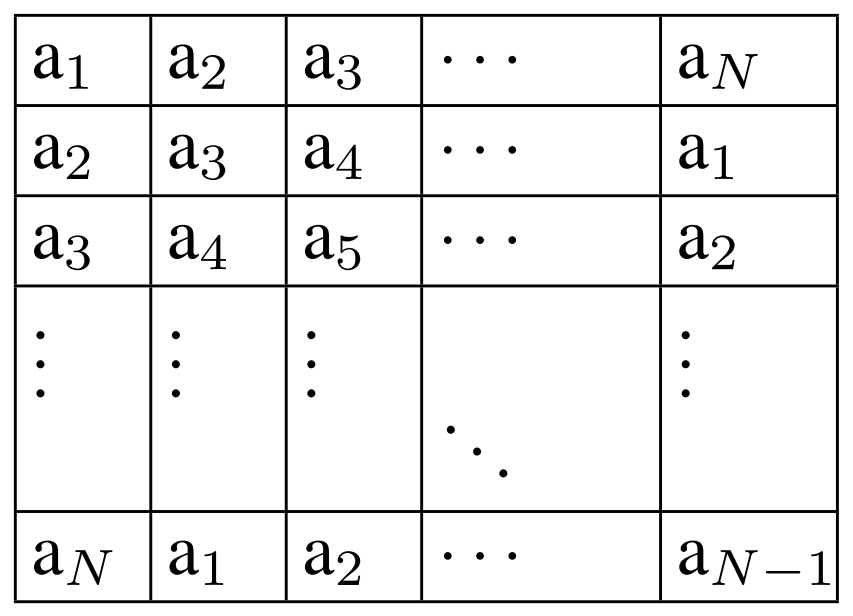}
\caption{\textbf{Filling the chart with cyclic permutations of orders makes the output state have off-diagonal terms $T\rho T$.}}
\label{f4}
\end{figure}

To determine the terms in final state generated by the quantum SWITCH, we can treat this task as a ball tossing game (see Fig.\ref{f2} for the example of 4 channels). We have N holes on the left- and right-hand-side of the ball representing the input state respectively (we label them with different numbers). On each side, we have N balls with different colors which represents Kraus operator from different channel, the aim is to fill in all the holes with these balls, and balls on the right-hand-side can only go to one of the holes on the right-hand-side of the input state and vice versa. So it is not hard to imagine how the number of possible combinations greatly increases as N scales up. We now define a quantity $N_{same}$ which denotes the colour-number pairs on the right-hand-side that match that on the left-hand-side. It is clear that when $N_{same}=N$, the term is a diagonal one. We did some case study for $N=3,4$ with different choices of causal orders and found out that there is a thing in common for optimal cases: we only have 1 cooling branch and all the rest are identical heating branches (provide same final state for the target system after measurement of control in a suitable basis), which means that the final entangled state generated by the SWITCH has the same off-diagonal terms when we set the input to be in a same thermal state $T$ as the reservoirs' initially. And we notice that this is exactly the case when we focus on the off-diagonal terms with $N_{same}=0$. So with the Latin-square-like chart in Fig.\ref{f3} (every element can only occur once in each row and column), we can always find a set of N different causal orders with N identical channels such that the off-diagonal terms generated by the N-SWITCH are all the same. There are many possible ways to fill in the chart when N is large, but there is always a simple strategy for arbitrary large N (see Fig.\ref{f4}). And this gives us the cyclic orders introduced in \cite{chiribella2021quantum}.

But even though there are many other ways to fill in the chart, we notice that when we set the input to be in a same thermal state $T$ as the reservoirs' initially, the quantum N-SWITCH based on those sets of N chosen causal orders provides the same off-diagonal terms ($T^{N}$ when N is odd and $T^{N+1}$ for N is even). But the performances of the ICO fridges based on these selections of causal orders decrease when N increases unfortunately. So for the next section we will focus on the cyclic orders and show the generalised N-SWITCH fridge based on them can attain greater advantages in thermodynamical tasks compared to the original one with two reservoirs.

\subsection{Quantum N-SWITCH for superposition of N cyclic orders of N identical thermalisations} \label{cyclicN}

Assisted by Fig.\ref{f4}, we can see that the Kraus operators of the generalised quantum N-SWITCH with cyclic orders are:
\begin{small}
\begin{equation}\label{10}
\begin{split}
    W_{a_{1}a_{2}\cdots a_{N}}=&|0\rangle\langle0|_c\otimes K_{a_{1}}K_{a_{2}}\cdots K_{a_{N}}+\\
    &|1\rangle\langle1|_c\otimes K_{a_{2}}K_{a_{3}}\cdots K_{a_{1}}+\cdots+\\
    &|N-1\rangle\langle N-1|_c\otimes K_{a_{N}}K_{a_{1}}\cdots K_{a_{N-1}},
\end{split}
\end{equation}
\end{small}With $K_{i} = \sqrt{\frac{1}{d}}AU_{i}$ (where $A$ is the square root of the diagonal matrix $T$) and $\sum_{i}K^{\dagger}_{i}K_{i} = I$.

For the ($i+1$)-$th$ row, we can denote that $K(i+1)= K_{a_{i+1}}K_{a_{i+2}}\cdots K_{a_{i}}$. So we can then write $W_{a_{1}a_{2}\cdots a_{N}}=\sum^{N-1}_{i=0}|i\rangle\langle i|_c\otimes K(i+1)$.

The quantum SWITCH of the N channels gives:
\begin{footnotesize}
    \begin{equation}\label{11}
    S(\underbrace{N^{T},\cdots,N^{T}}_{\text{N terms}})(\rho_c\otimes\rho)=\sum_{a_{1},\cdots,a_{N}} W_{a_{1}\cdots a_{N}}(\rho_c\otimes\rho)W_{a_{1}\cdots a_{N}}^{\dagger}.
\end{equation}
\end{footnotesize}

We initialise the state of the control system to be in $|\psi\rangle=\frac{1}{N}\sum_{k=0}^{N-1}|k\rangle$. For the diagonal terms of Eqn.(\ref{2}), they come from selecting 2 identical rows from the chart. Take the $k$-$th$ term as an example:

\begin{figure}[H]
    \centering
	\includegraphics[width=0.4\textwidth]{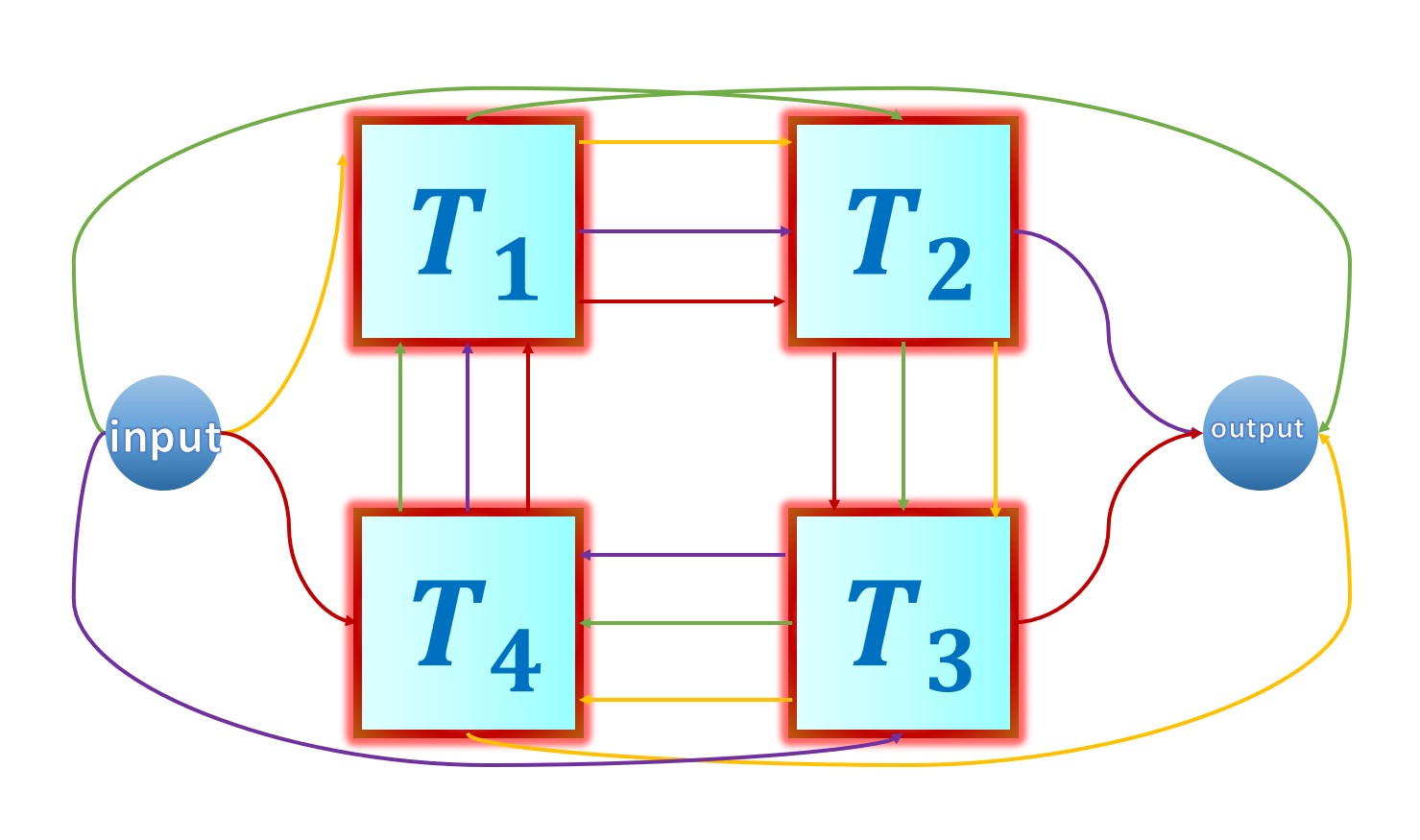}
	\caption{\textbf{A visual representation of 4 thermalising channels acting on an input system in a superposition of 4 cyclic causal orders.}}
	\label{f5}
\end{figure}

\begin{small}
\begin{equation}\label{12}
\begin{aligned}
&|k\rangle\langle k|_c\otimes K(k+1)\rho K^{\dagger}(k+1)\\
&=|k\rangle\langle k|_c\otimes\frac{1}{d^{N}}\sum_{a_{1}\cdots a_{N}} AU_{a_{k}}\cdots AU_{a_{k-1}}\rho U^{\dagger}_{a_{k-1}}A^{\dagger}\cdots U^{\dagger}_{a_{k}}A^{\dagger},\\
&= |k\rangle\langle k|_c\otimes T.
\end{aligned}
\end{equation}
\end{small}The second equality is given by application of the depolarizing channel for N times with the action of the depolarizing channel is:
\begin{equation}\label{13}
\aleph^{D}(\rho) = \text{Tr}[\rho]\frac{I}{d} = \frac{1}{d^2}\sum^{d^2}_{i}U_{i}\rho U^{\dagger}_{i}.
\end{equation}For the off-diagonal terms, they actually come from selecting 2 different rows (let's say ($i+1$)-$th$ and ($j+1$)-$th$ rows) from the chart. For the ($i+1$,$j+1$) pair:
\begin{footnotesize}
\begin{equation}\label{14}
\begin{split}
    &|i\rangle\langle j|_c\otimes K(i+1)\rho K^{\dagger}(j+1)\\
    &=|i\rangle\langle j|_c\otimes\frac{1}{d^{N}}\sum_{a_{1}\cdots a_{N}} AU_{a_{i+1}}\cdots AU_{a_{i}}\rho U^{\dagger}_{a_{j}}A^{\dagger}\cdots U^{\dagger}_{a_{j+1}}A^{\dagger},\\
    &= |i\rangle\langle j|_c\otimes T\rho T. 
\end{split}
\end{equation}
\end{footnotesize}The second equality is attained by application of the depolarizing channel for $(N-k)$ times (where $k=|i-j|$) and the fact that the operators $U_{i}$ form an orthonormal basis for the set of $d\times d$ matrices, i.e $\sum_{i}\text{Tr}[U_{i}M]U^{\dagger}_{i}=\sum_{i}\text{Tr}[MU^{\dagger}_{i}]U_{i}=M$ where M is an arbitrary $d\times d$ matrix. And this holds for all $i,j=0,1,...,(N-1)$ with $i\ne j$. See Appendix \ref{icooffcyclic} for the detailed derivation of this part.

We then have: 
\begin{footnotesize}
    \begin{equation}\label{15}
    (\underbrace{N^{T},\cdots,N^{T}}_{\text{N terms}})(\rho_c\otimes\rho)= \frac{1}{N}I\otimes T+\frac{1}{N}(\sum^{N-1}_{i\ne j,=0} \dyad{i}{j})\otimes T\rho T.
\end{equation}
\end{footnotesize}
So the quantum N-SWITCH with $N$ cyclic orders, produces only $T\rho T$ for the off-diagonal terms of the final entangled state of the control and target systems, and this will go back to the control qubit case when $N=2$.

For the ICO fridge, the working system shares the same thermal state as the reservoirs initially, and by following the general measurement strategy (see Appendix \ref{Mstra}) we can always construct a measurement basis such that there is only 1 cooling branch with the target system being in state:
\begin{equation}\label{16}
    \rho_{c} = \frac{\frac{1}{N}\{T+(N-1)T^3\}}{p_{c}},
\end{equation}
with $p_{c} = \frac{1}{N}\text{Tr}[T+(N-1)T^3]$, after the control system is measured and 
$(N-1)$ identical heating branches with the target system being in state:
\begin{equation}\label{17}
\rho_{h} = \frac{\frac{1}{N}(T-T^3)}{p_{h}},
\end{equation}with $p_{h} = \frac{1}{N}\text{Tr}[T-T^3]$, after the control system is measured. So
\begin{equation}\label{18}
p_{H} = (N-1)p_{h} = \frac{N-1}{N}\text{Tr}[T-T^3]. 
\end{equation}

\subsection{Building a superior control system for quantum N-SWITCH from qubits} \label{qubitcon}

\begin{figure}
\includegraphics[width=0.5\textwidth]{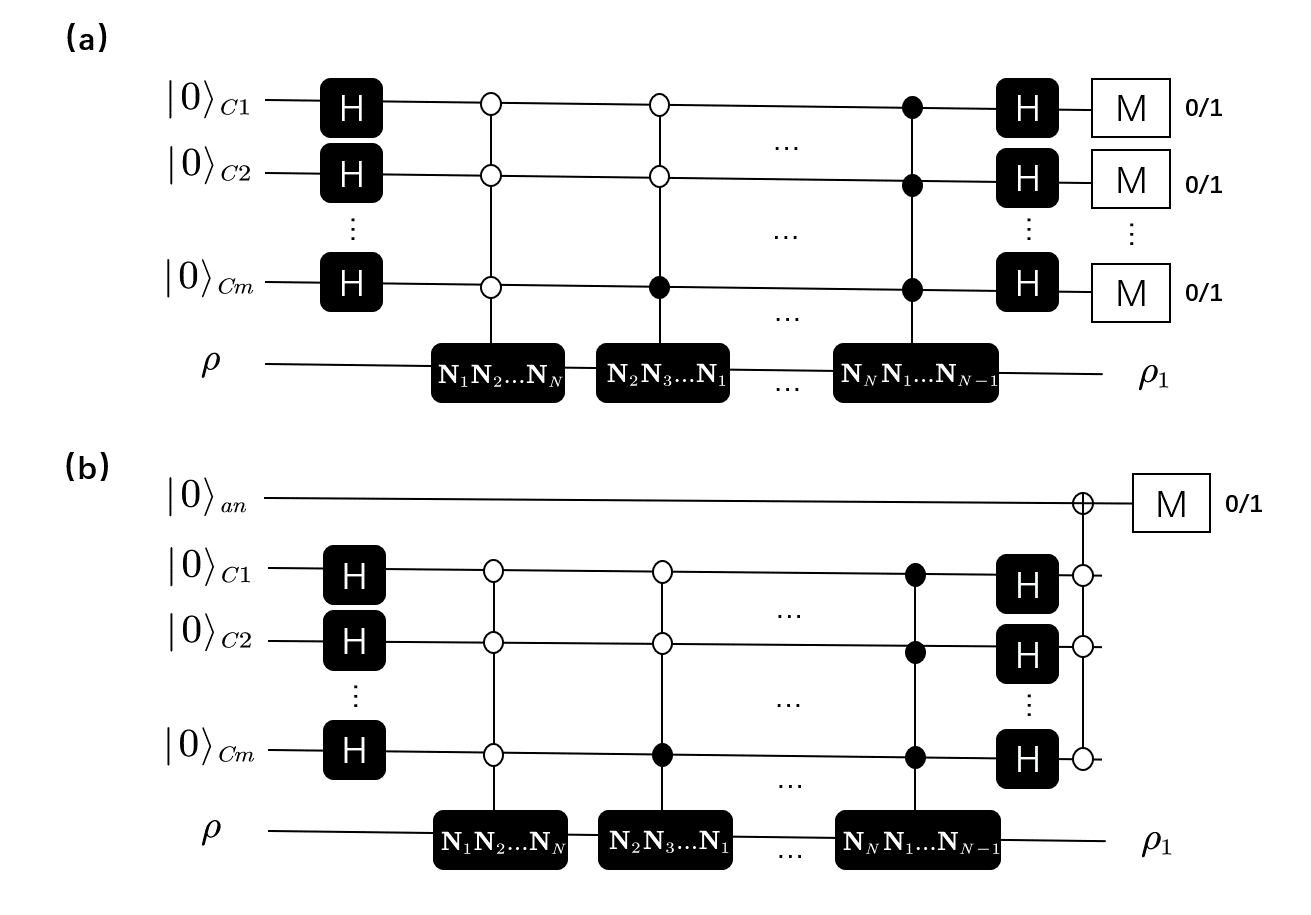}
\caption{ \textbf{(a) Quantum circuit which produces same final marginal state for target system as the quantum SWITCH of $N=2^m$ thermalising channels. (b) Simplified quantum circuit with $N=2^m$: measurement of the ancillary qubit is enough to determine whether or not the cooling branch is obtained.} 
}\label{circuit}
\end{figure}
Considering the feasibility of experiments, we discuss how to use qubits instead of quNit to construct the control system.
When the number of channels is $N=2^m$, the control quNit can be replaced by a $m$-qubit system. In this case we initialise the control state as $|\psi_{c}\rangle=|+\rangle^{\otimes m}$ instead of $|\psi_{c}\rangle=\frac{1}{2^m}\sum^{2^m-1}_{i=0}|i\rangle$,  where $|+\rangle=\frac{1}{\sqrt{2}}(|0\rangle+|1\rangle)$. As shown in Fig.\ref{circuit} (a), we have a $m$-qubit state $|0\rangle^{\otimes m}$ and a working qubit with state $\rho$ initially. The quantum state of control system becomes $|+\rangle^{\otimes m}$ after the actions of $m$ Hardamard gates. 
The ICO process with superposition of $N$ cyclic causal orders is realised by sequential multi-qubit-controlled operations.

Different from the quNit case we mentioned above, the choice of measurement basis for the control system can be simplified in this m-qubit control system scheme. As presented in Fig.\ref{circuit} (a),
the measurement of the control system can be conducted in computational basis for each qubit after the action of $m$ parallel Hadamard gates, which corresponds to the measurement basis $\{ |+\rangle^{\otimes m},  |+\rangle^{\otimes m-1}|-\rangle,...,  |-\rangle^{\otimes m}\}$. The quantum state of working system becomes:
\begin{small}
\begin{equation}\label{19}
    \langle +|^{\otimes m}_cS(N_1,...,N_N)(\rho_c\otimes\rho) |+\rangle^{\otimes m}_c=\frac{1}{N}\{T+(N-1)T^3\}.
\end{equation}
\end{small}and 
\begin{equation}\label{20}
\begin{aligned}
\langle G|_cS(N_1,...,N_N)(\rho_c\otimes\rho) |G\rangle_c
=\frac{1}{N}(T-T^3),
\end{aligned}
\end{equation}where $|G\rangle\in \{ |+\rangle^{\otimes m-1}|-\rangle, |+\rangle^{\otimes m-2}|-\rangle|+\rangle,\\ ...,  |-\rangle^{\otimes m}\}$. Obviously, the results are the same as the generalised quNit case where $(N-1)$ identical branches corresponding to heating. 

Moreover, we can just construct a POVM (positive operator-valued measurement) which will tell us whether we get the cooling branch or not by introducing an ancillary qubit: $|+\rangle^{\otimes m}\langle +|^{\otimes m}$ and $I-|+\rangle^{\otimes m}\langle +|^{\otimes m}$.  Even though the cost of resetting the whole control system including the ancillary qubit is just the same as the previous scheme (see section \ref{comico}), this framework provide more experimental conveniences because we only need to record one instead of m bits of information each time when we have a m-qubit control.
  As shown in Fig.\ref{circuit} (b), the ancillary qubit is entangled with the joint quantum state $(H^{\otimes m}\otimes I) S(N_1,..,N_N)(\rho_c\otimes\rho)(H^{\otimes m}\otimes I)$ by using a multi-qubit Toffoli gate. When the $m$-qubit state are in $|0\rangle^{\otimes m}$, the ancillary qubit is flipped to $|1\rangle$, otherwise the ancillary qubit remains unchanged.

After measuring the ancillary qubit in computational basis and tracing out the control state, we can get:
\begin{equation}\label{21}
\begin{aligned}
\rho^{(0)}_1=\frac{1}{N}\{T+(N-1)T^3\},\\
\rho^{(1)}_1=\frac{N-1}{N}(T-T^3),
\end{aligned}
\end{equation}where $\rho^{(0)}_1$ and $\rho^{(1)}_1$ are states of working system before normalization corresponding to the measurement outcomes $1$ and $0$ for ancillary qubit respectively.

\subsection{Extracting heat from ultra cold reservoirs with a D-dimensional working substance} \label{Dwork}
Results in \cite{felce2020quantum} place no restriction on the dimension of the working system. We consider replacing the working qubit with a higher-dimensional system and show that this can further boost the heat extracting ability of the working system of the ICO fridge within the ultracold temperature region (see secion \ref{comico} for more details about how to quantitatively determine what range of $r$ is corresponding to low temperature region).

The thermal state for a $D$-dimensional quantum system is
\begin{equation}\label{22}
T_D=\frac{1}{Z_{T_D}}\sum_{i=0}^{D-1}e^{-\beta_T\Delta_i}|i\rangle\langle i|,
\end{equation}where $\Delta_i$ is the eignenergy in eigenstate $|i\rangle$. Without losing any generality, we set $\Delta_0=0$. Then the Hamiltonian of the quDit system becomes $H_D=\sum_{i=1}^{D-1}\Delta_i|i\rangle\langle i|$. 

For simplicity, we consider the case when $D-1$ excited state are degenerated.  That is $\Delta_i=\Delta$ for all $i\ne 0$. Now the Hamiltonian of the quDit system is given as 
\begin{equation}\label{23}
H_D=\sum_{i=1}^{D-1}\Delta|i\rangle\langle i|.
\end{equation}so the probability of attaining heating branches is
\begin{footnotesize}
    \begin{equation}\label{24}
		\begin{aligned}
			&p^D_H
			=\frac{N-1}{N}\{1-\frac{1+(D-1)r^3}{[1+(D-1)r]^3}\}\\
			&=\frac{N-1}{N}\{\frac{(D-2)(D-1)D r^3+3(D-1)^2r^2}{[1+(D-1)r]^3}\\
			&+\frac{3(D-1)^2r^2+3(D-1)r}{[1+(D-1)r]^3}\}.
		\end{aligned}
	\end{equation}
\end{footnotesize}If $r$ is close to $0$, we have
\begin{footnotesize}
	\begin{equation}\label{25}
		p^D_H\approx
		\frac{3(N-1)}{N}(D-1)r.
	\end{equation}
\end{footnotesize}Compared with the case for qubit working system within the low temperature region ($r\to0$), the probability of using degenerated quDit working system with $N$ channel increases as $\frac{p_H^D}{p_h^2} \approx 3(D-1)\frac{N-1}{N}$. We also find that given an arbitrary $r$ that is not equal to 0, there is always a positive integer $M$, stratifying $D\ge M$, can makes $p_h\approx \frac{N-1}{N} $. This indicates that the upper bound of the probability of attaining heating branches is $1$ for ICO fridge with $N$ channels. For the weighted energy change,
\begin{small}
\begin{equation}\label{26}
		\begin{aligned}
			&\Delta \tilde{E}^{h}_D=p_H^D[\text{Tr}(\rho_hH_D)-\text{Tr}(T_DH_D)]\\
			&=\frac{(1-r^2)r}{[1+(D-1)r]^4}\frac{(N-1)}{N}(D-1)\Delta.
		\end{aligned}
	\end{equation}
\end{small}If $r\to 0$, we have $\Delta \tilde{E}^{h}_D=-\Delta\tilde{E}^{c}_D\approx\frac{(N-1)(D-1)r}{N}\Delta\approx\frac{1}{3}p^D_H\Delta$. This result suggests that within the low temperature regime ($r\to 0$), the weighted energy change is increased by a factor of $2(D-1)\frac{N-1}{N}$ compared to the 2 reservoirs qubit working system ICO fridge\cite{felce2020quantum} when we make use of $N$ reservoirs and a $D$-dimensional quantum systems.

\subsection{Working cycle for refrigerator with indefinite causal order}\label{wcyc}

\begin{widetext}

\begin{figure}
    \centering
	\includegraphics[width=0.8\textwidth]{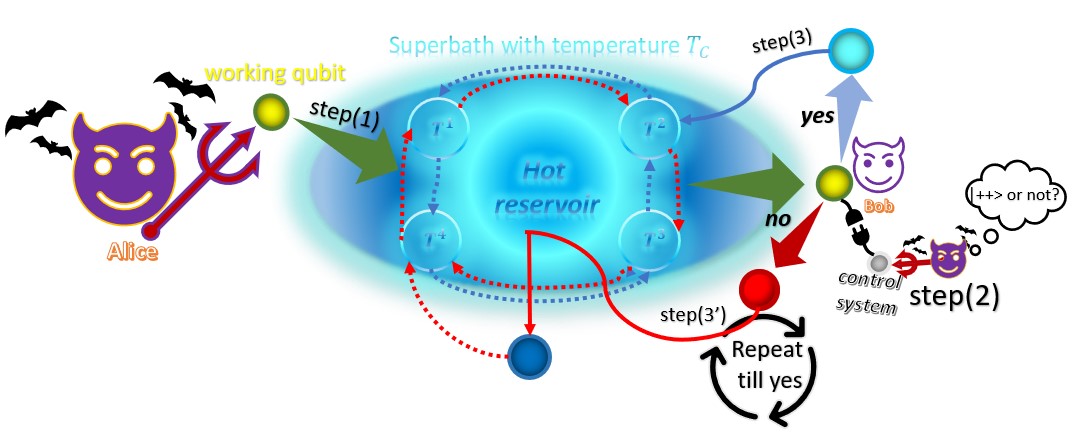}
	\caption{\begin{small}
	    \textbf{Working cycle for the ICO fridge. In step (1) Alice thermalises the working system with $N$ reservoirs sharing the same temperature ($N=4$ in the figure) in an superposition of $N$ cyclic orders assisted by the control system. Then in step (2), she measures the control system in a suitable coherent basis, depending on whether the measurement result is $\ket{e_{0}}=\frac{1}{N}\sum_{k=0}^{N-1}|k\rangle$, Bob goes to step (3) or (3'). For step (3) (see blue lines), Bob thermalises the cooled down working system with one of the $N$ cold reservoirs classically followed by the classical thermalisation of all the cold reservoirs with one another. At the end all the cold reservoirs and working system share the same lower temperature and thermal state. But when attaining heating branches (measurement result of the control is not $\ket{e_{0}}$, Bob goes to step (3')(see red lines). He first needs to thermalise the working system with the hot reservoir classically and then with one of the cold reservoirs followed by classical thermalisation of all the cold reservoirs with one another. Step (3') needs to be repeated till we get the cooling branch, and at the end of each cycle we need to reset the control based on the measurement result and erase the register which is used to store the measurement result of the control. The cold and hot reservoirs can originate from the same superbath with temperature T, so in the beginning they share the same temperature, but they get temperature difference when more and more cycles are performed, and at some points the fridge stops operating and can't further cool down the cold reservoirs(see section \ref{Lowestico})}.
	\end{small}}
	\label{f9}
\end{figure}

\end{widetext}

Before comparing the performances of ICO fridge with different N-SWITCH, let's first briefly go through the refrigeration cycle of the ICO fridge (see Fig.\ref{f9} for the case of 4 reservoirs)in the Maxwell-demon-like scenario as we described in the first section: The demon Alice prepares a entangled control-target state by thermalising the working system with $N$ reservoirs sharing the same temperature in a superposition of $N$ cyclic orders assisted by the control system. Then Alice measures the control system, if the result is $\ket{e_{0}}=\frac{1}{N}\sum_{k=0}^{N-1}|k\rangle$, which means it is the cooling branch, then Bob thermalises the working system (classically) with one of $N$ cold reservoirs, followed by the classical thermalisation of all the cold reservoirs with one another (see the blue lines in Fig.\ref{f9}), so all the cold reservoirs including the working system end up to be at the same temperature(and share the same thermal state). Otherwise, if the measurement result for the control is other than $\ket{e_{0}}$, Bob should thermalise the working system with the hot reservoir with inverse temperature $\beta_{H}$ classically, followed by a classical thermalisation between the working system and one of the cold reservoirs and  classical thermalisation of all the cold reservoirs with one another (see red lines in Fig.\ref{f9}). And this process should be repeated until Alice gets $\ket{e_{0}}$. The $N$ reservoirs and the hot reservoir can be taken from the same superbath with a certain temperature and be placed in thermal isolation at the beginning, and as more and more cooling cycles performed, the hot reservoir gradually heated up while the cold reservoirs become cooler and cooler, but at some point the fridge stops operating (can't cool down the cold reservoirs any more), see section \ref{Lowestico} for more details about this part.

\subsection{Performance of the quantum Switch fridge} \label{comico}

\subsubsection{Weighted energy change}\label{wec}
To quantify the performance of the ICO fridges with different numbers of reservoirs, besides coefficient of performance defined in \cite{felce2020quantum} which evaluates how efficient the ICO fridge is, another important quantity is the weighted energy change defined as $\Delta \tilde{E}_{h} = p_{H}\Delta E_{h}=-\Delta\tilde{ E}_{c} = -p_{c}\Delta E_{c}$ which quantifies \textit{on average} the amount of heat that can be transferred by the working system per cycle (in analogy to power in standard refrigeration machine). Since $r=e^{-\Delta\beta}$, so to determine what range of $r$ corresponds to low temperature region, we should

\begin{figure}[H]
    \centering
	\includegraphics[width=0.5\textwidth]{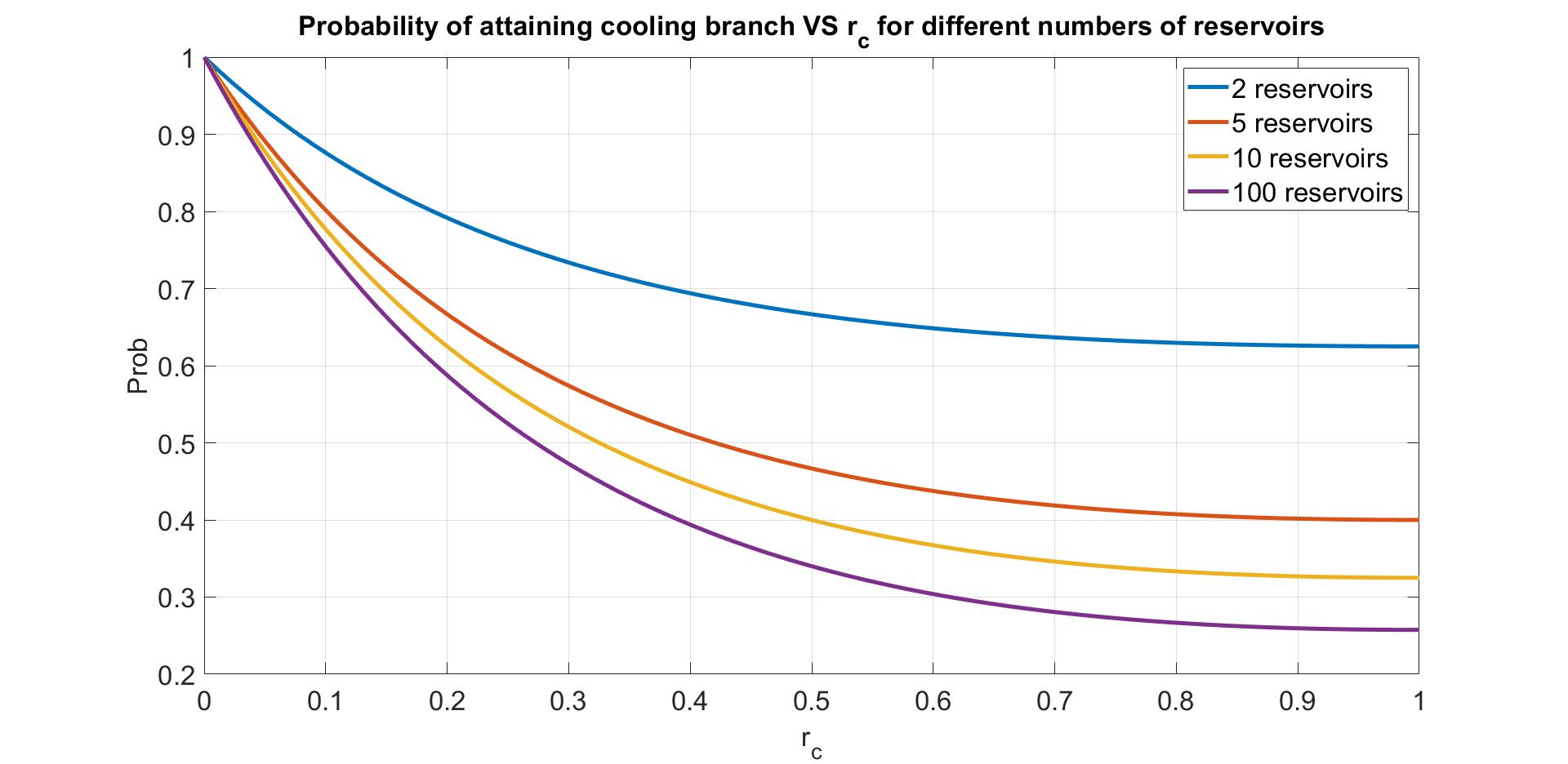}
	\caption{\textbf{Probability of obtaining the cooling branch (y-axis) at different temperatures (x-axis) for different numbers of reservoirs.}}
	\label{f7}
\end{figure}
refer to the specific physical systems (with different $\Delta$). For example, the hyperfine qubit used in trapped ion platform (IonQ, Honeywell), the energy difference between the excited and ground state of the qubit corresponds to a photon with frequency $3\sim 10$ GHz. So $\Delta\approx 1.99\sim 3.33\times 10^{-24}J$. Take $\Delta\approx 1.99\times 10^{-24}J$ as an example, we see that when $r$ ranges from 0 to 0.99, the effective temperature ranges from 0 to 14.34$K$. But for the platform using Rydberg atom,  $\Delta$ is around 5 $eV$. So only when $r$ is extremely small (smaller than $10^{-100}$), the corresponding effective temperature is low enough to be regards as cold (251.76$K$).

\begin{widetext}

\begin{figure}
    \centering
	\includegraphics[width=1\textwidth]{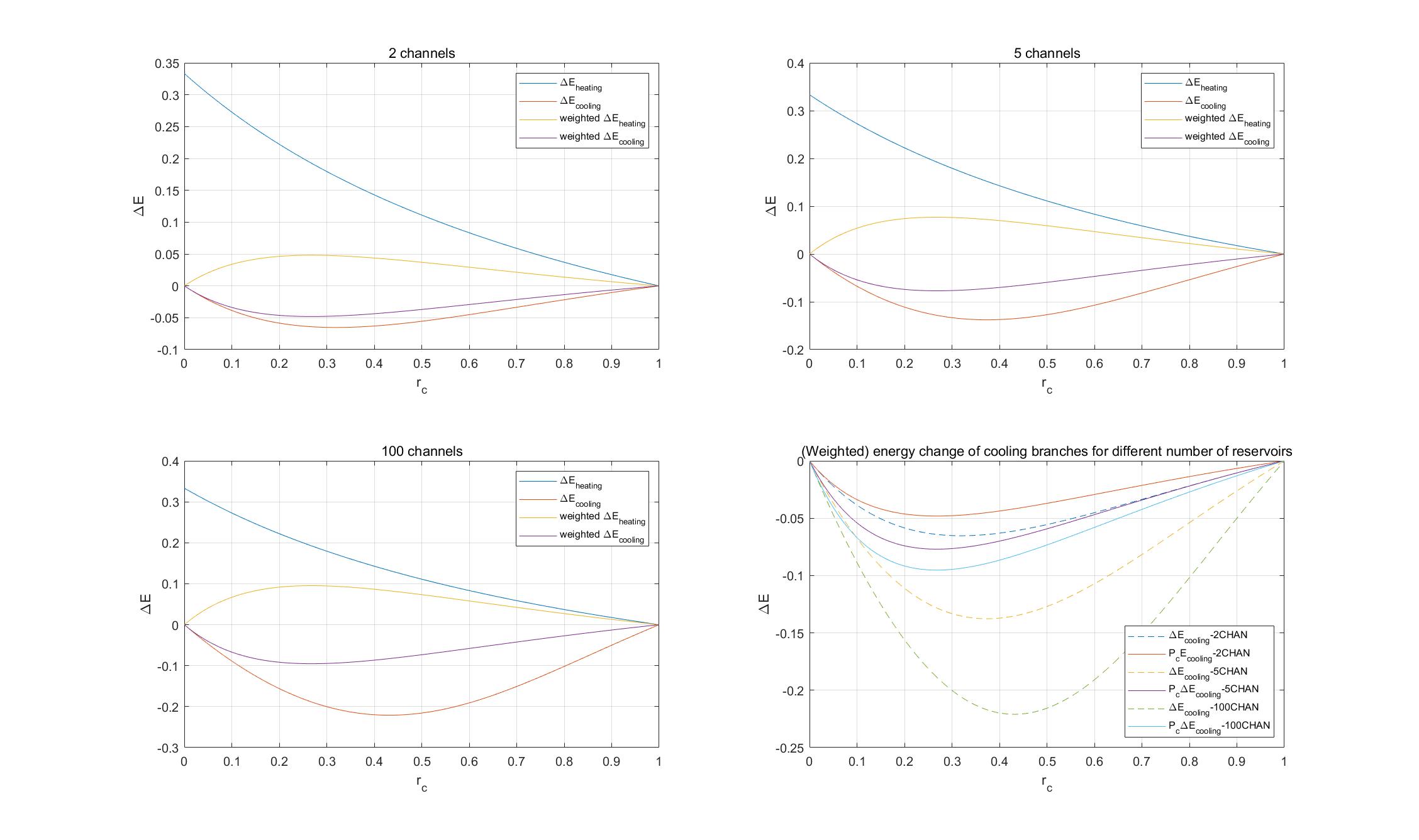}
	\caption{\textbf{How the weighted-energy-changes vary with the number of thermalising channels for the ICO fridge. The blue (orange) lines denote the energy change of the working system when attaining heating (cooling) branch. And the yellow (purple) lines denote weighted energy change} $\tilde{E}_{h}$ ($\tilde{E}_{c}$) \textbf{which illustrates on average how much heat can be extracted by the working system per cycle. As we can see weighted energy change can be almost doubled when $N$ is large enough.}}
	\label{f8}
\end{figure}

\end{widetext}

For the ICO fridge with 2 reservoirs in the system described in \cite{felce2020quantum}, as you can see in Figs.\ref{f7} and \ref{f8}, for the low-temperature regime, even though the probability of getting cooling branch is high, but the amount of heat that can be extracted by the working system when attaining cooling branch decreases rapidly as $T$ goes down (see orange curve in Fig.\ref{f8}). So it is difficult for the ICO fridge with 2 reservoirs to cool the cold reservoirs down to some ultracold temperatures (the number of cycles needed to run is large even for the ideal case when we get the cooling branch each run). By increasing the number of reservoirs in the fridge (while still using qubit working system), as we can see from Fig.\ref{f8} (dash lines), we greatly increase the amount of heat that can be extracted by the working system when attaining the cooling branch with the cost of decreasing the probability of getting the cooling branch, but within the low temperature regime, as temperature of the reservoirs decreases, the differences between probabilities of getting cooling branches for different number of reservoirs becomes much smaller (see Fig.\ref{f7}).

\begin{widetext}

\begin{figure}
    \centering
	\includegraphics[width=1\textwidth]{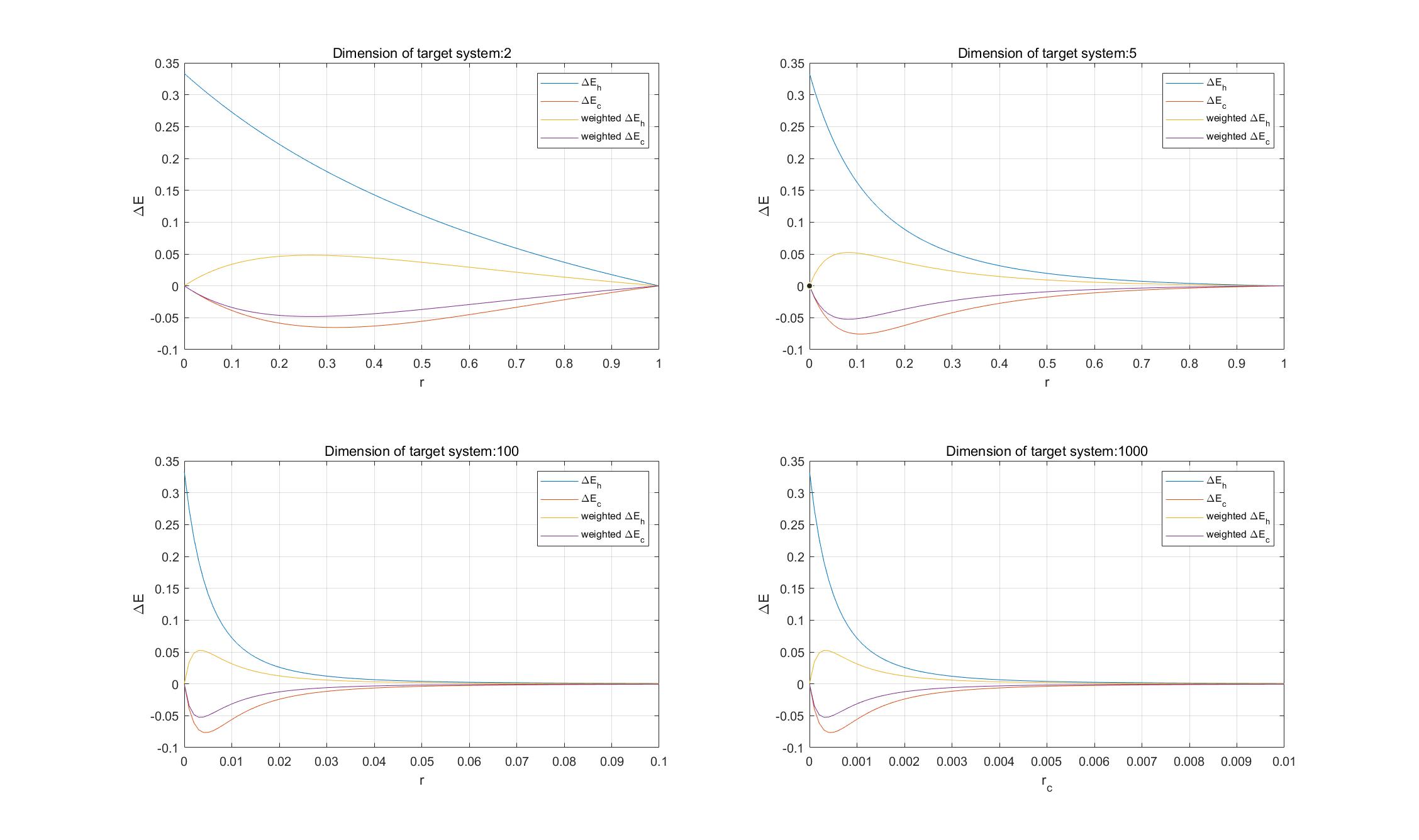}
	\caption{\textbf{How the weighted-energy-changes varies with the dimension of the target system (2 reservoirs). As the figure shown, utilizing a working system with more dimensions enables us to attain global maximal weighted-energy-changes, even in the ultracold temperature regime, but the cost is this fridge can only operate effectively within an very narrow temperature domain. } }
	\label{Wico2D}
\end{figure} 

\end{widetext}

By Eqns.(\ref{16}) and (\ref{17}) we have 
\begin{small}
\begin{equation*}
    \rho_{c} = \frac{\frac{1}{N}\{T+(N-1)T\rho T\}}{p_{c}},
\end{equation*}
\end{small}and 
\begin{small}
\begin{equation*}
    \rho_{h} = \frac{\frac{1}{N}\{T-T\rho T\}}{p_{h}},
\end{equation*}
\end{small}with
\begin{small}
\begin{equation*}
    p_{c}=\frac{1}{N}\text{Tr}[T+(N-1)T\rho T],
\end{equation*}
\end{small}and
\begin{small}
\begin{equation*}
    p_{H}=(N-1)p_{h}=\frac{N-1}{N}\text{Tr}[T-T\rho T].
\end{equation*}
\end{small}So we have $\Delta\tilde{E}_{h} = p_{h} (\text{Tr}[\rho_{h}H]-\text{Tr}[TH])$ and $\tilde{\Delta E_{c}} = p_{c} (\text{Tr}[\rho_{c}H]-\text{Tr}[TH])$. For ICO fridge, the working system is initialised to have the same thermal state as the reservoirs $\rho=T$. Since $\rho_{h}=\frac{T-T\rho T}{\text{Tr}[T-T\rho T]}$ is N-independent, so $\Delta E_{h}$ is fixed for different N. What plays a role in changing $\Delta\tilde{E}_{h}$ while changing N is the change of $p_{H}=(N-1)p_{h}=\frac{N-1}{N}\text{Tr}[T-T\rho T]$, it is obvious that when N is large enough $\frac{p^{(N)}_{H}}{p^{(2)}_{H}}\approx 2$. So when N is large, $\Delta\tilde{E}_{h}$ can be approximately doubled (so does $\Delta\tilde{E}_{c}$ since energy is conserved for the process on average, see Fig.\ref{f8}). And we should see that weighted energy change provides us with a more objective way to compare the cooling ability of the working system of the ICO fridges with different numbers of reservoirs. Moreover, we also see that how the power of the ICO fridge can be further boosted by using a D-dimensional working system in the extremely low temperature regime (as shown in Fig.\ref{Wico2D}). But the price is the ICO fridge can only work effectively in a very narrow temperature domain, the larger the D, the more narrow this domain is. And we should notice that even the probability of getting cooling branch decreases dramatically as temperature increases when D is large for a fixed N, in the region where this ICO fridge can operate effective, the probability of attaining the cooling branch can still be arbitrarily close to 1 when temperature is low enough. The application of the ICO fridge with D-dimensional working system can cool down the cold reservoirs to some extremely low temperatures with much fewer cycles for cold reservoirs initialised at low temperature compared to the qubit working system case. We should also notice that even though the increase of weighted energy change is at most 2 times compared to the 2-reservoir case when N is large, the increase of amount of heat that can be extracted by the working system when attaining cooling branch is not bounded in this sense, and lower the temperature, the more obvious the enhancement is for large N. And as you can also see in Fig.\ref{Wico2D} (also Fig.\ref{Wico100D} in Appendix \ref{perDw} for the 100 reservoirs case), for cold reservoirs starting from a low enough temperature, the increase of the dimension of the working system can further increase the amount of heat that can be extracted when getting the cooling branch for some narrow low temperature regions.

\subsubsection{Work cost to reset the register (and control system)}
To run the ICO fridge in cycle, we need to properly reset the control system each run. To do so we need a register (memory) to record the measurement result of the control system, and since we assume the control system has a trivial zero Hamiltonian, the energy cost of resetting the control to the initial state $|\psi\rangle=\frac{1}{N}\sum_{k=0}^{N-1}|k\rangle$ is zero if we have information\cite{del2011thermodynamic}
(stored in the register) about its final state after the measurement of it. Measurement can be conducted in a reversible fashion, so the energy cost can be made arbitrary small ideally\cite{maruyama2009colloquium}. Then the energy cost arises from the process of erasing the register. Notice that we only need a classical instead of a quantum memory, so the Shannon entropy (instead of von Neumann entropy) for the register $\rho_{M}$ in the projective measurement case (with measurement operators $\{P_{k}\}$) is
\begin{equation*}
     S(\rho_{M})=-\sum p_{k}\text{ln}p_{k}.
\end{equation*}with $p_{k}=\text{Tr}[P_{k}\rho_{M}]$. In the case of generalised N-SWITCH with $N$ identical thermalising channels, since we have 1 cooling branch and $(N-1)$ identical heating branch, the Shannon entropy of the register is 
\begin{equation}\label{27}
     S(\rho_{M})=-p_{c}\text{ln}p_{c}-(N-1)p_{h}\text{ln}p_{h}.
\end{equation}Do be careful here, even though we have only 2 different branches, for the case of getting the heating one, we still need to know exactly which state the control system becomes in order to reset it to the initial state $|\psi\rangle=\frac{1}{N}\sum_{k=0}^{N-1}|k\rangle$ without energy cost (by a unitary operation). So the Shannon entropy is $ S(\rho_{M})=-p_{c}\text{ln}p_{c}-(N-1)p_{h}\text{ln}p_{h}$ instead of  $S(\rho_{M})=-p_{c}\text{ln}p_{c}-(N-1)p_{h}ln((N-1)p_{h})=S(\rho_{M})=-p_{c}\text{ln}p_{c}-p_{H}\text{ln}p_{H}$. And that's why in the alternative scheme with a POVM described in the section \ref{qubitcon}, the final total entropy of the register for the ancillary qubit and the control system is the same as Eqn.(\ref{27}). For the register of the ancillary qubit, its Shannon entropy is
\begin{equation}\label{28}
    S(\rho_{M^{'}})=-p_{c}\text{ln}p_{c}-p_{H}\text{ln}p_{H}.
\end{equation}when we get $\ket{1}$ for the measurement of the ancillary qubit, we don't need to do anything except erasing the 2-dimensional register. But when we get $\ket{0}$, besides the erasure of the register, since we have no information about the state of control and the probability of getting each heating branch is identical, so we can only describe the state of the control as a fully mixed state with dimension $2^{m}-1$. So the von Neumann entropy of the control is

\begin{equation}\label{29}
\begin{split}
     S(\rho_{Con})&=-\text{Tr}[\rho_{Con}\text{ln}\rho_{Con}]\\
     &=-\sum_{i}^{2^m-1}\frac{1}{2^m-1}\text{ln}(\frac{1}{2^m-1}).
\end{split}
\end{equation}We should see $S(\rho_{M})=S(\rho_{M^{'}})+p_{H}S(\rho_{Con})$. Then the work cost of resetting the register (and the control system when we attain $\ket{0}$ in the POVM case) when they are coupled with a reservoir with inverse temperature $\beta_{R}$ is
\begin{equation}\label{30}
    \Delta W_{E}=\frac{1}{\beta_{R}}S(\rho_{M}).
\end{equation}for both schemes. But for the POVM case, it provides much more experimental conveniences for the simplified multi-qubit control system case ($N=2^m$) as we discussed previously.

\subsubsection{Coefficient of performance}\label{coeffp}
It is a important quantity which illustrates how efficient the ICO fridge can operate. The coefficient of performance (COP) is on average, the amount of heat that can be extracted by the working system from the cold reservoirs per cycle divided by the work

\begin{figure}[H]
	\includegraphics[width=0.5\textwidth]{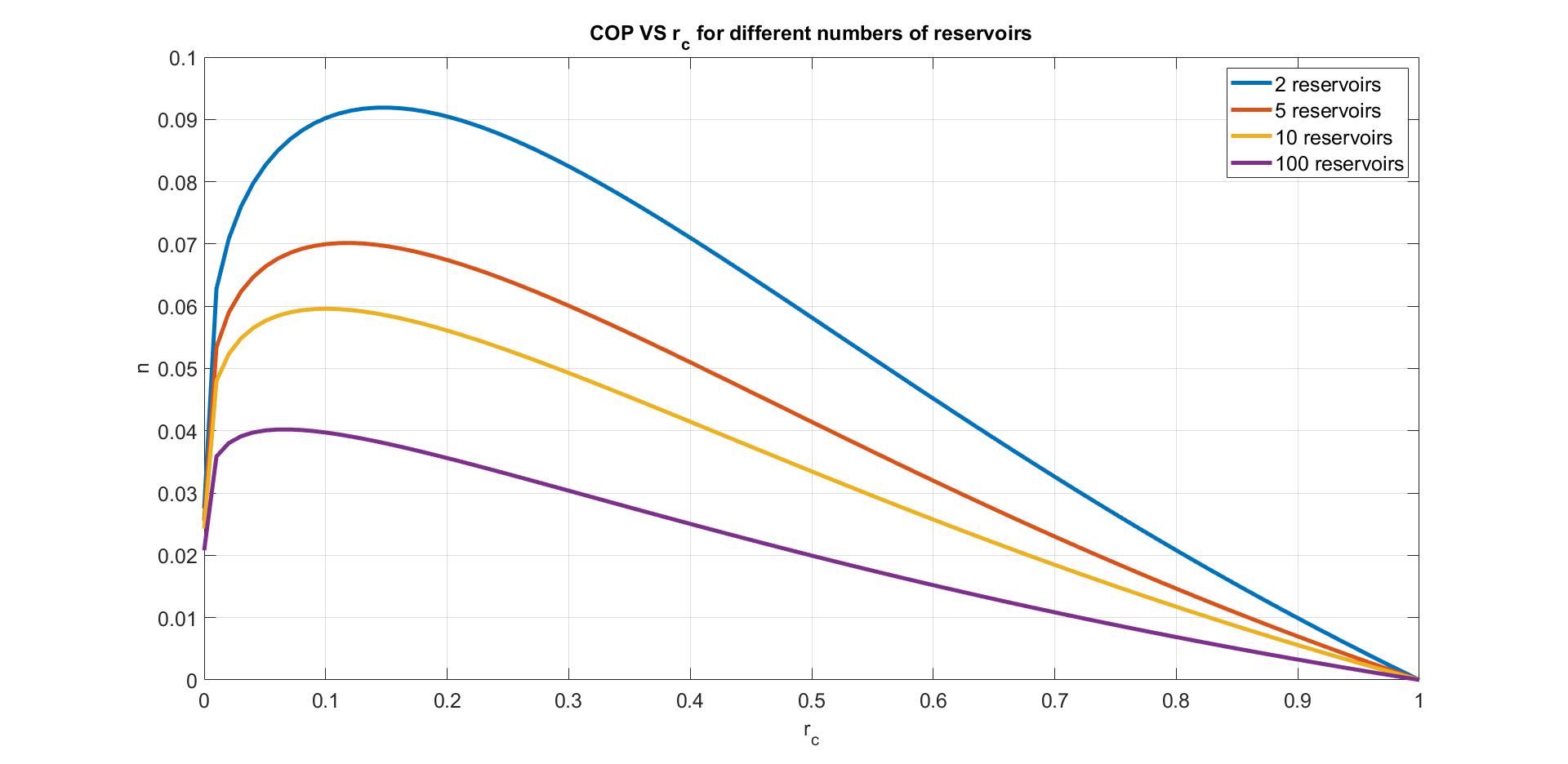}
	\caption{\textbf{Coefficient of performance (divided by $\Delta \beta_{R}$) VS $r_c$ for ICO fridges with different numbers of reservoirs for the ideal case when $T_{c}=T_{H}$.}}
	\label{f10}
\end{figure}
\begin{figure}
    \centering
	\includegraphics[width=0.4\textwidth]{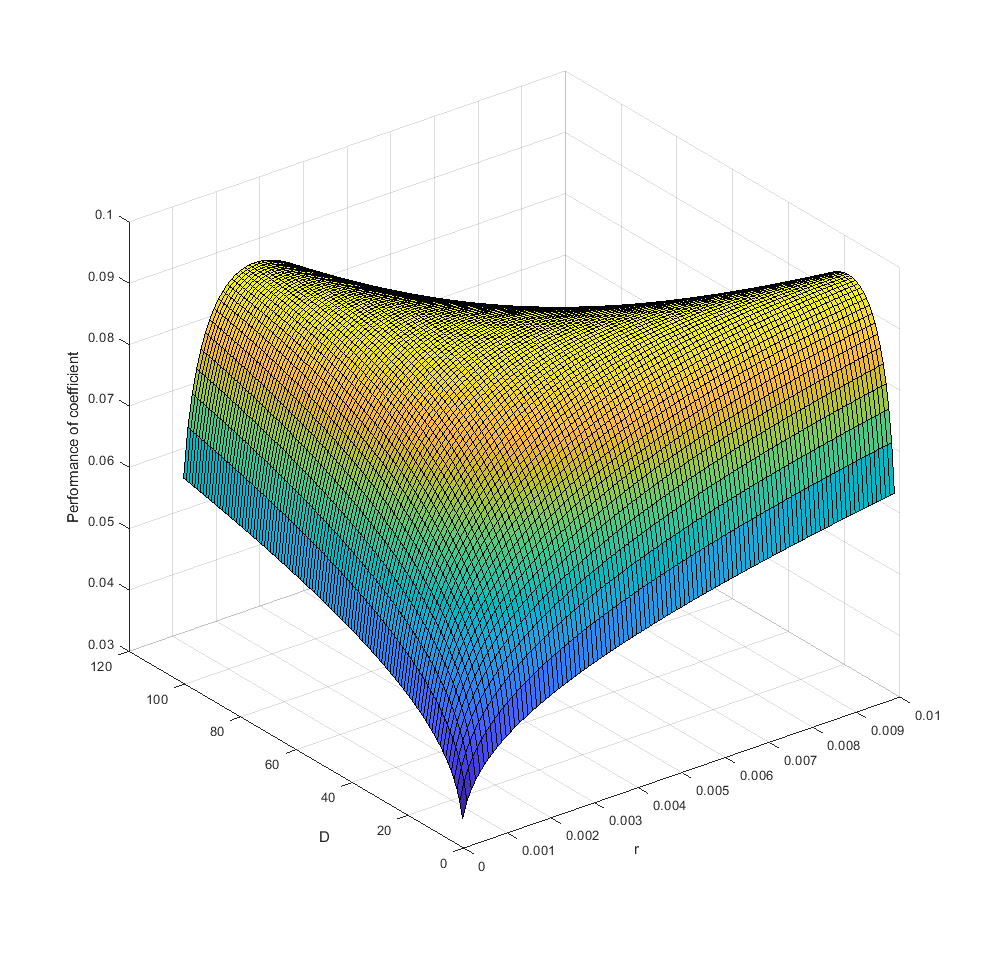}
	\caption{\textbf{Coefficient of performance (divided by $\Delta \beta_{R}$) by using $D$-level quantum working system. For region of arbitrary low temperature, we can always find a working system with a suitable dimension such that the COP of the fridge attains the global maximum.}}
	\label{f11}
\end{figure}

cost. When attaining cooling branch, the amount of heat that can be extracted by the working system is $-(\text{Tr}[\rho_{c}H]-\text{Tr}[TH])$, and when we attain the heating branch, we need to thermalise it with the hot reservoir and then cold reservoirs, the amount of heat transferred from the working system to the cold reservoirs is $\text{Tr}[T_{H}H]-\text{Tr}[TH]$ ($T_{H}$ is the thermal state of the hot reservoir). And since the probability of getting cooling branch is $p_{c}$, so on average we should repeat the cycle $\frac{1}{p_{c}}$ times which means to attain the cooling branch once, we need to (on average) reset the register for $\frac{1}{p_{c}}$ times and the amount of heat transferred from the working system to the cold reservoirs is $(\frac{1}{p_{c}}-1)(\text{Tr}[T_{H}H]-\text{Tr}[TH])$. So the coefficient of performance can be written as
\begin{align*}
     &\frac{-(\text{Tr}[\rho_{c}H]-\text{Tr}[TH])}{\frac{1}{p_{c}}\Delta W_{E}}\\
     &+\frac{-(\frac{1}{p_{c}}-1)(\text{Tr}[T_{H}H]-\text{Tr}[TH])}{\frac{1}{p_{c}}\Delta W_{E}},\\
     &=\frac{-\Delta\tilde{E}_{c}-p_{h}(\text{Tr}[T_{H}H]-\text{Tr}[TH]))}{\Delta W_{E}},\\
     &\geq \frac{-\Delta\tilde{E}_{c}-p_{h}(\text{Tr}[\rho_{h}H]-\text{Tr}[TH]))}{\Delta W_{E}},\\
     &=\frac{-\Delta\tilde{E}_{c}-\Delta\tilde{E}_{h}}{\Delta W_{E}},\\
     &=0.
\end{align*}as we can see from the last few lines, when the temperature of the resetting reservoirs $T_{H}$ is equal to the effective temperature of the working system when attaining heating branch, the ICO fridge stops functioning (we will discuss this in detail next section). And for the optimal case, where $T_{H}=T$, the COP achieves the highest values, in this sense the term in numerator in the expression of COP is just simply $-\Delta\tilde{E}_{c}=\Delta\tilde{E}_{h}$ which is weighted energy change as we discussed above. Just like most of the standard refrigeration machines, an ICO fridge with working system with greater cooling ability (when $N$ is large) bears a lower efficiency as shown in Fig.\ref{f10}. But we notice that (for a fixed number of reservoirs) by manipulating the dimension of the working system, we can always construct an ICO fridge with the global maximal COP at arbitrary low temperature as you can see in Fig.\ref{f11}. 

With the generalised N-SWITCH of $N$ identical thermalising channels and D-dimensional working system, we can construct ICO fridge for wider use. For example, for the low temperature region, ICO fridge with larger $N$ can cool down the cold reservoirs to some ultracold temperatures with much fewer cycles compared to the 2-reservoir one. And for the low-temperature region, the 2-reservoir ICO fridge also bears low efficiency, in this sense, making use of a working system with a suitable number of dimensions can always give us an ICO fridge with the global maximal efficiency at arbitrary low temperature. But there is always a trade-off between efficiency and cooling ability for the working system for the ICO fridge.

\subsubsection{Computer simulation to evaluate fridge performance}\label{sim}
In this section, we use computer program to simulate the ICO fridges assisted by different N-SWITCH, and show how the temperature of cold reservoirs will change when sufficient number of cooling cycles are performed. With the help of this program, we can clearly see that the ICO fridge with more reservoirs (weighted energy change of its working system is larger) can cool down the cold reservoirs to some given temperatures with much fewer cycles compared to the $N=2$ one starting at the same temperature. But the ICO fridge with more reservoirs bears lower efficiency even it needs much fewer cycles to cool down the cold reservoirs to some given temperature due to the fact that the cost for erasing the high-dimensional register is high.
\begin{figure}
    \centering
	\includegraphics[width=0.5\textwidth]{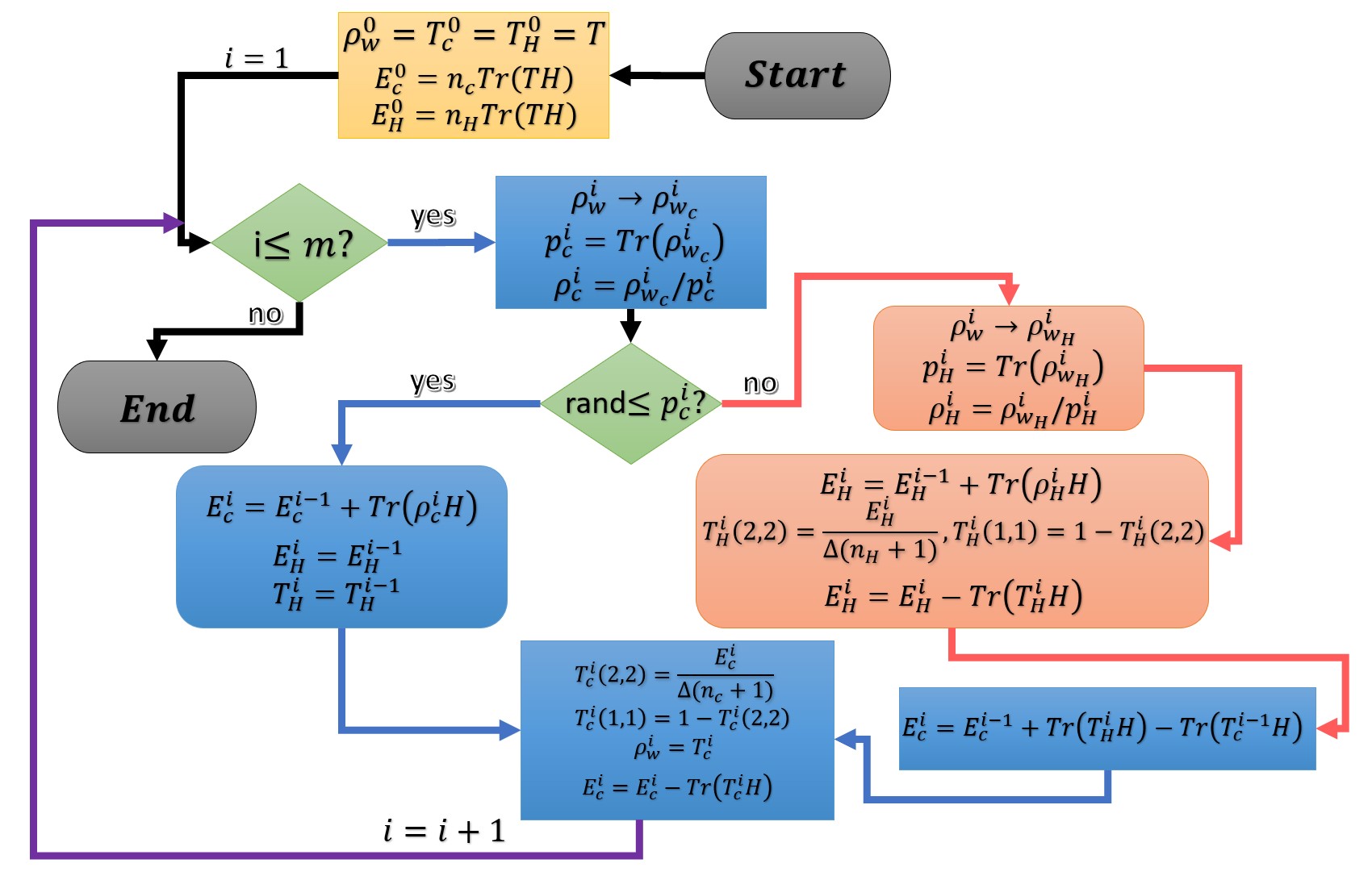}
	\caption{\textbf{Flowchart for the ICO fridge simulation.} }
	\label{program}
\end{figure}

We assume the $N$ cold reservoirs and hot reservoir are from the same superbath with thermal state $T$ initially, $n_c$($n_H$) is the total number of particles (qubits) in the $N$ cold reservoirs (hot reservoir). We follow the cooling cycle described in section \ref{wcyc}. At the beginning, the working system has the same thermal state as the cold reservoirs' and the energy for the cold reservoirs (hot reservoir) is $E^{0}_{c}=n_c\text{Tr}(TH)$($E^{0}_{H}=n_H\text{Tr}(TH)$). Since we don't know the probability of getting cooling (or heating branch) before we calculate the trace of the final state of the working system after measurement of the control system, we can first assume that we attain cooling branch and update the state of the working system , then we can determine the probability of attaining cooling branch(see Eqn.(\ref{16})). For each cycle, we generate an uniformly distributed random number in the interval (0,1), when it is larger than $p_c$, it means that we attain heating branch this run, otherwise we stay in the cooling branch.

For cooling branch, in the $i$-$th$ cycle, the total energy of the $N$ cold reservoirs plus the working system is the total energy of these reservoirs in $(i-1)$-$th$ cycle plus the energy of the working system when attaining cooling branch. So it should be updated as $E^{i-1}_{c}+\text{Tr}(\rho^{i}_c H)$. The total energy and thermal state of the hot reservoir remain the same in this branch. Since for simplicity, we have Hamiltonian $H=\Delta\dyad{e}{e}$ for the thermal qubits, so the (2,2) element of the thermal state times $\Delta$ denotes the energy of each qubit(on average) with such thermal state. Then the (2,2) element of the thermal state sharing by all the qubits in the cold reservoirs and working qubit after the classical thermalisation between the working system and one of $N$ cold reservoirs followed by the classical thermalisation of all the cold reservoirs with one another is $T^{i}_c(2,2)=\frac{E^{i}_c}{\Delta (n_c+1)}$. And the last step $E^{i}_{c}=E^{i}_{c}-\text{Tr}(T^{i}_c H)$ is to update the total energy of the cold reservoirs without working system after the process.

When attaining one of $(N-1)$ heating branches, the state of the working system should be updated to the one in Eqn.(\ref{17}). The total energy of hot reservoir plus the working system in the $i$-$th$ cycle is then  $E^{i-1}_{H}+\text{Tr}(\rho_H H)$. The (2,2) element of the thermal state sharing by all the qubits in the hot reservoirs and working qubit after the classical thermalisation between the working system and hot reservoir is $T^{i}_H(2,2)=\frac{E^{i}_H}{\Delta (n_H+1)}$. Step $E^{i}_{H}=E^{i}_{H}-\text{Tr}(T^{i}_H H)$ is to update the total energy of the hot reservoir without working system after the process. And the total energy of the cold reservoirs plus the working system after it brings back some heat from the hot reservoir is $E^{i-1}_{c}+\text{Tr}(T^{i}_H H)$. The remaining steps are the same as those in cooling branch. The cooling cycle will be repeated until $m$ cycles are performed.

In order to see obvious temperature change of the cold reservoirs, the number of cycles needed to be performed should be of the similar order of the total number of particles (qubits) in the $N$ cold reservoirs. In Fig.\ref{2r0.01} and \ref{10r0.01}, we set $n_c=n_H=1000000$ and run the cooling cycle for 2000000 and 1000000 times for the ICO fridge with 2 and 10 cold reservoirs starting from $r_c=e^{-\Delta\beta_c}=0.01$ respectively. As we can compare, to cool down the cold reservoirs from $r_c=0.01$ to be around $r_c=0.0042$, 80$\%$ more cycles needed to be performed for ICO fridge with 2 cold reservoirs compared to the one with 10 cold reservoirs. And the ratio of their weighted energy changes is $\frac{\Delta\tilde{E^{10}}}{\Delta\tilde{E^{2}}}=1.8$. This result holds for any starting temperature and attainable final temperature of the cold reservoirs(see Fig.\ref{2r0.5} and \ref{10r0.5} for cold reservoirs start from $r_c=0.5$). And as the results in section \ref{wec} suggest, $\frac{\Delta\tilde{E^{10}}}{\Delta\tilde{E^{2}}}\approx 2$ for large enough $N$, so the cycles needed to be performed to cool down the cold reservoirs by applying ICO fridge with $N$ (sufficiently large) cold reservoirs to some attainable temperatures is around half of that of ICO fridge with 2 cold reservoirs starting from the same temperature(can compare Fig.\ref{2r0.01}, \ref{2r0.5}, \ref{100r0.01}, \ref{100r0.5}).

But even though the heat extracting ability of  ICO fridge with more cold reservoirs is greater, so fewer cycles are  needed to be performed to cool down the cold reservoirs to some attainable temperatures, it bears lower efficiency.
\begin{figure}
    \centering
	\includegraphics[width=0.5\textwidth]{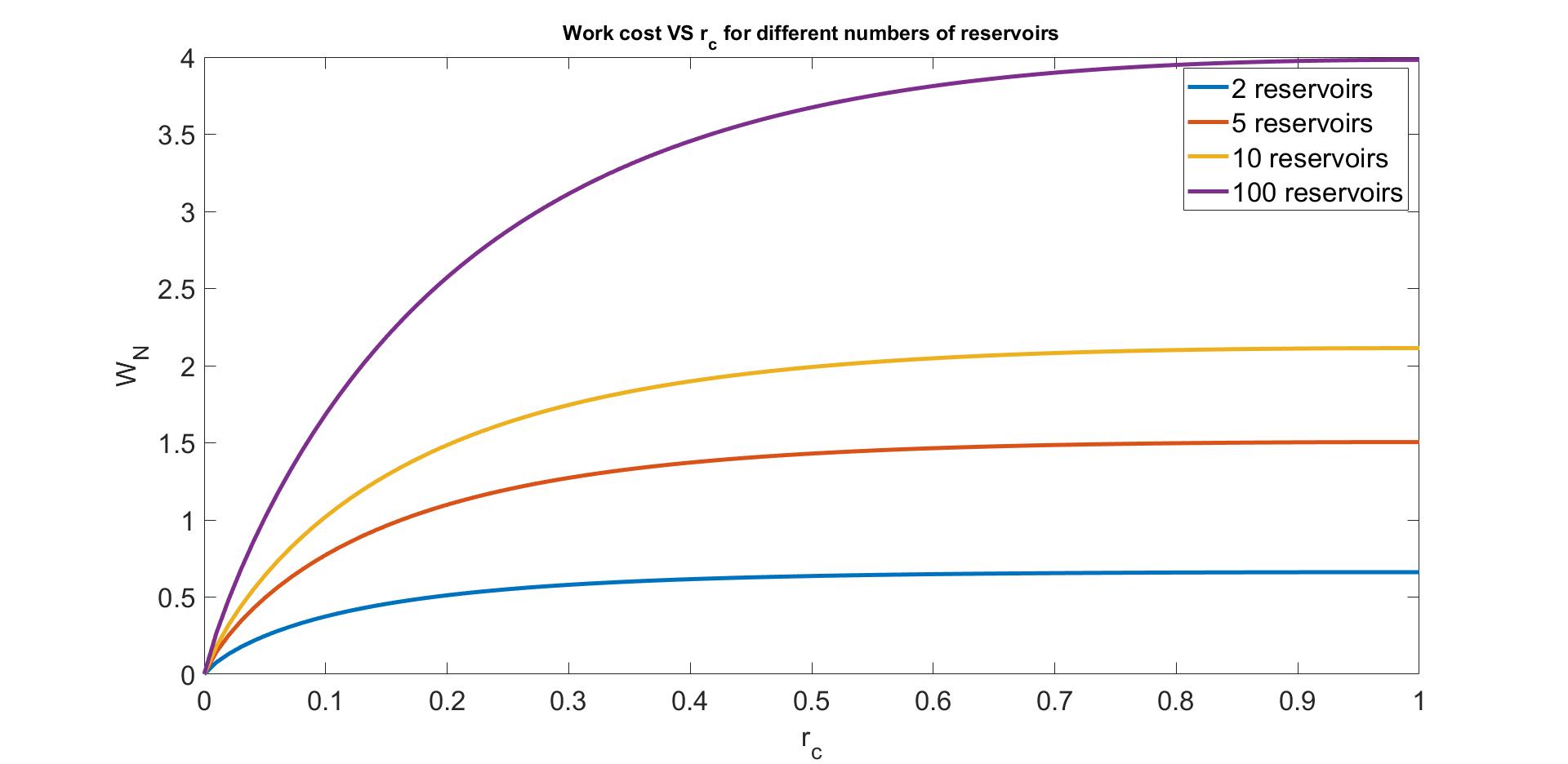}
	\caption{\textbf{Work cost(divided by $k_{B}T_{R}$) per cycle for erasing the registers with different dimensions for ICO fridges with different numbers of cyclic orders. Suppose we couple the registers with the reservoir with temperature $T_{R}$ for erasure.} }
	\label{costera}
\end{figure}
For example, the ICO fridges with 2 reservoirs starting from $r_c=0.01$ need around 80$\%$ more performed cycles to cool down the cold reservoirs to be around $r_c=0.0042$ compared to the one with 10 cold reservoirs. But for $r_c\in (0.0042,0.1)$, the ratio of work cost per cycle for ICO fridge with 10 and 2 reservoirs is ranging from 2.04 to 3.2 if we assume the registers are all coupled with the resetting reservoir with temperature $T_R$. The higher the temperature of the cold reservoirs, the larger this ratio is. This result explains why the heat extracting ability for the working system in ICO fridge with more cold reservoirs is greater but this fridge bears lower efficiency(smaller coefficient of performance). It should not surprise us since there is always a trade-off between power and efficiency for most of the thermal machines. 

\subsection{Lowest temperature cold reservoirs achievable with ICO fridge} \label{Lowestico}
The authors of \cite{felce2020quantum} provides a “positive refrigeration condition” for the ICO fridge, that is when the temperature of the hot reservoirs $T_{H}$ is equal to the effective temperature of the working system when attaining heating branch, the ICO fridge can't further cool down the cold reservoirs. Refer to situation discussed in Fig.\ref{f9}, the hot reservoir and all $N$ cold reservoirs are from the same superbath with a fixed temperature $T$, and initially we should let all of the reservoirs be thermally isolated from each other. As more and more cycles are performed, the cold reservoirs keep exchanging heat with the hot reservoir assisted by the ICO fridge. Even though $n_c$ (number of particles in all the cold reservoirs) and $n_{H}$ (number of particles in the hot reservoir) all tend to be infinity in thermodynamic limit, it is still reasonable to assume $\frac{n_{H}}{n_c}=k$(a positive constant). In this sense, we can derive the lowest temperature that the cold reservoirs can achieve with ICO fridge starting from a certain temperature, and it depends only on $k$ and the staring temperature as we will see.

Assume the maximum amount of heat that can be transferred from the cold reservoirs to the hot one is $Q_{max}$, $T'$ is the thermal state of the qubit from the cold reservoirs at the lowest achievable temperature and $\rho'_{h}$ is the normalised thermal state of the working qubit (initially in $T'$) after thermalising with the $N$ cold reservoirs in an ICO and attain one of the $(N-1)$ heating branch. And we know from the results of section \ref{cyclicN} that by applying the generalised N-SWITCH in cyclic orders, all the normalised thermal state of working qubit in heating branch is $N$ independent. 

We know that when $\rho'_{h}=T_{H}$ the fridge stop working in refrigeration mode. So we have $E^{f}_{cold}=n_c\text{Tr}[TH]-Q_{max}=n_c\text{Tr}[T'H]$ and $E^{f}_{H}=n_{H}\text{Tr}[TH]+Q_{max}=n_{H}\text{Tr}[\rho'_{h}H]$, then we can write

\begin{figure}[H]
    \centering
	\includegraphics[width=0.4\textwidth]{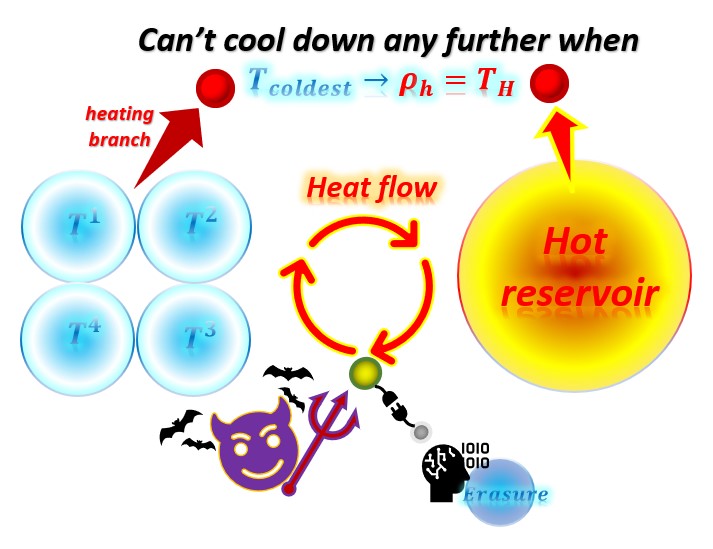}
	\caption{\textbf{Condition when the fridge stops operating. After the working system thermalised by the cold reservoirs in an indefinite causal order and we attain one of $N-1$ heating branch, if the thermal state of the working system at this point is the same as the one of the hot reservoir, the working system can't further release heat to the hot reservoir such that can't further cool down the cold ones.}}
	\label{f12}
\end{figure}

\begin{figure}
    \centering
	\includegraphics[width=0.5\textwidth]{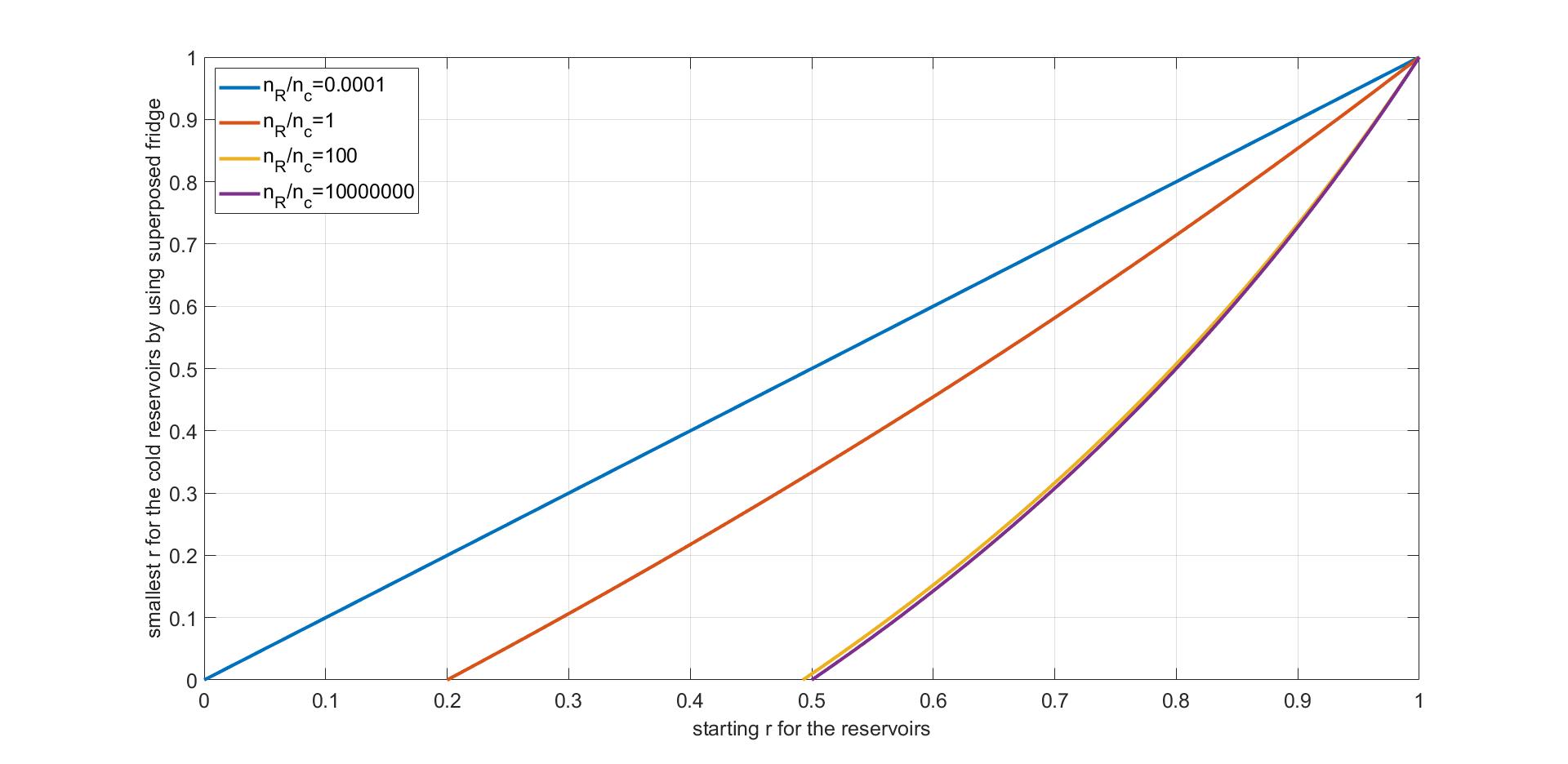}
	\caption{\textbf{Smallest r for the cold reservoirs can be obtain with the ICO fridge starting from a certain r. We can see that when the size of hot reservoir is relative big compare to the cold ones, the cold reservoirs can always be cooled down to arbitrarily close to absolute zero in principle starting from some not too high temperatures. But when all the cold reservoirs start from some relative high temperatures, the lowest temperature that the cold reservoirs can achieve is highly depending on the starting temperature.}}
	\label{f13}
\end{figure}

\begin{small}
\begin{equation*}
    n_c[\frac{r}{(1+r)}-\frac{r'}{(1+r')}]=n_{H}[\frac{2r'+1}{3(r'+1)}-\frac{r}{(1+r)}],
\end{equation*}
\end{small}and we can get the relation between $r$ and $r'$
\begin{small}
    \begin{equation}\label{31}
    r'=\frac{k-(2k+3)r}{kr-3-2k}.
\end{equation}
\end{small}where $k=\frac{n_{H}}{n_c}$. As we can see in Fig.\ref{f13}, when $k$ is small, the ICO fridge can hardly cool down the reservoirs, Fig.\ref{2chancant} gives an example for $k=0.1$ with cold reservoirs starting from $r_c=0.1$. But when $k$ is getting larger and larger, the highest initial temperature which allows the ICO fridge to cool down the reservoirs to be sufficiently close to absolute zero is becoming higher and higher, but
there is still a upper bound of it (around $r=0.5$) which suggests that the lowest achievable temperature is lower-bounded by some values larger than absolute zero for reservoirs start at $r_i>0.5$ no matter how large $k$ is. Fig.\ref{0.8lowest} gives an example for $k=1000000$ with cold reservoirs starting from $r_c=0.8$, as we can see, even though around $10n_c$ times cycles are performed, the smallest $r_c$ can be attained is around 0.5 and the cold reservoirs can't be further cooled down which agrees with the predictions we make in this section (Fig.\ref{13}). Take the hyperfine working qubit with $\Delta\approx 1.99\times 10^{-24}J$ in trapped ion platform as an example, when cold reservoirs start from $r_c=0.999603$ (the corresponding effective temperature is $T^{i}_c\approx 288.33K$), the lowest achievable temperature for the cold reservoirs by making use of ICO fridge is $T^{f}_c\approx 120.10K$. As the results in this section suggest, for cold reservoirs start from $r_c\lesssim 0.5$, the lower bound for the lowest attainable tempeature is absolute zero. For the hyperfine qubit working system, when $r_c=0.5$, $T_c\approx 0.21\sim 0.69K$, but for Rydberg atom working system, the effective temperature is around $8.4\times 10^{4}K$ when $r_c=0.5$. So physical implementation using Rydberg atom exhibits much greater advantage in this sense, since ICO fridge utilising this working system can cool down the cold reservoirs starting from some very high temperatures to temperature which is arbitraily close to absolute zero in principle.

The results in this section place a bound on the lowest achievable temperature for the cold reservoirs starting at some fixed temperatures by utilizing the ICO fridge, and it is N-independent but it is related to the ratio of the numbers of particles in hot and the $N$ cold reservoirs and the starting temperature of the cold reservoirs.

\section{Controlled-SWAPs circuit protocol without ICO}\label{ConSWAP}

\subsection{Quantum-controlled thermalising channels}\label{conthem}
\begin{figure}\label{spt}
    \centering
	\includegraphics[width=0.45\textwidth]{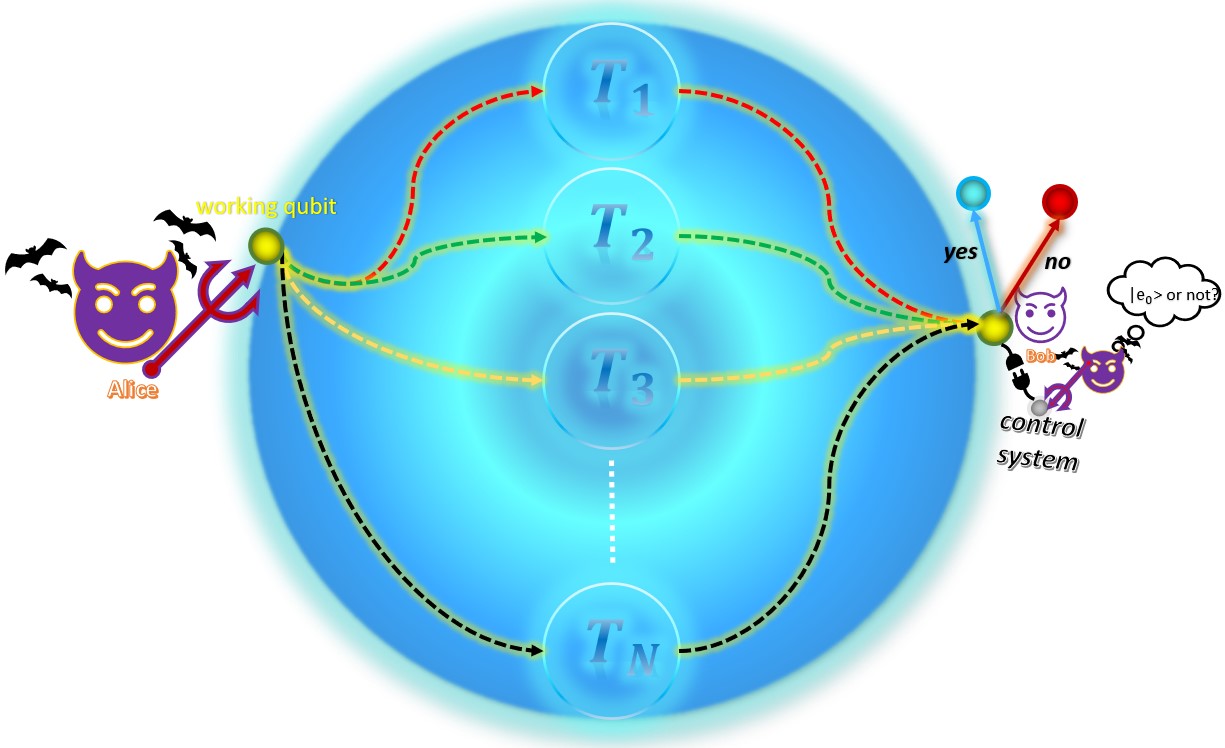}
	\caption{\textbf{Thermalisation in superposing trajectories. Similar to the procedure described in Fig.\ref{f9}, but the only different is that in step (1) Alice just thermalises the working system with one of $N$ reservoirs with the same temperature assisted by the control system.}}
\end{figure}
In \cite{felce2020quantum}, the author introduces a unitary quantum circuit acting on the control and target systems, as well as the thermal reservoir qubits, which can produce the same final marginal state for the working system as the Quantum SWITCH of two thermalising channels.
In this section, we generalise the framework to N reservoirs with the same temperature and show that this quantum cooling scheme outperforms the quantum switch scheme (where we don't have access to the reservoir qubits which are quantum correlated with the target system) for all temperatures and N. We go on to present an alternative unitary circuit in which causal indefiniteness plays no role, and show that this produces the same control-target-reservoir state as the circuit with ICO, while bearing much lower complexity.

At first we should take the quantum-controlled interactions between the working qubit and the reservoir\cite{wood2021operational} into account. We consider a composite reservoir comprising $N$ subsystems in a product state $\rho_{R} = \bigotimes^{N-1}_{i=0}\rho_{R_{i}}$.The whole system including the control system is also in a product state initially $\rho=\rho_{R}\otimes\rho_{C}\otimes\rho_{t}$. The interaction between the reservoirs and the working system controlled by the control degree of freedom is then
\begin{widetext}

\begin{figure}
    \centering
	\includegraphics[width=0.7\textwidth]{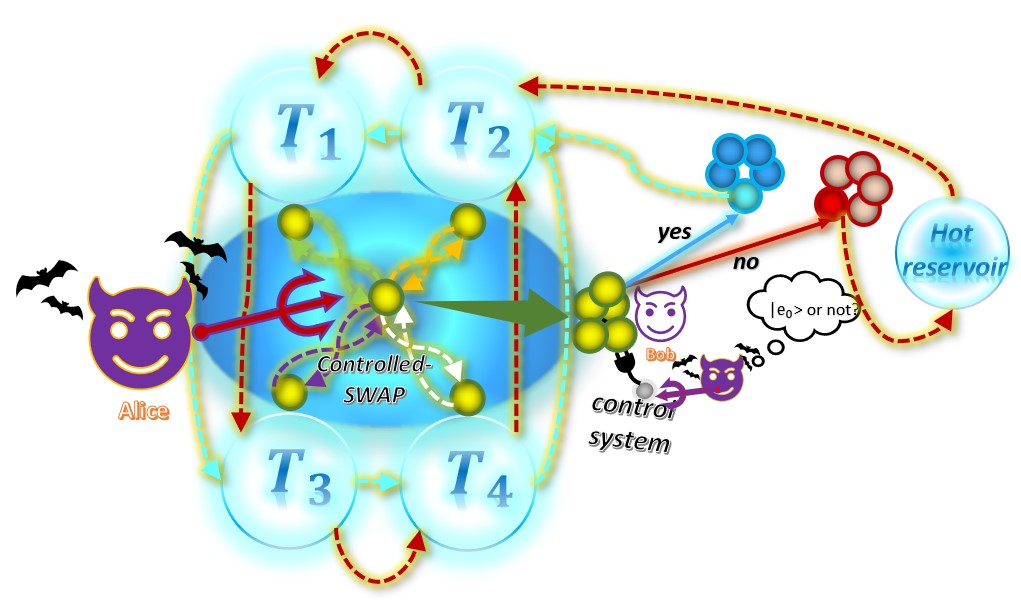}
	\caption{\begin{small}
	    \textbf{Controlled-SWAPs of thermal qubits. Similarly, most of the steps of this quantum cooling scheme are the same as the one in Fig.\ref{f9}. But in step (1) Alice swaps the working qubit coherently with one of $N$ reservoir qubits($N=4$ in this figure), each of them is randomly taken from the one of $N$ cold reservoirs with the same temperature. When attaining cooling branches, $N+1$ qubits (one target qubit and $N$ reservoir qubits) are all cooled down, even though the reservoir qubits have different margin state compared to the target qubit except when $N=2$, they can still serve as cooling resources. In step (3), Bob thermalises all the cooled down qubits with one of the $N$ cold reservoirs classically followed by the classical thermalisation of all the cold reservoirs with one another. At the end all the cold reservoirs and working mediums (including the $N$ reservoir qubits) share the same lower temperature and thermal state. Actually the amount of heat that can be reduced by the working mediums for the cold reservoirs is the same if Bob thermalises the qubits with the cold reservoir one after another, see Appendix \ref{msp}). But if heating branch is attained, Bob needs to thermalise all the working qubits with the hot reservoir classically and then with one of the cold reservoirs followed by classical thermalisation of all the cold reservoirs with one another. All the other steps are the same as those in \ref{f9}.}
	\end{small}}
	\label{cd}
\end{figure}

\end{widetext}

\begin{footnotesize}
    \begin{equation}\label{32}
\begin{split}
    &U=U_{R_{0}t}\otimes I_{R_{1}}\otimes\cdots \otimes I_{R_{N-1}}\otimes\dyad{0}{0}_{C}\\
    &+I_{R_{0}}\otimes\cdots U_{R_{k}t}\cdots \otimes I_{R_{N-1}}\otimes\dyad{k}{k}_{C}+\cdots\\
    &+I_{R_{0}}\otimes\cdots\otimes U_{R_{N-1}t}\otimes\dyad{N-1}{N-1}_{C},
\end{split}
\end{equation}
\end{footnotesize}this is the quantum-controlled interaction between the working system and the reservoir $R_{i}$  via a unitary $U_{R_{i}t}$ depending on the state of the control system.

The scheme of quantum-controlled thermalisation introduced in \cite{wood2021operational} is sensitive to the full process on the Kraus decomposition of the individual thermalising channel, but we notice that the swapping operation decribed in \cite{felce2020quantum} is a particular target-reservoir interaction leading to a specific Kraus representation of the thermalising channel, and it obviously satisfies $Tr_{R_{i}}\{U_{R_{i}t}\rho_{R_{i}t}U^{\dagger}_{R_{i}t}\}=\rho^{\beta_{i}}_{t}$ (the working system can be fully thermalised by the reservoir and attains the same thermal state as the reservoir's with effective inverse temperature $\beta_{i}$). And this unitary can be expressed as
\begin{equation}\label{33}
    U_{R_{i}t}=\sum_{k,l}\dyad{k}{l}_{t}\dyad{l}{k}_{R_{i}},
\end{equation}
with the corresponding Kraus operators $M_{kl}=\ket{l}_{t}\bra{k}_{t}$. And for the qubit working system case it is just the SWAP gate for the interacting two parties.

\subsection{Quantum-controlled-SWAPs of thermal qubits}\label{quanSWAP}

In our case since all the reservoirs(plus the working system) are at the same temperature initially, so we have $\rho_{R_{i}}=\sum_{l}c^{\beta}_{l}\ket{l}\bra{l}=T=\frac{1}{r+1}\dyad{0}{0}+\frac{r}{r+1}\dyad{1}{1}$. Taking the initial state of the control to be $\ket{e_{0}}=\frac{1}{\sqrt{N}}\sum^{N-1}_{k=0}\ket{k}$, we have the final state (including the qubits from reservoirs) when attaining cooling branch
\begin{small}
\begin{equation}\label{34}
\begin{split}
    &\bra{e_{0}}U\rho U^{\dagger}\ket{e_{0}}\\
    &=\frac{1}{N^2}(\sum^{N-1}_{i=0}U_{R_{i}t}\rho_{Rt}U^{\dagger}_{R_{i}t}+\sum^{N-1}_{i=0}U_{R_{i}t}\rho_{Rt}\sum^{N-1}_{i'\neq i}U^{\dagger}_{R_{i'}t}).
\end{split}
\end{equation}
\end{small}

To take a closer look at the unitary evolution (swapping) of the control-target-reservoirs (qubits) system, we need to first purify the same thermal state shared by the working system and reservoir qubits initially
\begin{equation}\label{35}
    \ket{T^{\beta}}=\sum^{0}_{a_{i}=0}\sqrt{p_{a_{i}}}\ket{a_{i},a_{i}},
\end{equation}where $p_{a_{i}}=\frac{1}{1+r}$ or $\frac{1}{1+r}$ and $i$ denotes from which reservoir the qubit is from (0 denotes the working qubit). So the initial target-reservoirs (qubits) system (including the purification ancilla qubits) is \begin{equation*}
    \ket{T^{\beta}_{0}}=\sum_{a_{0},a_{1}\cdots a_{N}}\sqrt{p_{a_{0}}p_{a_{1}}\cdots p_{a_{N}}}\ket{a_{0}a_{0}\cdots a_{N}a_{N}}.
\end{equation*}
we can define $U^{k}_{SW}$ as
\begin{equation*}
    U^{k}_{SW}\ket{a_{0}a_{0}\cdots a_{N}a_{N}}=\ket{a_{k}a_{0}\cdots a_{0}a_{k}\cdots a_{N}a_{N}},
\end{equation*}
so the control-target-reservoirs (qubits) system evolves unitarily as
\begin{tiny}
\begin{equation}\label{36}
\begin{split}
    &\ket{e_0}\otimes\ket{T^{\beta}_{0}}\\
    &\to \ket{T^{\beta}_{f}}= \frac{1}{\sqrt{N}}\sum_{a_{0}\cdots a_{N}}\sqrt{p_{a_{0}}\cdots p_{a_{N}}}\sum^{N-1}_{k=0}\ket{k}\otimes U^{k+1}_{SW}\ket{a_{0}a_{0}\cdots a_{N}a_{N}},
\end{split}
\end{equation}
\end{tiny}
we then have
\begin{small}
    \begin{equation}\label{37}
    \bra{e_0}Tr_{anc,R_{1}\cdots R_{N}}[\ket{T^{\beta}_{f}}\bra{T^{\beta}_{f}}]\ket{e_0}=\frac{1}{N}(T+(N-1)T^3).
\end{equation}
\end{small}as the final local Gibbs state (before normalization and it is the same as the one produced by the N-SWITCH ICO fridge) of the working system when attaining cooling branch. And the reservoir qubits share the same local Gibbs state (before normalization)
\begin{footnotesize}
    \begin{equation}\label{38}
    \rho^{c}_{R}=\frac{(N-1)(N-2)(r^3+1)+N(r+1)^3}{N^2(r+1)^3}T+\frac{2(N-1)}{N^2}T^3.
\end{equation}
\end{footnotesize}see Appendix \ref{mgs} for more details about the derivation of this part. Similarly, by following the general measurement strategy in Appendix \ref{Mstra}, we can see that the working qubit has the same final local Gibbs state $\frac{1}{N}(T-T^3)$ when we attain one of the $N-1$ identical heating branches(measurement result of the control is other than $\ket{e_{0}}$).

When attaining the cooling branch (see Fig.\ref{cd}), Bob thermalises the working qubit with the cold reservoirs, and Alice passes the reservoir qubits to Bob one after another and let Bob thermalise them with the cold reservoirs. At each step, the target (or reservoir) qubit discards from the quantum correlated $N+1$ qubits system, and the local Gibbs states of the remaining qubits remains the same as in Eqn.(\ref{38}). Or Alice can give Bob all the cooled down qubits and let him thermalise the $N+1$ working qubits with the cold reservoirs at once, the amount of energy decrease in total for the cold reservoirs is the same as the case when Bob thermalises the qubits with the cold reservoirs one after another, this is because the thermalisation of 1 of $N+1$ subsystems of the quantum correlated system is local-Gibbs-state-preserving for the remaining qubits (see Appendix \ref{msp} for more discussion about this part).

\begin{figure}\label{sw1b1}
    \centering
	\includegraphics[width=0.5\textwidth]{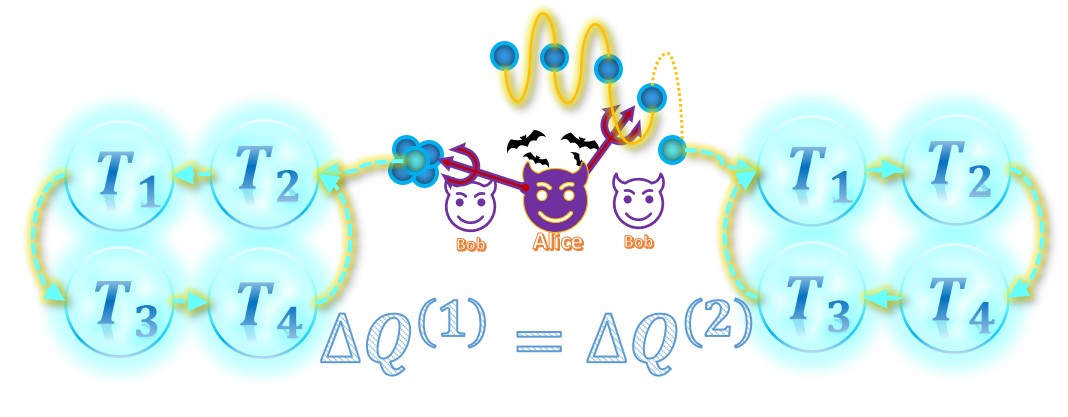}
	\caption{\textbf{Thermalising all the cooled down qubits at once with the cold reservoirs is thermodynamical equivalent to thermalise them one after another in quantum cooling task. The amount of heat that can be reduced by the working mediums for the cold reservoirs when attaining cooling branch is the same, no matter Bob thermalises all the qubits at once or one after another with the cold reservoir in step (3) of Fig.\ref{cd}, because thermalisation of one of $N+1$ quantum correlated working mediums in quantum-controlled-SWAPs scheme with reservoir (with thermal state $T$) is a local-Gibbs-state-preserving process for the remaining correlated subsystems, see Appendix \ref{msp}.}}
\end{figure}\medskip

\subsection{Advantages of Quantum-controlled-SWAPs ciruit-based fridge}\label{SWAPadv}

\begin{widetext}

\begin{figure}
    \centering
	\includegraphics[width=1\textwidth]{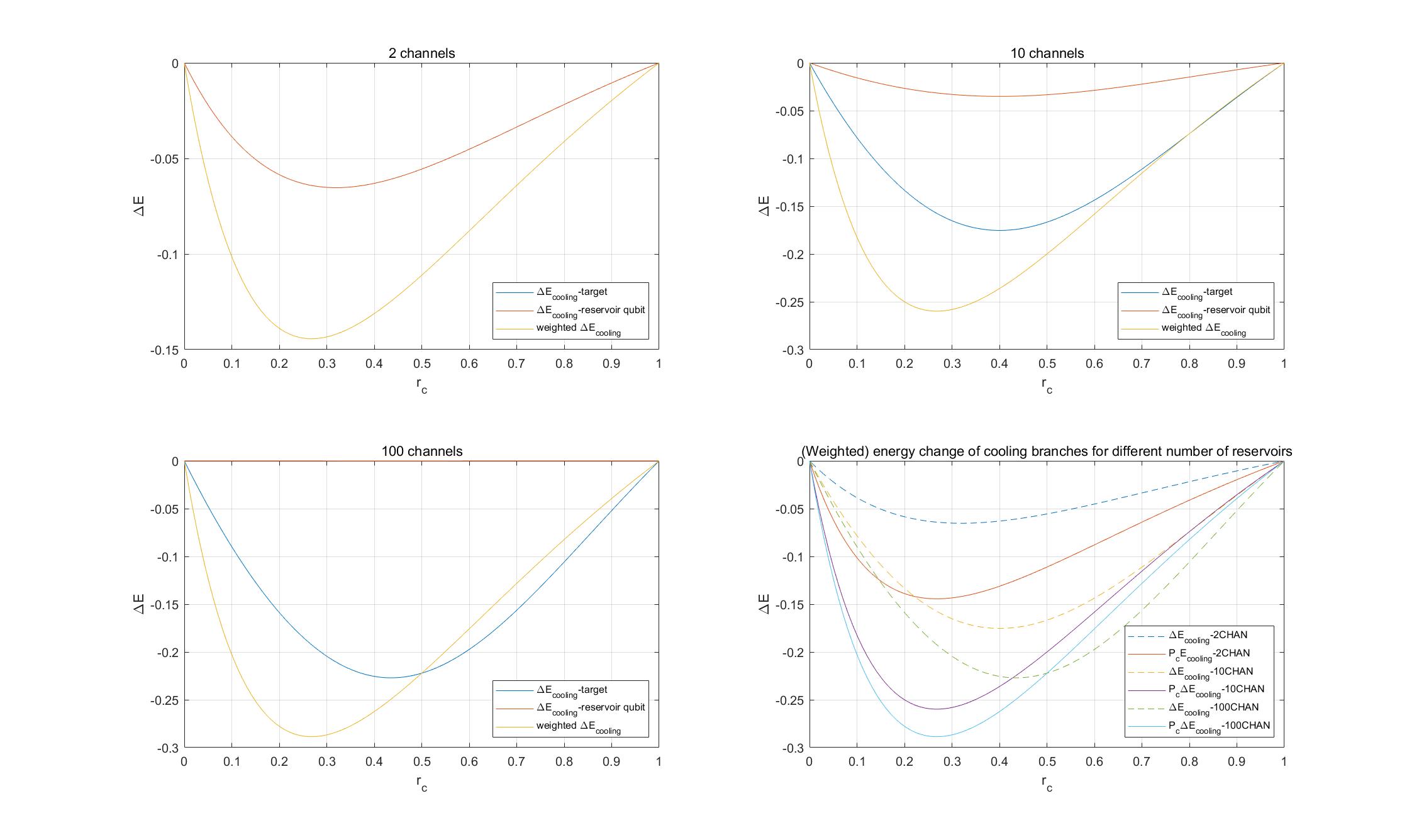}
	\caption{\textbf{How the weighted energy changes vary with the number of thermalising channels for controlled-SWAPs circuit based fridge without ICO. In the $N=2$ case the reservoir qubits share the same local Gibbs state as the target system when attaining cooling branch. Even though when $N$ becomes larger and larger, the heat extracting ability of each reservoir qubit decrease, but on average the total heat that can be extracted by all the working mediums per cycle increases greatly. The dash lines in the last plot denote the amount of heat that can be transferred by the target qubit when attaining cooling branch, as we can compare, when we utilise the reservoir qubits, the weighted energy change is even larger than the solely energy change of the target system (can see Fig.\ref{f8} for the comparison of these two terms).} }
	\label{f17}
\end{figure}

\end{widetext}

As we can see in Fig.\ref{f17}, when $N$ goes up, the heat that can be extracted by each reservoir qubit when we attain cooling branch decreases (the special case is $N=2$ when all the reservoir qubits share the same local Gibbs state as working system), but on average, the amount of heat that can be extracted by all the qubits (target qubit and all the reservoir qubits) increases greatly. Similar to the case of ICO fridge (the only working medium is the target qubit), there is an upper bound for the total amount of heat that can be extracted on average, which makes sense since the heat that can be extracted by each reservoir qubit decreases rapidly when attaining cooling branch as $N$ increases. And as we can calculate, when attaining cooling branch
\begin{small}
    \begin{equation}\label{3t}
\begin{split}
    N(\text{Tr}[\frac{\rho^{c}_{R}}{\text{Tr}(\rho^{c}_{R})}H]-\text{Tr}[TH])=2(\text{Tr}[\rho_{c}H]-\text{Tr}[TH]).
\end{split}
\end{equation}
\end{small}where $\rho_{c}$ is the final local Gibbs state of target system after normalization when attaining cooling branch. In summary, as we can compare from Fig.\ref{f17} and Fig.\ref{f8}, the weighted energy change and COP (for the ideal case when $T_c=T_{H}$) for the fridge based on quantum-controlled-SWAPs circuit is \textit{tripled} in general compared to those of the ICO fridge with N-SWITCH while target qubit is the only working medium. 

\begin{figure}
\centering
	\includegraphics[width=0.4\textwidth]{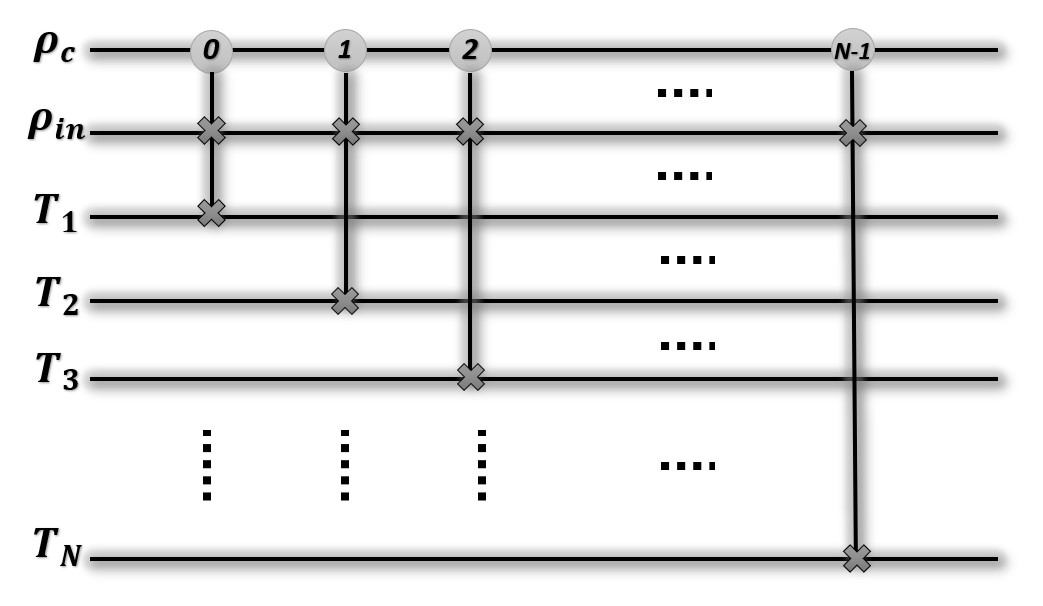}
	\caption{\textbf{Controlled-SWAPs circuit without ICO. As we can compare with Fig.\ref{f16}, the circuit complexity is much lower than the one with indefinite causal order, but it produces the same cooling resources for quantum cooling in the quantum-controlled-SWAPs scheme.}}
	\label{f15}
\end{figure}

\begin{figure}
\centering
	\includegraphics[width=0.4\textwidth]{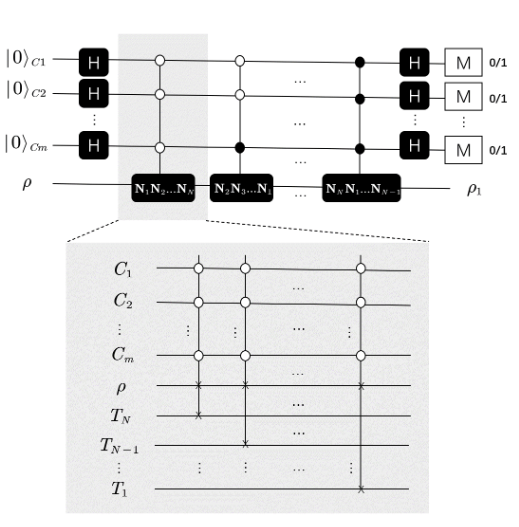}
	\caption{\textbf{Controlled-SWAPs circuit with ICO for the multi-qubit control system.}}
	\label{f16}
\end{figure}

Moreover, an very obvious advantage of this quantum refrigerator compared to the ICO one is its low circuit complexity. Actually by applying the controlled-SWAPs circuit with superposition of cyclic causal orders in Fig.\ref{f16} we can get the same final state (for the whole control-target-reservoirs (qubits) system, not just the local Gibbs state for the target system) as the one without ICO. So the fridges based on these 2 circuits have the same performance when the reservoir qubits are utilised. But the circuit complexity of the one without ICO is way lower than the one with ICO especially when $N$ is large, the gates needed for the one without ICO is $\frac{1}{N}$ of the controlled-SWAPs circuit with ICO.

Similar to the exchange between the authors of \cite{kristjansson2020resource,abbott2020communication}, reasonable comments can be made for the results we showed in previous section suggesting that ICO fridge assisted by generalised N-SWITCH with more alternative causal orders and reservoirs can boost the heat extracting ability of the working system on average. Because for ICO fridges with different numbers of reservoirs, the numbers of thermalising channels applied in sequence for each run of cycle are different, so it is not very clear whether the increase of the amount of heat that can be extracted on average for the cooling task origins from the more complex interference pattern arises from superposition of more causal orders or just simply because the working system thermlises with more reservoirs each run(but in Maxwell-demon-like scenario describe in section \ref{Maxdem}, you will see that there are things $N=2$ ICO fridge can't do no matter how many runs we repeat the cycle but it is attainable by utilising ICO fridges with sufficiently large $N$). However, for the quantum-controlled thermalisation in controlled-SWAPs scheme, we can be sure that the enhancement of the heat extracting ability of the work systems arises from the more complex interference pattern instead of interacting with more reservoirs each run because the working qubit interacts with \textit{only one} reservoir qubit (swaps with one of $N$ reservoir qubits depending on the state of the control) no matter how many reservoirs with same temperature we have.

Another important thing worth noticing is that even though controlled-SWAPs circuit-based thermalisation process produces the same marginal final target state as the quantum SWITCH of thermalising channels as described in the previous section, the accessible resources are different for these two quantum cooling schemes. As the results in this section suggest, the total quantum correlations generated during the process should be the multi-party ones within the reservoirs-target-control system instead of the two-party ones between the composite control-target system. In general, thermalisation can be regarded as thermal randomization which makes use of the randomness of the thermal states in a reservoir (in thermal equilibrium)\cite{maruyama2009colloquium}. For a target state in contact with a reservoir, it will changes gradually after interacting(colliding) with the reservoir, and sufficiently many interactions make it indistinguishable from the thermal state of the reservoir. Similarly, the quantum correlations generated during the quantum-controlled thermalisation (or in an indefinite causal order) process between the cold reservoirs and the control-target system will become untraceable (thermalise away) for the full thermalisation case. 

In the quantum-controlled-SWAPs scheme, for each run of the cooling cycle, Alice randomly takes one qubit from each cold reservoir, and after the quantum-controlled (or ICO) thermalisation and the follow-up measurement of the control system, all the qubits (working system and those from the reservoirs) are cooled down when we attain cooling branch (even though they don't share the same local Gibbs state when $N\neq 2$). In this sense, all the qubits can serve as working mediums in the cooling process. At the end of each cycle, Alice puts all the qubits from the reservoirs back and again randomly takes one qubit from each reservoir, then starts a new cycle. This is very different from the case when we prepare more than one working system and run the cycle for all of them simultaneously, because in the latter situation we need the same number of bits in the register (memory) to store the measurement results.But we should notice that the operational requirements are different for the thermalisation scheme of quantum-controlled-SWAPs and the quantum SWITCH of thermalising channels. The controlled-SWAPs one requires precise quantum control over the working and reservoirs qubits, so the reservoir qubits which are quantum correlated with the target system are fully accessible for Alice since she knows exactly which qubit (from each reservoir) the working system interacts with. But for quantum SWITCH of thermalising channels, the final accessible entangled state is just the control-target one.  

\begin{figure}
	\includegraphics[width=0.5\textwidth]{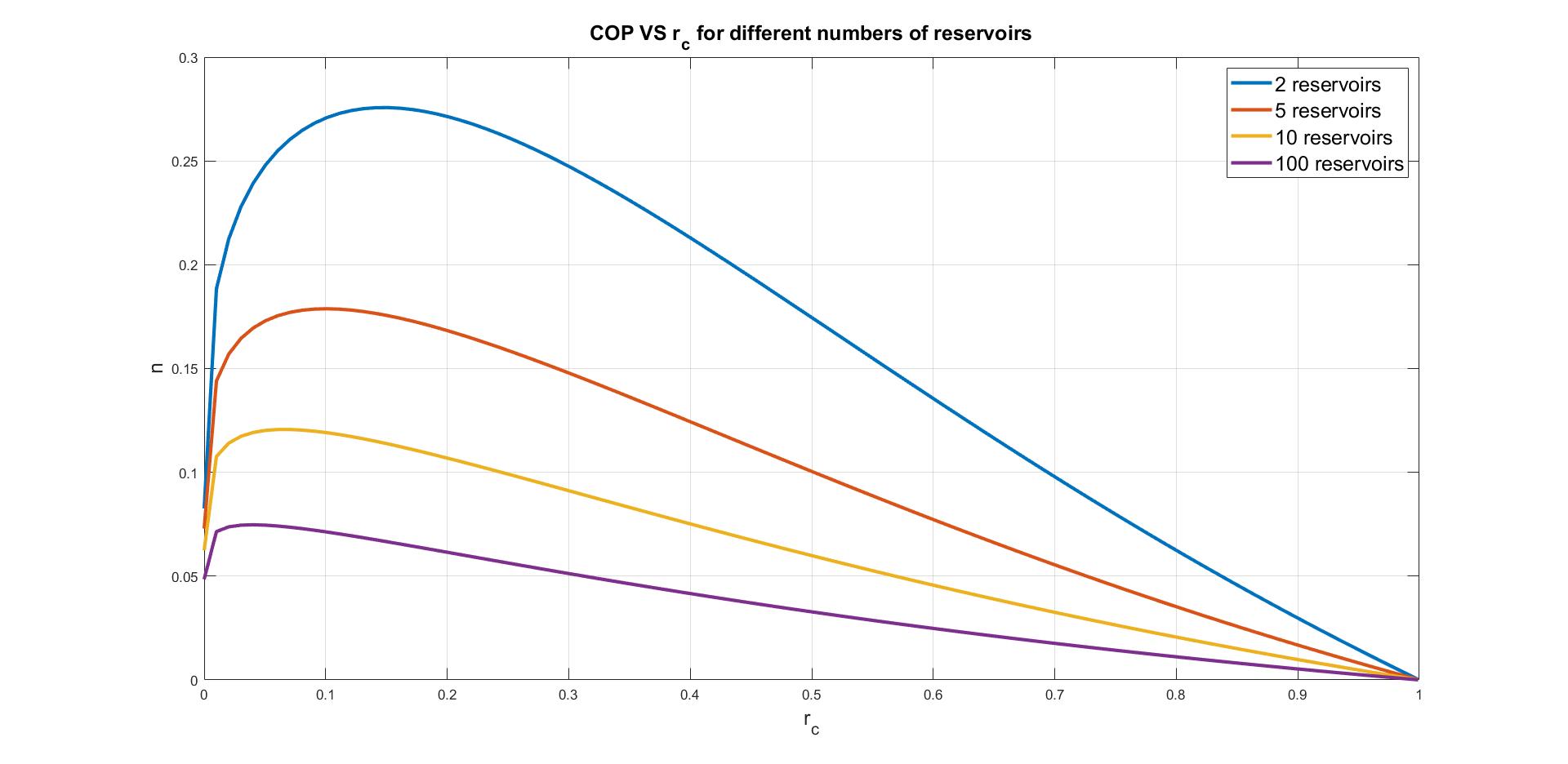}
	\caption{\textbf{COP(divided by $\Delta \beta_{R}$) VS $r_{c}$ for controlled-SWAPs circuit based fridges without ICO with different numbers of reservoirs for the optimal case when $T_{c}=T_{H}$. And it is tripled compared to the fridge assisted by generalised N-SWITCH but only the control-target system is accessible, as we can compared with Fig.\ref{f12}}.}
	\label{f18}
\end{figure}

\section{Experimentally simulatable quantum cooling protocol with coherently-controlled thermalising channels}\label{expfri}
Before the discussion about quantum-controlled thermalisation in \cite{wood2021operational}, authors of \cite{abbott2020communication} came up with scenarios where causal indefiniteness plays no role but similar advantages can be achieved in communication with noisy channels (although  \cite{kristjansson2020resource} suggests since in this scheme the target system only passes through one of the noisy channels instead of a sequence of them, so in a more fair comparison, the advantage provided by quantum switch is still the greatest in this kind of tasks). But different from the protocol with in
\begin{figure}[H]
    \centering
	\includegraphics[width=0.5\textwidth]{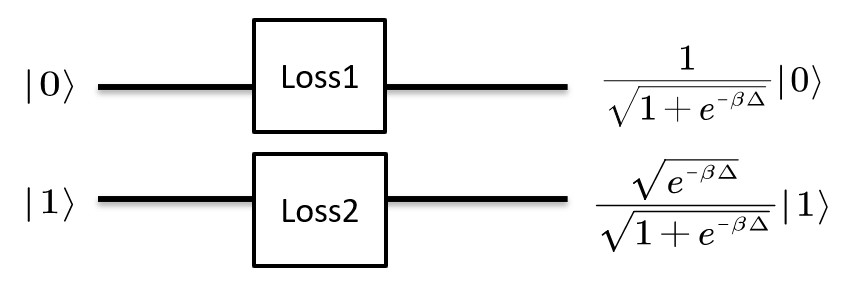}
	\caption{\textbf{thermalising channel can be decomposed as a fully depolarizing channel followed by an amplitude damping operation as shown.}}
	\label{f20}
\end{figure}

definite causal order, the new scheme of communication with superposing trajectories is implementation-dependent (additional transformation

\begin{figure}
\centering
	\includegraphics[width=0.5\textwidth]{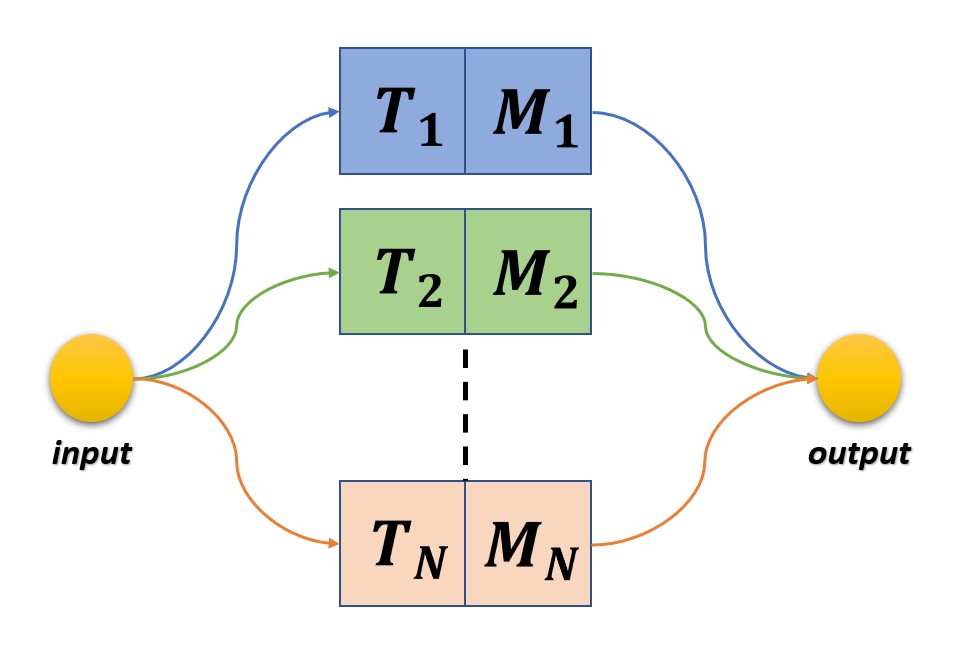}
	\caption{\textbf{Quantum cooling via coherently-controlled thermalising channels which is implementation-dependent, transformation matrices are needed for the evaluation of the fridge based on this scheme.}}
	\label{f19}
\end{figure}
matrices\cite{abbott2020communication} or access to vacuum extension\cite{chiribella2019quantum} with specific Kraus operators are needed).

Following the same spirit, here we demonstrate an experimental simulatable quantum cooling protocol with coherent control of the of $N$ identical thermalising channels, and we notice that it can  outperform the one with quantum N-SWITCH (in the case when Alice only has access to the final entangled control-target system) for some particular implementations. Even though the fridge with thermalisation in superposing trajectories is not always giving the better performance for all the implementations of thermalising channels compared with the ICO one (the ICO scheme is implementation-independent\cite{abbott2020communication}), the result sheds the light on the search of thermal machine assisted by thermalisation in superposition of quantum trajectories which has better performance but much lower complexity compared to the one with ICO. 

\subsection{Quantum cooling with implementation-dependent coherently-controlled thermalising channels}\label{qcimld}
In order to evaluate the composite control-target-environments state's unitary evolution in the framework introduced in \cite{abbott2020communication}, besides the purification of the initial thermal state of the working system (Eqn.(\ref{35})), we also need to purify the thermalising channels via Stinespring dilations\cite{stinespring1955positive}. By introducing an environment in an initial state $\ket{\epsilon_{a_{i}}}^{e_{a_{i}}}$ for the i-th thermalising channel with Kraus operators $\{K^{i}_{a_{i}}\}$, the unitary operation acts on the target-environment gives $\ket{\psi_{in}}\otimes\ket{\epsilon_{a_{i}}}^{e_{a_{i}}}\to \sum_{a_{i}}K^{i}_{a_{i}}\ket{\psi_{in}}\otimes\ket{a_{i}}^{e_{a_{i}}}$, where $\ket{a_{i}}^{e_{a_{i}}}$ are orthogonal states of the environment used to purify the $i$-$th$ thermalising channel, the states of environments ($\ket{\epsilon_{a_{i}}}^{e_{a_{i}}}$) for the different thermalising channels are initially uncorrelated. Thus the control system controls the action of the purified unitary extensions of the channels and the initially thermalised target system.

Under the control of the control system, the composite purified control-target-environments state evolves to
\begin{widetext}

\begin{footnotesize}
\begin{equation}\label{39}
\begin{split}
    &\ket{e_{0}}\otimes(\frac{1}{\sqrt(r+1)}\ket{0,0}+\sqrt{\frac{1}{(r+1)}}\ket{1,1})\otimes\ket{\epsilon_{a_{0}}}^{e_{a_{0}}}\otimes\cdots\ket{\epsilon_{a_{N-1}}}^{e_{a_{N-1}}}\\
    &\to\frac{1}{\sqrt{N}}\ket{0}\otimes\sum_{a_{0}}(K^{0}_{a_{0}}\otimes I)(\frac{1}{\sqrt(r+1)}\ket{0,0}+\sqrt{\frac{1}{(r+1)}}\ket{1,1})\ket{a_{0}}^{e_{a_{0}}}\otimes\ket{\epsilon_{a_{1}}}^{e_{a_{1}}}\cdots\otimes\ket{\epsilon_{a_{N-1}}}^{e_{a_{N-1}}}+\cdots+\\
    &\frac{1}{\sqrt{N}}\ket{k}\otimes\sum_{a_{k}}(K^{k}_{a_{k}}\otimes I)(\frac{1}{\sqrt(r+1)}\ket{0,0}+\sqrt{\frac{1}{(r+1)}}\ket{1,1})\ket{\epsilon_{a_{0}}}^{e_{a_{0}}}\otimes\ket{\epsilon_{a_{1}}}^{e_{a_{1}}}\otimes\cdots\otimes\ket{a_{k}}^{e_{a_{k}}}\cdots\otimes\ket{\epsilon_{a_{N-1}}}^{e_{a_{N-1}}}\\
    &+\cdots+\frac{1}{\sqrt{N}}\ket{N-1}\otimes\sum_{a_{N-1}}(K^{N-1}_{a_{N-1}}\otimes I)(\frac{1}{\sqrt(r+1)}\ket{0,0}+\sqrt{\frac{1}{(r+1)}}\ket{1,1})\ket{a_{0}}^{e_{a_{0}}}\otimes\ket{\epsilon_{a_{1}}}^{e_{a_{1}}}\otimes\cdots\otimes\ket{a_{N-1}}^{e_{a_{N-1}}}.
\end{split}
\end{equation}   
\end{footnotesize}

\end{widetext}
after tracing out all the environments including the ancillary system used to purify the thermal state of target system, the final control-target state $\rho^{ct}_{out}$ is then
\begin{footnotesize}
    \begin{equation}\label{40}
    \rho^{ct}_{out}=\frac{1}{N}[I^{c}_{d\times d}\otimes N^{T}(\rho^{t}_{in})+\sum^{N-1}_{k=0}\sum^{N-1}_{k'\neq k}\dyad{k}{k'}_{c}\otimes M_{k}\rho^{t}_{in}M^{\dagger}_{k'}].
\end{equation}
\end{footnotesize}where $M_{k}=\sum_{a_{k}}\langle\epsilon_{a_{k}}|a_{k}\rangle K^{k}_{a_{k}}$ is the transformation matrix for $k$-$th$ thermalising channel with a specific implementation(we use M instead of T to denote transformation matrix becasue we already assigned $T$ as thermal state). But not all transformation matrices are obtainable from some implementations of thermalising channel $N^{T}$, the constraint is $M_{N^{T}}=\{M:Tr(M^{\dagger}N^{T}M)\leq\frac{1}{d}\}$ for the d-dimensional thermalising channel $N^{T}$ (can refer to Appendix A of \cite{abbott2020communication}). We focus on the 2-dimensional thermalizng channel (qubit working system) for the discussion of this part.

\subsection{Implementation of thermalising channel and the performance of the superposed refrigerator }\label{impfri}

The implementation of a thermalising channel can be decomposed as a fully depolarizing channel followed by an amplitude damping channel with some specific parameters (depend of the thermal state of the thermalizng channel). With the amplitude damping operation defined as $A=\sqrt{T}=\sqrt{\frac{1}{1+e^{-\Delta\beta}}}\dyad{0}{0}+\sqrt{\frac{e^{-\Delta\beta}}{1+e^{-\Delta\beta}}}\dyad{1}{1}$ in our case.

In this sense, for the thermalising channel with a fixed effective temperature $T$, the implementation of $N^{T}$ depends on how we realise the fully depolarizing channel. If we employ the uniform randomization over 4 (qubit case) unitary channels strategy, we get sets of Kraus operators for each thermalising channel $\{K^{i}_{a_{i}}=\frac{1}{\sqrt{2}}AU_{a_{i}}\}_{a_{i}}$. Especially, when we take $K^{i}_{0}=\frac{1}{\sqrt{2}}AI$ for all N identical channels and the initial environment states to be $\ket{\epsilon_{a_{0}}}^{e_{a_{0}}}=\ket{\epsilon_{a_{1}}}^{e_{a_{1}}}=\cdots=\ket{\epsilon_{a_{N-1}}}^{e_{a_{N-1}}}=\ket{0}$. By following the general measurement strategy in Appendix \ref{Mstra}, we can always construct a measurement basis such that there is only 1 cooling branch with the target system being in state:
\begin{equation}\label{41}
\rho_{c} = \frac{\frac{1}{N}\{T+(N-1)A\rho A^{\dagger}]}{p_{c}},
\end{equation}
with $p_{c} =\frac{1}{N}\text{Tr}[T+(N-1)A\rho A^{\dagger}]$, after the control system is measured and 
$(N-1)$ identical heating branches with the target system to being in state:
\begin{equation}\label{42}
\rho_{h} = \frac{\frac{1}{N}(T-A\rho A^{\dagger})}{p_{h}},
\end{equation}
with $p_{h} = \frac{1}{N}\text{Tr}[T-A\rho A^{\dagger}]$, after the control system is measured. So
\begin{equation}\label{43}
p_{H} = (N-1)p_{h} = \frac{N-1}{N}\text{Tr}[T-A\rho A^{\dagger}].
\end{equation}The coherent-controlled thermalisation can also be regarded as a supermap of the vacuum-extended channels\cite{chiribella2019quantum} (corresponding to a particular implementation of thermalising channel).

By decomposing thermalising channel as $N^{T}=\frac{A(\rho+X\rho X+Y\rho Y+Z\rho Z)A^{\dagger}}{4}$ and if one has access to the vacuum extensions with Kraus operator $\tilde{A_{0}}=\frac{1}{\sqrt{2}}AI\oplus1$, $\tilde{A_{1}}=\frac{1}{\sqrt{2}}AX\oplus1$, $\tilde{A_{2}}=\frac{1}{\sqrt{2}}AY\oplus i$, $\tilde{A_{3}}=\frac{1}{\sqrt{2}}AZ\oplus i$, the same results above can be reproduced (see section about `Perfect communication through asymptotically many paths' in \cite{chiribella2019quantum}).

\begin{widetext}

\begin{figure}[H]
    \centering
	\includegraphics[width=1\textwidth]{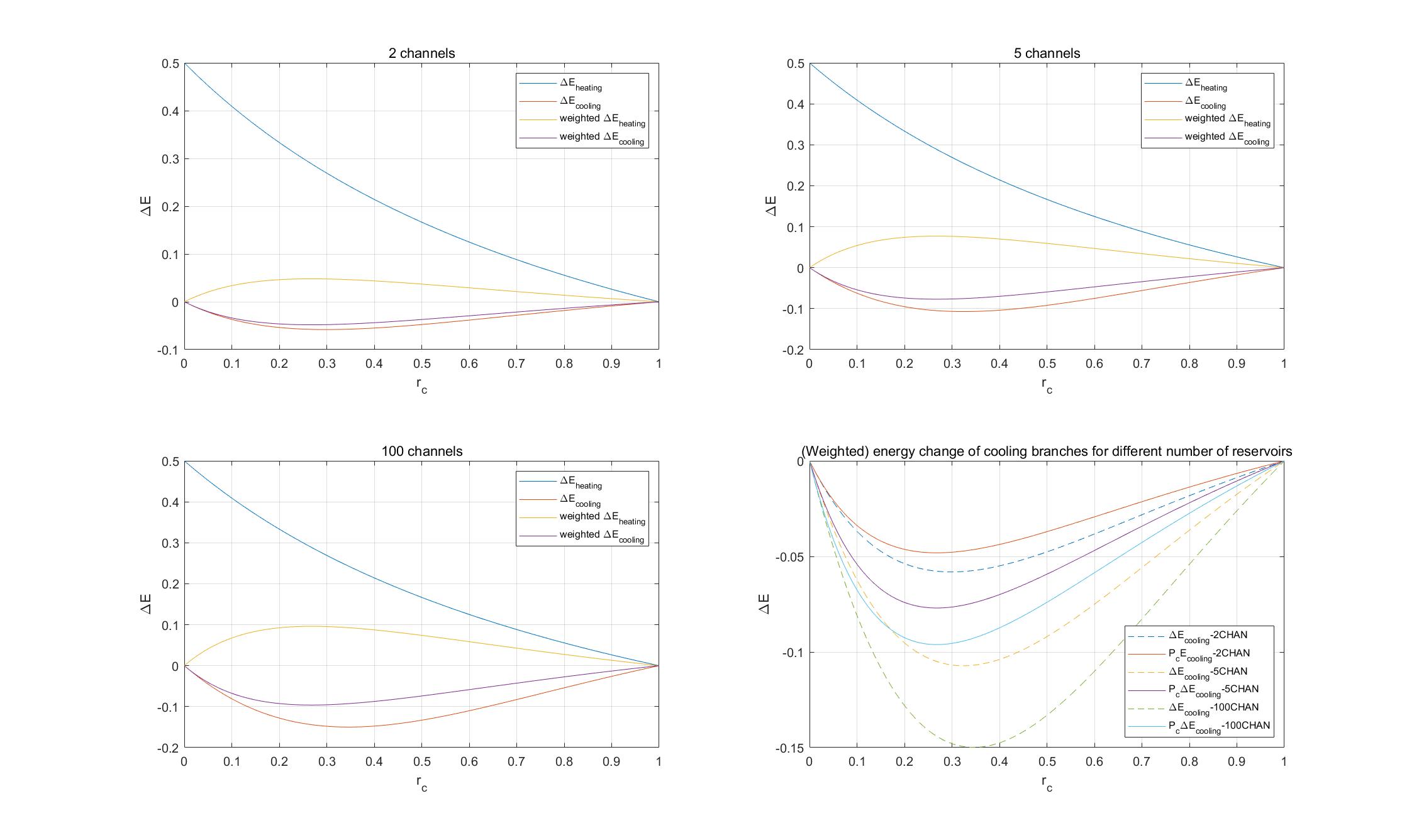}
	\caption{\textbf{How the weighted energy changes vary with the number of thermalising channels for the fridge with thermalisation in superposing trajectories.}}
	\label{f21}
\end{figure}

\end{widetext}

With target system initially set to be at the same thermal state as the reservoirs' and construct a quantum fridge based on this scheme, we can see the weighted energy changes of the fridge with thermalisation in superposing trajectories vary in a similar fashion as the ICO fridge when the number of thermalising channels increases (as you can compare Fig.\ref{f21} and Fig.\ref{f8}). But the COP of the fridge with
thermalisation in superposing trajectories is around 30$\%$ higher than the ICO fridge's in general.

Moreover, since when $\rho'_{h}=T_{H}$ the fridge stop working in refrigeration mode. So we have $E^{f}_{cold}=n_c\text{Tr}[TH]-Q_{max}=n_c\text{Tr}[T'H]$ and $E^{f}_{H}=n_{H}\text{Tr}[TH]+Q_{max}=n_{H}\text{Tr}[\rho'_{h}H]$, then we can write
    \begin{equation*}
    n_c[\frac{r}{(1+r)}-\frac{r'}{(1+r')}]=n_{H}[\frac{1}{2}-\frac{r}{(1+r)}],
\end{equation*}for the fridge with thermalisation in superposing trajectories

\begin{figure}
    \centering
	\includegraphics[width=0.5\textwidth]{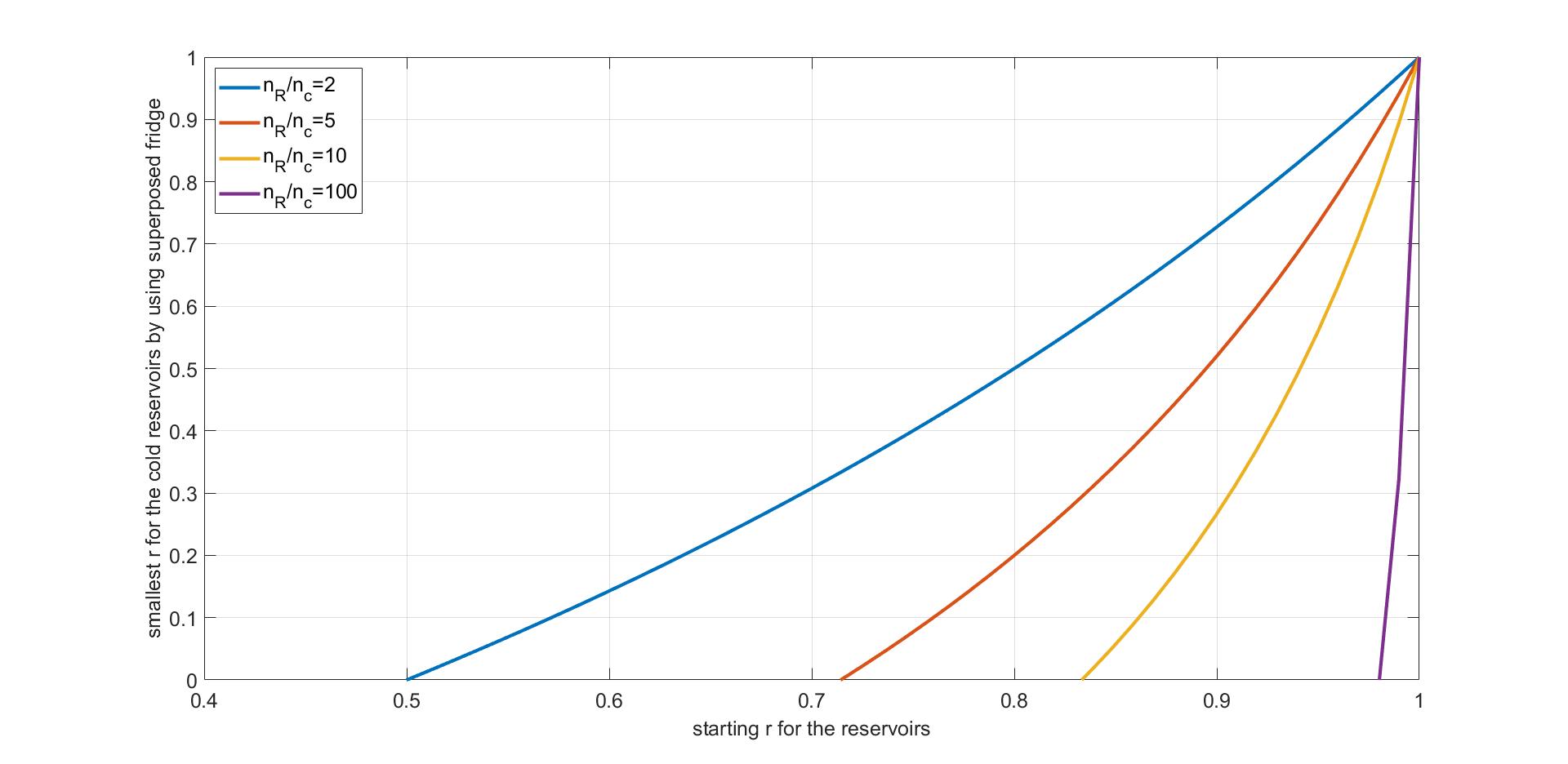}
	\caption{\textbf{Smallest r for the cold reservoirs can be obtain with the fridge assisted by thermalisation in superposing trajectories starting from a certain r. Different from the ICO fridge, the cold reservoirs in this scheme can always be arbitrary close to absolute zero in principle no matter how hot their starting temperature is  when the size of the hot reservoir is sufficiently big compared to the cold ones. }}
	\label{f23}
\end{figure}

and we can get the relation between $r$ and $r'$ which is different from the one of ICO fridge
\begin{equation}\label{44}
    r'=\frac{k-(k+2)r}{kr-2-k}.
\end{equation}where $k=\frac{n_{H}}{n_c}$.
As we can see in Fig.\ref{f23}, when the
size of the hot reservoir is sufficiently large compared to the composite cold reservoirs, the lower bound for the lowest temperature can be achieved by the cold reservoirs is always absolute zero no matter what temperature the cold reservoirs start from, which is completely different from the ICO fridge.

\section{Maxwell-demon-like scenario with thermalisation in a superposition of quantum trajectories}\label{Maxdem}
Different from the previous section when we focus on the heat exchange between the cold and hot reservoirs, in this section let's aim at the Maxwell-demon-like scenario for a sample of k particles ($k\ll$ number of particles in the reservoirs, so the change of energy and thermal state of the reservoirs can be regarded as negligible during the whole process) which initially share the same temperature as the reservoirs' assisted by thermalisation in superposition of quantum trajectories.

\begin{figure}
	\includegraphics[width=0.5\textwidth]{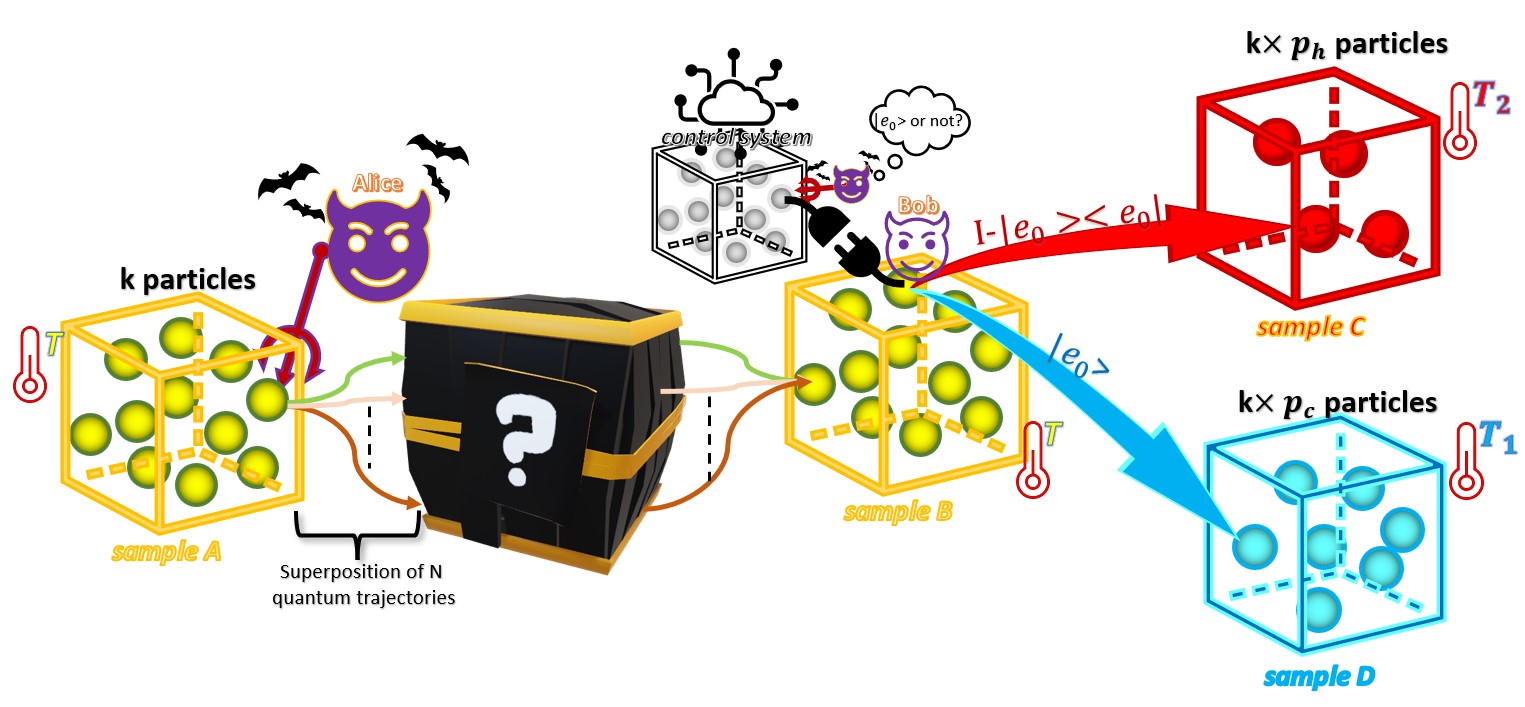}
	\caption{\textbf{Maxwell-demon-like scenario with thermalisation in superposition of quantum trajectories. Similar to the situation described in Fig.\ref{f1}, the only different is that Alice not just can thermalise the working system with two reservoirs with identical temperature in an indefinite causal order, she can now thermalise the working system with $N$ reservoirs in a superposition of $N$ quantum trajectories ($N$ cyclic orders or superposing trajectories). }}
	\label{maxdemblack}
\end{figure}

As Fig.\ref{maxdemblack} shown, the demon Alice can thermalise all the particles in sample A in a superposition of $N$ quantum trajectories(indefinite orderly or in superposing trajectories). In this sense, she can prepare $N$ pairs of entangled control-target states, in the schemes (N-SWITCH with cyclic orders or $N$ parallel trajectories) we discussed previously, the follow-up measurement of control system yield 2 different results (1 cooling branch and $N-1$ identical heating branches), based on the measurement results, Bobs put particles in different boxes (samples C and D).

\begin{figure}
    \centering
	\includegraphics[width=0.5\textwidth]{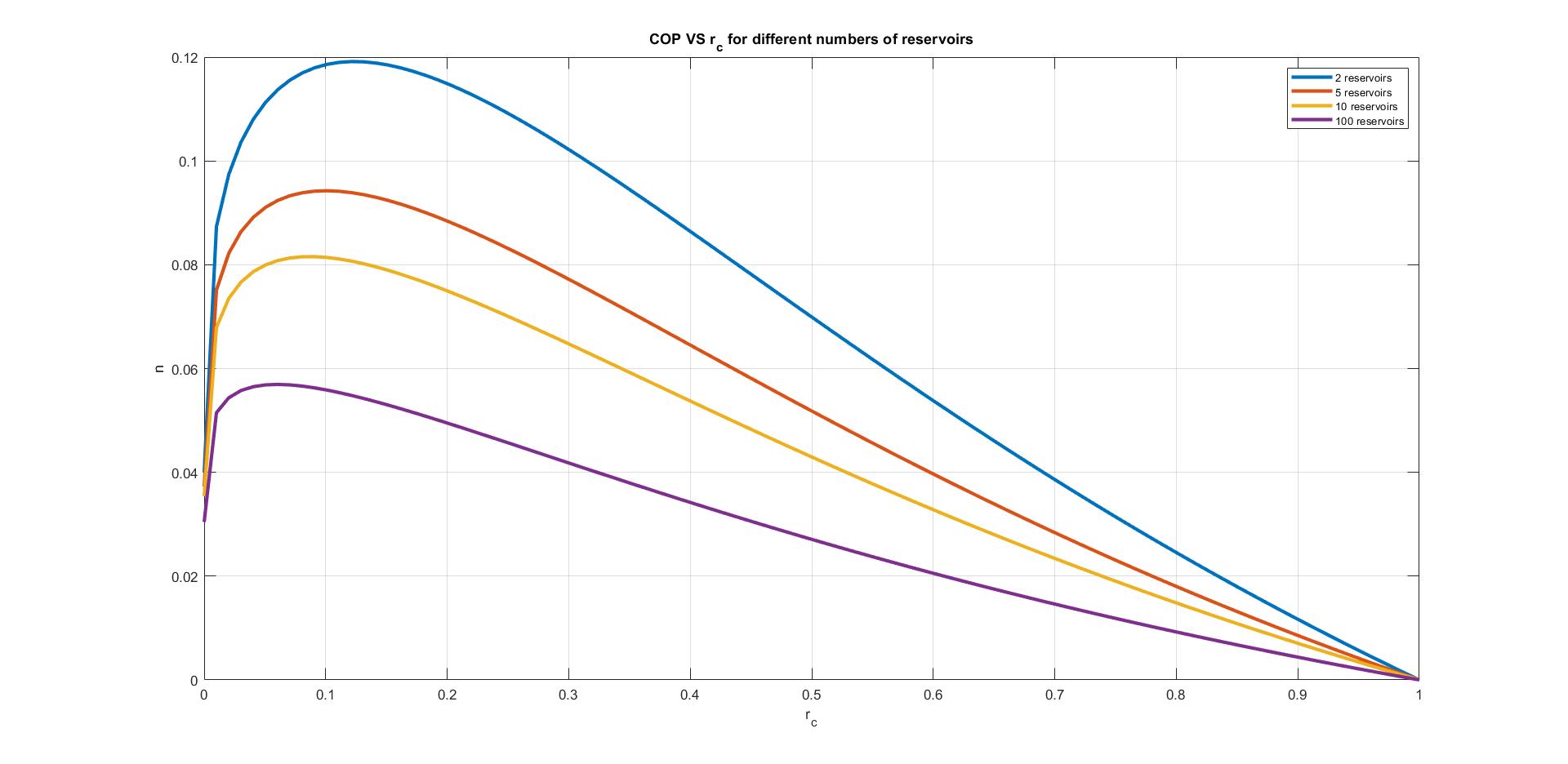}
	\caption{\textbf{COP(divided by $\Delta \beta_{R}$) VS $r_c$ for fridge with thermalisation in superposing trajectories with different numbers of reservoirs. It is around $30\%$ better than the ICO fridge in general.}}
	\label{f22}
\end{figure}

And as the results in previous sections suggest, the number of particles in each sample varies with the number of reservoirs and dimensions of the target system. Moreover, the total amount of energy transferred between these two samples varies in a similar fashion. 

\begin{widetext}

\begin{figure}
    \centering
	\includegraphics[width=0.7\textwidth]{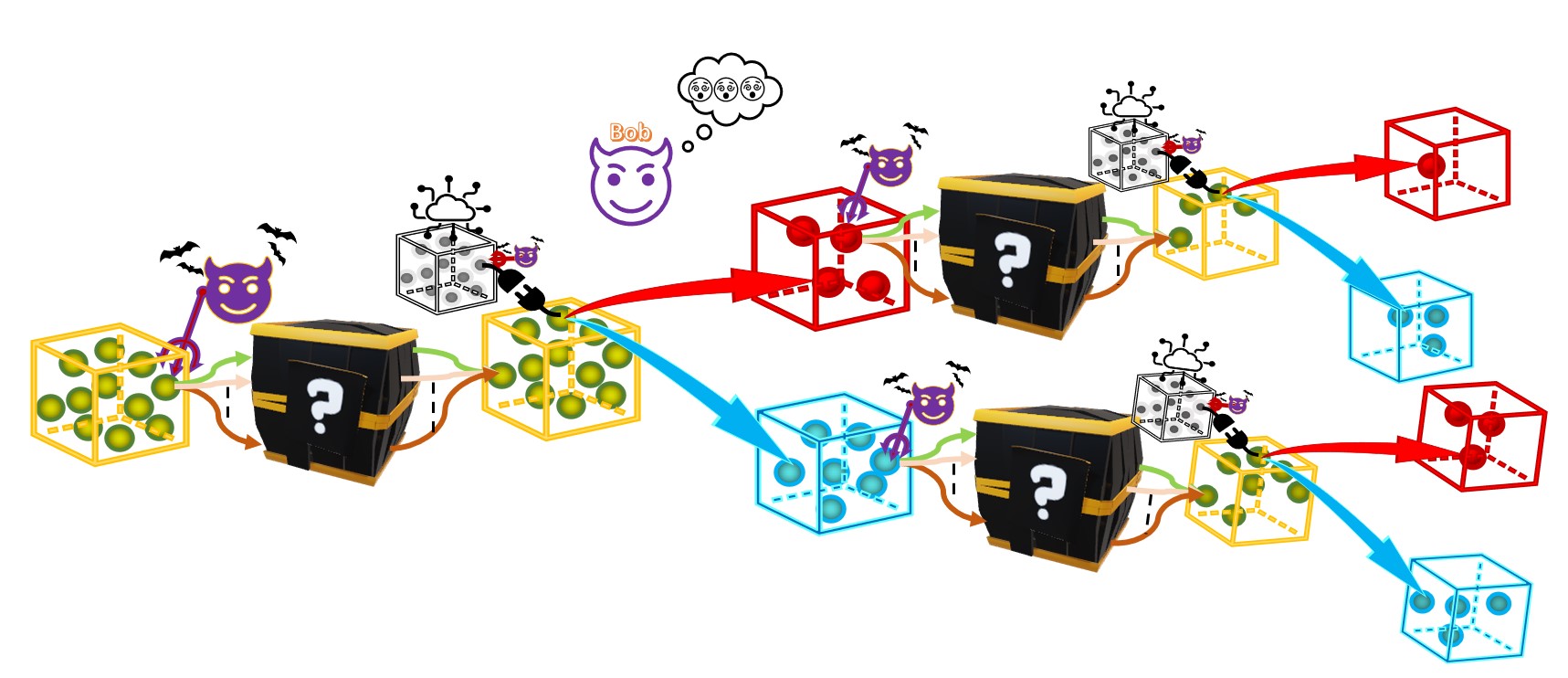}
	\caption{\textbf{Multiple runs scheme. Bob can conduct the thermalisation in superposition of quantum trajectories and sorting process for the sample assisted by Alice for multiple times. Even though the total amount of heat that transferred is the same not matter how many times you repeat the procedure, the energy distribution of the particles is more spread.}}
	\label{maxdemmulti}
\end{figure}

\end{widetext}

We focus on the thermalisation in a superposition of $N$ cyclic orders in the following discussion. The X-axis of Fig.\ref{multi0.1} is defined as $\frac{U_{final}-U_{0}}{U_{0}}$ which represents the ratio of the energy change and the initial energy of the particle in each box. We run a sample of 10000 particles start from $r=0.1$ (relative low temperature regime) the same as the reservoirs, as Fig.\ref{multi0.1} shown (in Appendix \ref{perDw}), for particles in sample D in Fig.\ref{maxdemblack}, the larger the $N$, the more extreme the temperature of the sample is. For the case of 100 reservoirs, around 75$\%$ of the particles become cooler after the process, and around 98$\%$ of the energy of these particles transfers to those in sample C. At the meanwhile, for the 2 reservoirs case, even though there are more particles become cooler (the difference of the particles number in sample D for cases of different numbers of reservoirs becomes much smaller for the low temperature region (see Fig.\ref{multi0.01} in Appendix \ref{perDw}), the energy extracted from the particles in sample D is much less than those in the 100 reservoirs case. For the 10000 particles start from $r=0.1$, the total energy is around $909\Delta$, for the case of 2 reservoirs, 38.4$\%$ of the total energy transferred between the samples, but for 100 reservoirs case the number is 72.5$\%$ (for the case when sample and cold reservoirs start at very low temperature ($r=0.01$), the difference is 54.5$\%$ vs 98$\%$). We can also repeat the process multiple times as illustrated in Fig.\ref{maxdemmulti}, in the case we focus on when $k\ll$ number of particles in the reservoirs, the reservoirs' temperature(and thermal state) can be regarded as fixed. But as shown in Fig.\ref{multi2N}(in Appendix \ref{perDw}), multiple runs of the process won't change the amount of energy transferred between the (cold and hot)samples, it does generate a more spread distribution of energy for the particles, particles with more extreme temperature can be attained. A worth mentioning phenomenon is heat jump\cite{cao2021experimental} in this multiple runs of cycle scheme. The lower the temperature of the input particle, the hotter it will be when attaining heating branch (see Fig.\ref{heatj}). We also notice that the larger the $N$, the more obvious this phenomenon can be. See Fig.\ref{heatj}, in the case of $N=100$ and start from $r=0.33$, the particle can even attain negative effective temperature ($Tr(\rho H)\le 0.5\Delta$ for (2 level) particle with positive 
effective temperature). Moreover, we can probabilistically generate a sample of particles at ultracold temperature assisted by thermalisations with reservoirs with fixed temperature (even the temperature is very high, the lower the temperature of the reservoirs, the higher the probability) when $N$ is sufficiently large. But it is not the case when $N=2$ because the amount of energy transferred in total between the samples is very limited (see Fig.\ref{2chan0.1}-\ref{100000chan0.99}).

\newpage
\section{Conclusions and outlook}\label{dis}
In this work, we show that ICO fridge assisted by the generalised N-SWITCH with N cyclic causal orders can boost the heat extracting ability of the working system, and by manipulating the dimension of the working system we can always attain global maximum COP for cold reservoirs at arbitrary low temperature. Although in the section about Maxwell-demon-like scenario we give examples of what tasks are unaccomplishable for the $N=2$ case but are achievable with sufficiently large N, an unavoidable question about the cooling task with ICO fridge is how we can be sure that the enhancement of the heat extracting ability is from the more complex interference pattern arises from superposition of more alternative orders instead of working system thermlises with more reservoirs each run. 

We show that in the controlled-SWAPs scheme first described in \cite{felce2020quantum}, there exists protocol where ICO plays no role can produce the same cooling resources as the one assisted by the generalised N-SWITCH with N cyclic orders for arbitrary N. And for this protocol, we can clearly see that enhancement of the heat extracting ability in the cooling task origins from the more complex interference patterns of thermalisation in superposition of more quantum trajectories instead of more performed thermalisations for the working system each run when N scales up. Quantum controlled-SWAPs operation gives rise to a particular implementation of thermalising channel, and actually it is a quantum coherently-controlled entangling gate between different thermal qubits. The accessibility of the reservoir qubits provides us with much greater advantages in cooling task since the reservoirs qubits are also cooled down when we attaining cooling branch. And as Fig.\ref{f17} shown, the heat extracting ability of each reservoir qubit decreases when N scales up, and the total amount of heat that can be transferred by all the qubits(working system and N reservoirs qubits) has a upper-bound when attaining cooling branch. This implies that the amount of total quantum correlations that can be generated between the thermal qubits via operation like thermalisation in superposition of quantum trajectories is limited. And The circuit complexity of the scheme without ICO is much lower so it is more accessible for the implementation of this type of quantum fridge with greater cooling power compared to the $N=2$ case. We also hope that the simulatable quantum cooling protocol without ICO will encourage the search of quantum fridge assisted by thermalisation in superposition of quantum trajectories with even better performance but lower circuit complexity.

Our results in this work focus on the ideal case of full thermalisation, and we notice that there is a recent paper\cite{wood2021operational} which modelises thermalising channel as sequential collisions between the working system and the reservoirs, making the evaluation of the partial and pre-thermalisation processes possible. It is more realistic to analyse quantum working system which is far from equilibrium interacting with some reservoirs for a finite time. Within this framework, the comparison between the performances of the ICO fridges with different number of reservoirs(different numbers of cyclic causal orders) and dimensions of working system will be more objective. And this may also shed the light on the construction of a more pratical quantum refrigerator assisted by thermalisation in superposition of quantum trajectories. We leave this exploration for future work.

\section{Acknowledgements}
The authors would like to thank David Felce, Felix Tennie, Benjamin Yadin, Chiara Marletto and Yunlong Xiao for the helpful discussions and comments. HN is supported by CQT PhD programme. TF is supported by SYSU PhD programme. VV’s research is supported by the
National Research Foundation and the Ministry of Education in Singapore and administered by Centre for Quantum Technologies, National University of Singapore.

\bibliographystyle{unsrtnat}
\bibliography{ref}

\appendix
\begin{widetext}
\newpage
\section{Deriving interference terms produced by the quantum N-SWITCH of N identical thermalising channels in N cyclic orders}\label{icooffcyclic}

Here we give a proof of why all the off-diagonal terms in the final entangled state of the control and target systems are just $T\rho T$ following the strategy we demonstrated above:
\begin{equation}\label{45}
	\frac{1}{d^{N}}\sum_{a_{1}\cdots a_{N}} AU_{a_{i+1}}\cdots AU_{a_{i}}\rho U^{\dagger}_{a_{j}}A^{\dagger}\cdots U^{\dagger}_{a_{j+1}}A^{\dagger} = T\rho T.
\end{equation}

Suppose $j=i+k$ and $i+s=N$, with $k\geq 2$(for the case $k=1$, it is easy to derive, and for the cases with $k\leq 1$ we can get the results by simply take the hermitian adjoint of those cases with $k\geq 1$ ). For convenient, denote $V_{x}=AU_{x}$, since $\sum_{x}K^{\dagger}_{x}K_{x} = I$, so $\sum_{x}V^{\dagger}_{x}V_{x} = d\times I$.

	$\begin{aligned}
		&\frac{1}{d^{N}}\sum_{a_{1}\cdots a_{N}} V_{a_{i+1}}\cdots V_{a_{i+s}} V_{a_{1}}\cdot (V_{a_{i}} \rho V^{\dagger}_{a_{i+k}}\cdot V^{\dagger}_{a_{i}})\cdots V^{\dagger}_{a_{1}}V^{\dagger}_{a_{i+k+s-k}}\cdots V^{\dagger}_{a_{n+k+2}}V^{\dagger}_{a_{i+k+1}},\\
		=& \frac{1}{d^{N-1}}\sum_{a_{1}\cdots a_{i+1}\cdots a_{N}}A \text{Tr}[A^{\dagger}\rho V^{\dagger}_{a_{i+k}}\cdots V^{\dagger}_{a_{i+2}}U^{\dagger}_{a_{i+1}}]U_{a_{i+1}}V_{a_{i+2}}\cdots V_{a_{i+s}}V_{a_{1}}\cdots V_{a_{i-1}}TV^{\dagger}_{a_{i-1}}\cdots\\
		&V^{\dagger}_{a_{1}}V^{\dagger}_{a_{i+k+s-k}}\cdots V^{\dagger}_{a_{i+k+2}}V^{\dagger}_{a_{i+k+1}},\\
		=& \frac{1}{d^{N-2}}\sum_{a_{1}\cdots a_{i+2}\cdots a_{N}}T\rho V^{\dagger}_{a_{i+k}}\cdots V^{\dagger}_{a_{i+2}}V_{a_{i+2}}\cdots (V_{a_{i+k+1}}\cdots V_{a_{N}}V_{a_{1}}\cdots V_{a_{i-1}}T\underbrace{V^{\dagger}_{a_{i-1}}\cdots V^{\dagger}_{a_{1}}V^{\dagger}_{a_{N}}\cdots V^{\dagger}_{a_{n+k+1}}}_{\text{($N-k$) terms}}),\\
		=& \frac{1}{d^{N-2-(N-k)}}\sum_{a_{i+2}\cdots a_{i+k}} T\rho (V^{\dagger}_{a_{i+k}}\cdots V^{\dagger}_{a_{i+2}}\underbrace{V_{a_{i+2}}\cdots V_{a_{i+k}}}_{\text{($k-2$) terms}})T,\\
		=& T\rho T.
	\end{aligned}$

The first equality is attained by application of the depolarizing channel once and for the second equality we make use of the the fact that the operators $U_{i}$ form an orthonormal basis for the set of $d\times d$ matrices, i.e $\sum_{i}\text{Tr}[U_{i}M]U^{\dagger}_{i}=\sum_{i}\text{Tr}[MU^{\dagger}_{i}]U_{i}=M$ where M is an arbitrary $d\times d$ matrix. And we get the third equality by applying depolarizing channel by $(N-k-1)$ times.
For the fourth equality, the fact $\sum_{x}V^{\dagger}_{x}V_{x} = d\times I$ helps.

\section{D-dimensional working system can further boost the heat extracting ability of ICO fridge within the low temperature region}\label{Dw}

Here we consider a $D$-dimensional working system  while the control system is initialised in $\dyad{+}{+}$.
The thermal state for the $D$-dimensional working system is
\begin{equation}\label{46}
	T_D=\frac{1}{Z_{T_D}}\sum_{i=0}^{D-1}e^{-\beta\Delta_i}|i\rangle\langle i|,
\end{equation}
where $\Delta_i$ is the eignenergy corresponding to eigenstate $|i\rangle$. Without losing generality, we set $\Delta_0=0$. Then the Hamiltonian of the quDit system becomes $H_D=\sum_{i=1}^{D-1}\Delta_i|i\rangle\langle i|$.

The quantum state for the working system after ICO process is given as 
\begin{equation}\label{47}
	T_D-T^3_D=\sum_{i=0}^{D-1}(\frac{1}{Z_{T_D}}e^{-\beta\Delta_i}-\frac{1}{Z^3_{T_D}}e^{-3\beta\Delta_i})|i\rangle\langle i|,
\end{equation}
The probability of attaining heating branch is
\begin{equation}
	p^D_h=\frac{1}{2}Tr(T_D-T^3_D)=
	\frac{1}{2}\sum_{i=0}^{D-1}(\frac{r_i}{Z_{T_D}}-\frac{r^3_i}{Z^3_{T_D}})
	\label{pd}
\end{equation}
where $r_i=e^{-\beta\Delta_i}=e^{-\frac{\Delta_i}{k_BT}}$. Given a multi-level particle at temperature $T$, the partition function $Z_{T_D}$ is a constant . We define

\begin{equation}\label{49}
	g(r_i)=\frac{r_i}{Z_{T_D}}-\frac{r^3_i}{Z^3_{T_D}},
\end{equation}
which is a cubic function with the variable $r_i$. The derivative of $g(r_i)$ is $g^{(1)}(r_i)=\frac{1}{Z_{T_D}}-\frac{3r^2_i}{Z^3_{T_D}}$. It is not hard to check that $g^{(1)}(r_i)>0$  if and only if  $r_i<\frac{Z_{T_D}}{\sqrt{3}}$. Since $1>r_i>0$, $g(r_i)$ is a monotonically increasing function for $0<r_i<\frac{Z_{T_D}}{\sqrt{3}}$. In the ultra-cold temperature region, we have $r_i\ll 1<Z_{T_D}$,
so $g(r_i)$ is still a monotonically increasing function.

We can define $r_0=1$ and $r_0\ge r_1\ge r_2 \ge r_3 \ge ... \ge r_i \ge .. \ge r_{D-1}$ ($\Delta_0 \le \Delta_1 \le \Delta_2 \le ... \le \Delta_{D-1}$).
Therefore, $g(r_i)>g(r_{i+1})$ holds for all $i$.
If $D-1$ energy levels are degenerated ($g_i=g_k$ for $i\ne 0$), the probability of attaining heating branch is 
\begin{equation}\label{50}
	p^D_h(\Delta_k)
	=\frac{1}{2}\{1-\frac{1+(D-1)r_k^3}{[1+(D-1)r_k]^3}\}.
\end{equation}
In general, we have  $p^D_h(\Delta_{D-1})\le p^D_h \le p^D_h(\Delta_1)$ within the low temperature region.  
Obviously we can always find a $D$ such that $p^D_h(\Delta_{D-1})$ is close to $\frac{1}{2}$. 

In practical, quantum systems are usually multi-level, the qubit system can be considered as a fraction of multi-level system. The ground state and first excited state of a multi-level quantum system can be used to construct a qubit. So one can calculate

\begin{equation}\label{51}
	p^{D}_h=
	\frac{1}{2}\sum^{D-1}_{i=0}[\frac{r_i}{\sum_{j=0}^{D-1}r_i}-\frac{r^3_i}{(\sum^{D-1}_{j=0}r_i)^3}]\ge p^D_h(\Delta_{D-1}),
\end{equation}

\begin{equation}\label{52}
	p^{D=2}_h=
	\frac{1}{2}\sum^{1}_{i=0}[\frac{r_i}{\sum_{j=0}^{1}r_i}-\frac{r^3_i}{(\sum^{1}_{j=0}r_i)^3}].
\end{equation}
Here we set $r_{D-1}=qr_1$, where $q$ is fixed for a given quantum system and $q\le 1$.  We have 
\begin{equation}\label{53}
	p^{D}_h\ge p^D_h(\Delta_{D-1})
	=\frac{1}{2}\{1-\frac{1+(D-1)q^3r_1^3}{[1+(D-1)qr_1]^3}\},
\end{equation}
\begin{equation}\label{54}
	p^{D=2}_h
	=\frac{1}{2}[1-\frac{1+r_1^3}{(1+r_1)^3}].
\end{equation}
When $p^{D}_h$ is larger than $p^{D=2}_h$, we have  $p^D_h(\Delta_{D-1})>p^{D=2}_h$. That is 

\begin{equation}\label{55}
	\frac{1+r^3_1}{(1+r_1)^3}>\frac{1+(D-1)q^3r^3_1}{(1+(D-1)qr_1)^3}.
\end{equation}
A special case is $(D-1)q\approx1$, the above equation becomes
\begin{equation}\label{56}
	1+r^3_1>1+q^2r^3_1,
\end{equation}
which is hold for arbitrary $r_1$. And by Eqn.(\ref{26}) in section \ref{Dwork}, we see that within the low temperature region where the D-dimensional working system can still effectively operate, the weighted energy change for heating branch increases as $p^{D}_h$ increases, such that the heat extracting ability of the ICO fridge can be further boosted within the low temperature region when we make use of a D-dimensional working system.

\section{Measurement strategy for arbitrary number of thermalising channels} \label{Mstra}

We now show that, for output states of the desired form (with off-diagonal elements of the form $T\rho T$ only) and for any number of cyclic causal orders being exploited, there exists a measurement on this output state which produces one of $N$ branches : the cooling branch of the form $T+(N-1)T^3$, and $N-1$ identical heating branches.\\

\noindent\textbf{Theorem 1.} For all $N$ there exists a measurement basis  $\{ P_{0,c} \otimes I_{\mbox{\tiny $ 2$ }\!\!,s}\, , \, Q_{c}  \otimes I_{\mbox{\tiny $ 2$ }\!\!,s}\}$ - where ${P_{j,c}}={\,\ket {\phi_j} \!\! \bra {\phi_j} }_c$ and $Q_c=\sum_{j=1}^{N-1}\!P_{j,c}$ are rank-$1$ and rank-$(N\!-\!1\!)$ projectors respectively acting on the $N$-dimensional Hilbert space $\mathcal{H}_c$ of the control system - such that one of the two possible outcomes of projective measurement of the output state is a heating branch with working system of the form $ T-T^3$. 

\noindent\textbf{Proof of Theorem 1.} 
The desired output state for $n$ channels and $n$ causal orders is 
\begin{equation}\label{57}
	S\!\left(\rho_c \! \otimes \! T \right) = I_{N,c}\otimes \! \frac{T}{N} + \big( \, n{\ket {\phi_0} \!\! \bra {\phi_0} }_c \! - \! I_{N,c} \,\! \big) \! \otimes \! \frac{T^3}{N},
\end{equation}
where $\ket{\phi_0}_c\!=\!\ket{\psi}_c\! =\! \frac{1}{\sqrt{N}} \sum_i^N {\ket{l_i}}_c$ for $i \in [1,N]$ is the state that the control system is initialised in. Now define 
\begin{equation}\label{58}
	\ket{\phi_i}_c=\frac{\sum_{j=1}^N B_{ij}\ket{l_j}_c}{\sqrt{\sum_{j=1}^N \!B_{ij}^{\,2}}}    
\end{equation}
for $i \in [1,N-1]$, where $B_{ij}$ are the elements of the matrix
\begin{equation*}
	B = 
	\begin{pmatrix}
		1 & -1 & 0 & 0 & 0& 0 &0 & \cdots & \cdots & 0 \\
		1 & 1 & -2 & 0 & 0 & 0 & 0 & \cdots & \cdots & 0 \\
		1 &  1 & 1 & -3 & 0 & 0 & 0 & \cdots & \cdots & 0 \\
		1 &  1 & 1 &  1 & -4 & 0 & 0 & \cdots & \cdots & 0 \\
		1 &  1 & 1 &  1 & 1 & -5 & 0 & \cdots & \cdots & 0 \\
		\vdots  & \vdots  & \vdots & &  &  \ddots & \ddots &  && \vdots \\
		1 & 1 & 1 & \cdots & \cdots & \cdots & 1 & (3-N) & 0 & 0 \\
		1 & 1 & 1 & \cdots & \cdots & \cdots & 1 & 1 & (2-N) & 0 \\
		1 & 1 & 1 & \cdots & \cdots & \cdots & 1 & 1 & 1 & (1-N) \\
	\end{pmatrix}.
\end{equation*}
It can be seen that the rows of matrix $B$ are orthogonal and hence $\{ \ket{\phi_i}_{\!c}\}$ for $i \in [1, N-1]$ are orthonormal. Since $\ket {\phi_0}_c$ is also orthonormal to these vectors, $\{ \ket{\phi_i}_{\!c}\}$ for $i \in [0, N-1]$ forms an orthonormal basis. Then upon projective measurement of the output state with respect to $\{ P_{0,c} \otimes I_{\mbox{\tiny $ 2$ }\!\!,s}\, , \, Q_{c}  \otimes I_{\mbox{\tiny $ 2$ }\!\!,s}\}$ where $P_{j,c}={\,\ket {\phi_j} \!\! \bra {\phi_j} }_c$ and $Q_c=\sum_{j=1}^{N-1}\!P_{j,c}$, we obtain 
\begin{equation}\label{59}
	\frac{Q_c \, \mathcal{S}(\rho_c \!\otimes \! T)\, Q_c^{\dagger}}{\text{tr}\!\left( Q_c^{\dagger}Q_c \, \mathcal{S}(\rho_c \! \otimes \! T) \right)} =  \frac{1}{N-1}\sum_{j=1}^{N-1} {\,\ket {\phi_j} \!\! \bra {\phi_j} }_c \otimes \frac{T-T^3}{\text{tr}(T-T^3)}, 
\end{equation}
with probability $p_-=\text{tr}\!\left( Q_c^{\dagger}Q_c\,\mathcal{S}(\rho_c \! \otimes \! T) \right) = \frac{N-1}{N}\text{tr}\left[ T-T^3\right]$, and 
\begin{equation}\label{60}
	\frac{P_{0,c} \, \mathcal{S}(\rho_c \!\otimes \! T)\, P_{0,c}^{\dagger}}{\text{tr}\!\left( P_{0,c}^{\dagger}P_{0,c} \, \mathcal{S}(\rho_c \! \otimes \! T) \right)} =  {\,\ket {\phi_0} \!\! \bra {\phi_0} }_c \otimes \frac{T+(N-1)T^3}{\text{tr}(T+(N-1)T^3)}. 
\end{equation}
with probability $p_+=\text{tr}\!\left( P_{0,c}^{\dagger}P_{0,c}\,\mathcal{S}(\rho_c \! \otimes \! T) \right) = \frac{1}{N}\text{tr}\left[ T+(N-1)T^3\right]$.

\section{Performance of the generalised N-SWITCH fridge with D-dimensional working system} \label{perDw}

\subsection{ICO fridges with different N-SWITCH for qubit working system} \label{Nicopow}

\begin{figure}
    \centering
	\includegraphics[width=1\textwidth]{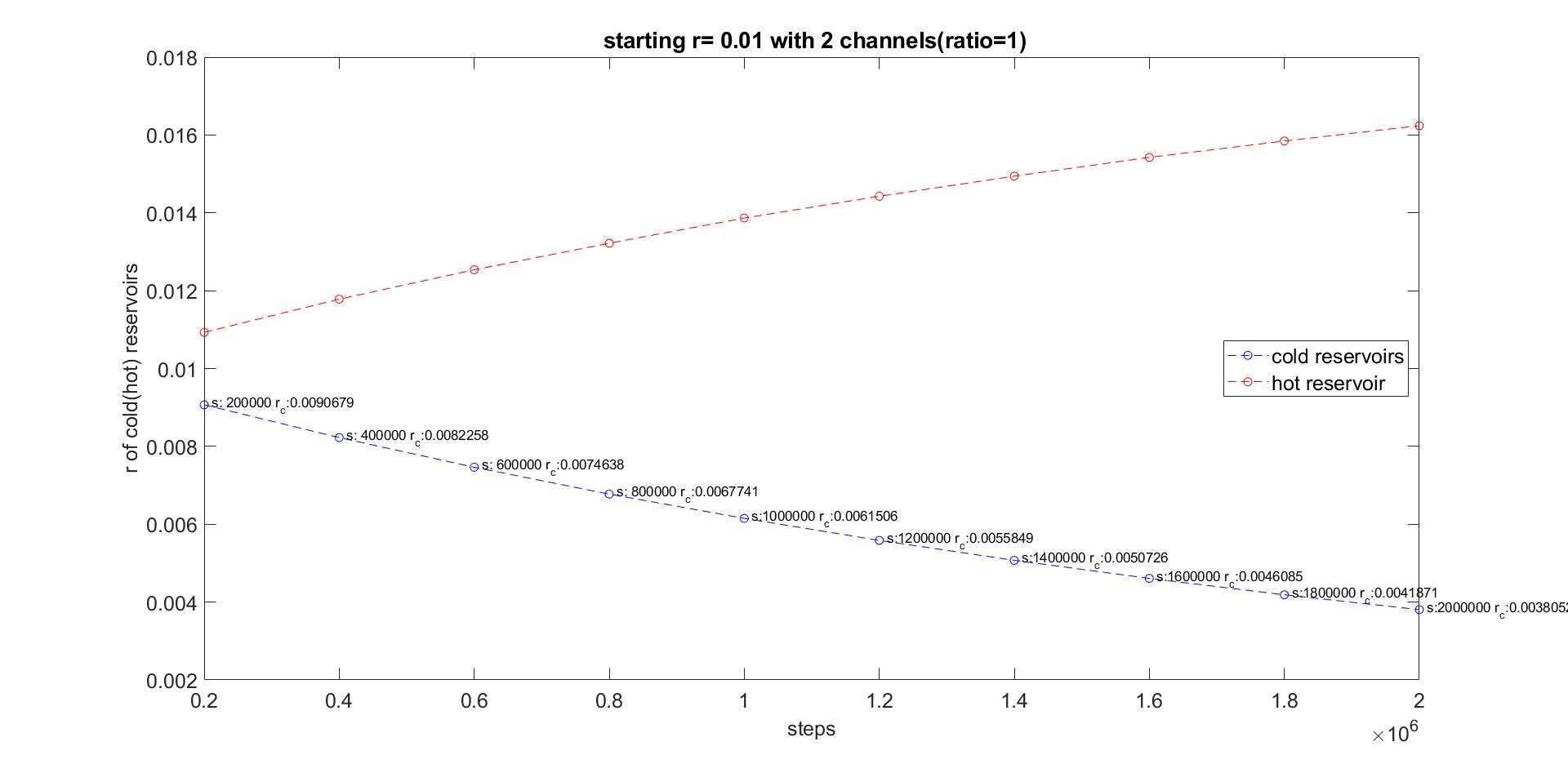}
	\caption{\textbf{2 reservoirs starting at $r_c=0.01$.} }
	\label{2r0.01}
\end{figure}

\begin{figure}
    \centering
	\includegraphics[width=1\textwidth]{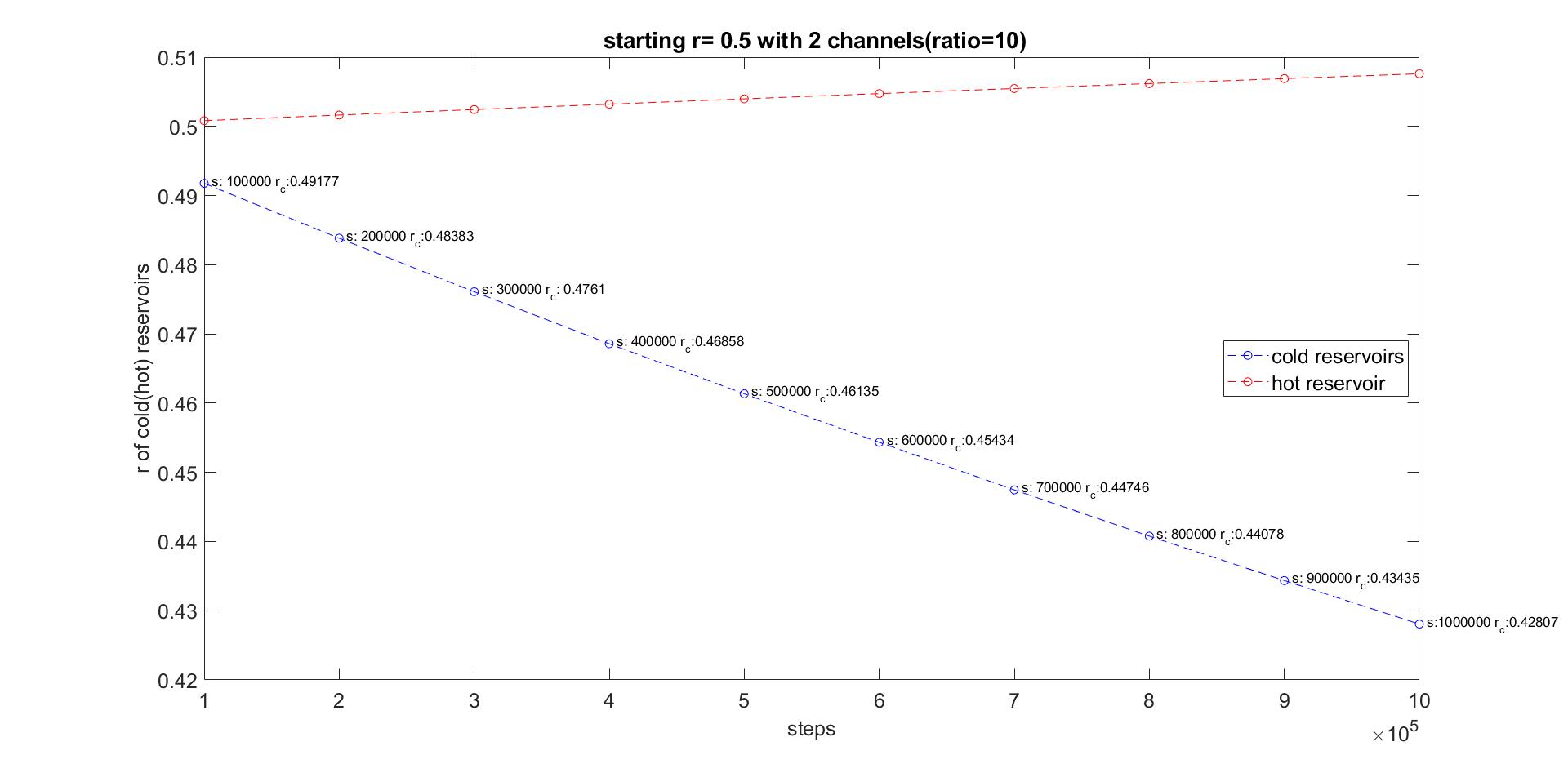}
	\caption{\textbf{2 reservoirs starting at $r_c=0.5$.} }
	\label{2r0.5}
\end{figure}

\begin{figure}
    \centering
	\includegraphics[width=1\textwidth]{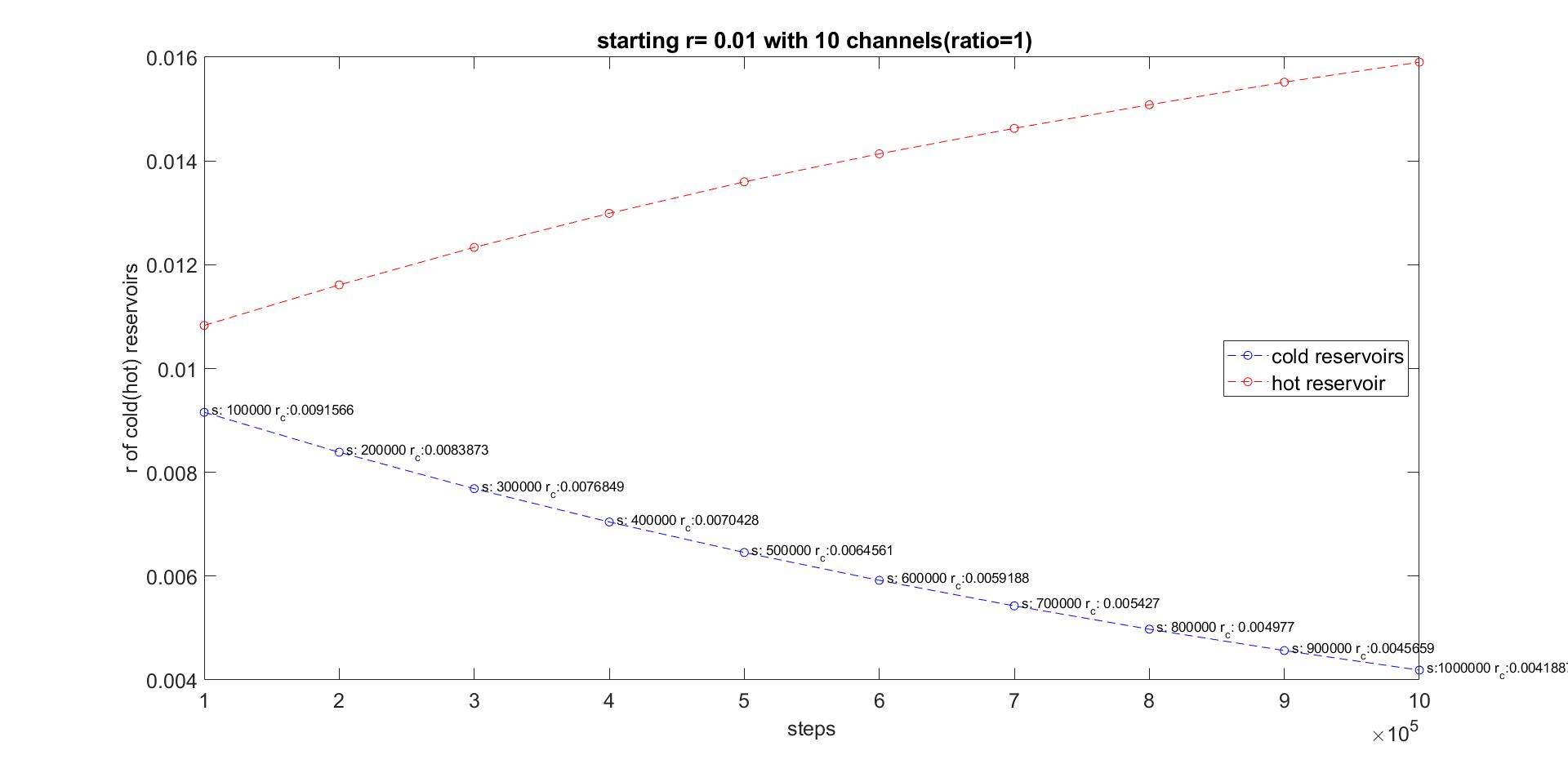}
	\caption{\textbf{10 reservoirs starting at $r_c=0.01$.} }
	\label{10r0.01}
\end{figure}

\begin{figure}
    \centering
	\includegraphics[width=1\textwidth]{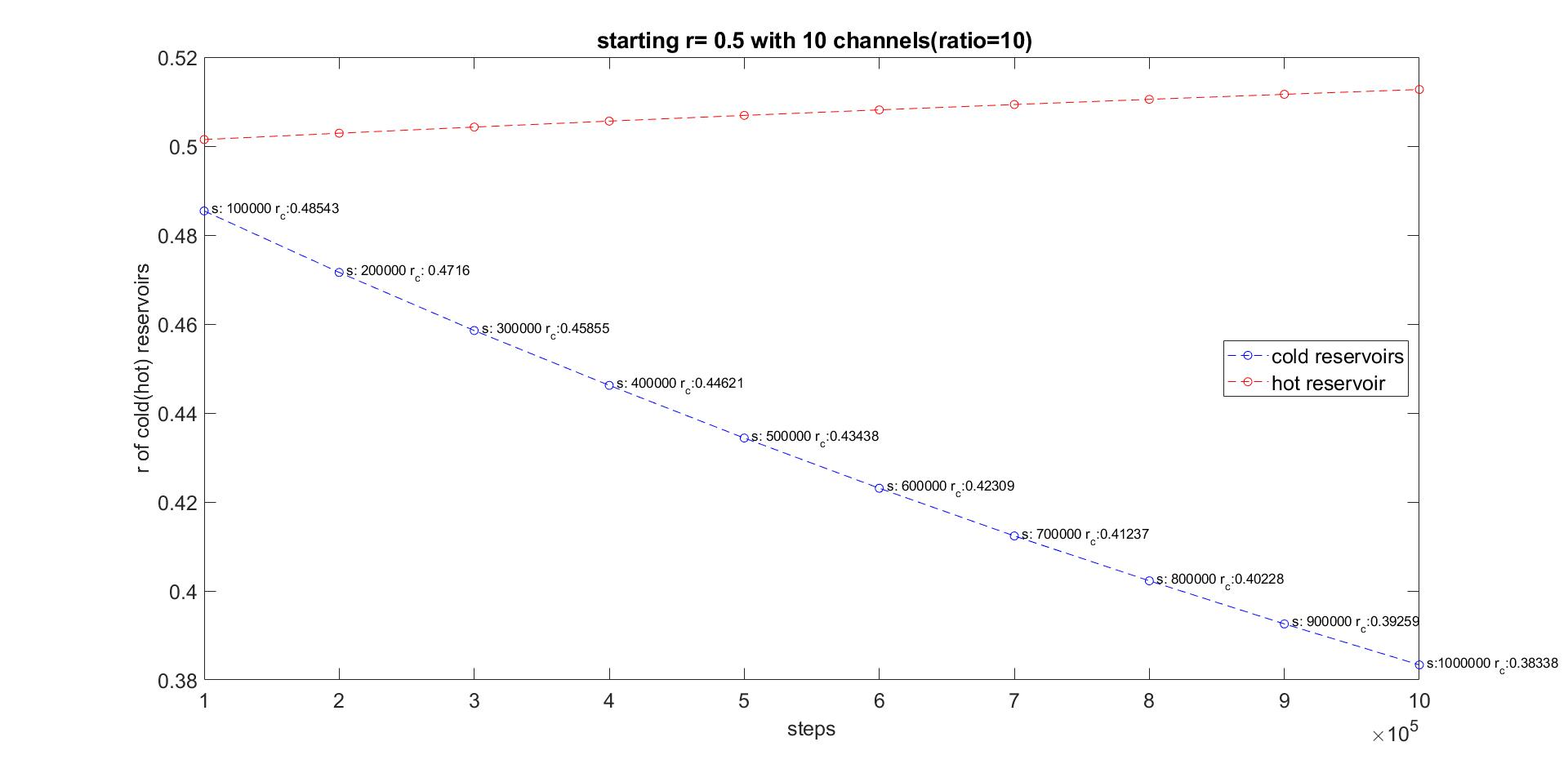}
	\caption{\textbf{10 reservoirs starting at $r_c=0.5$.} }
	\label{10r0.5}
\end{figure}

\begin{figure}
    \centering
	\includegraphics[width=1\textwidth]{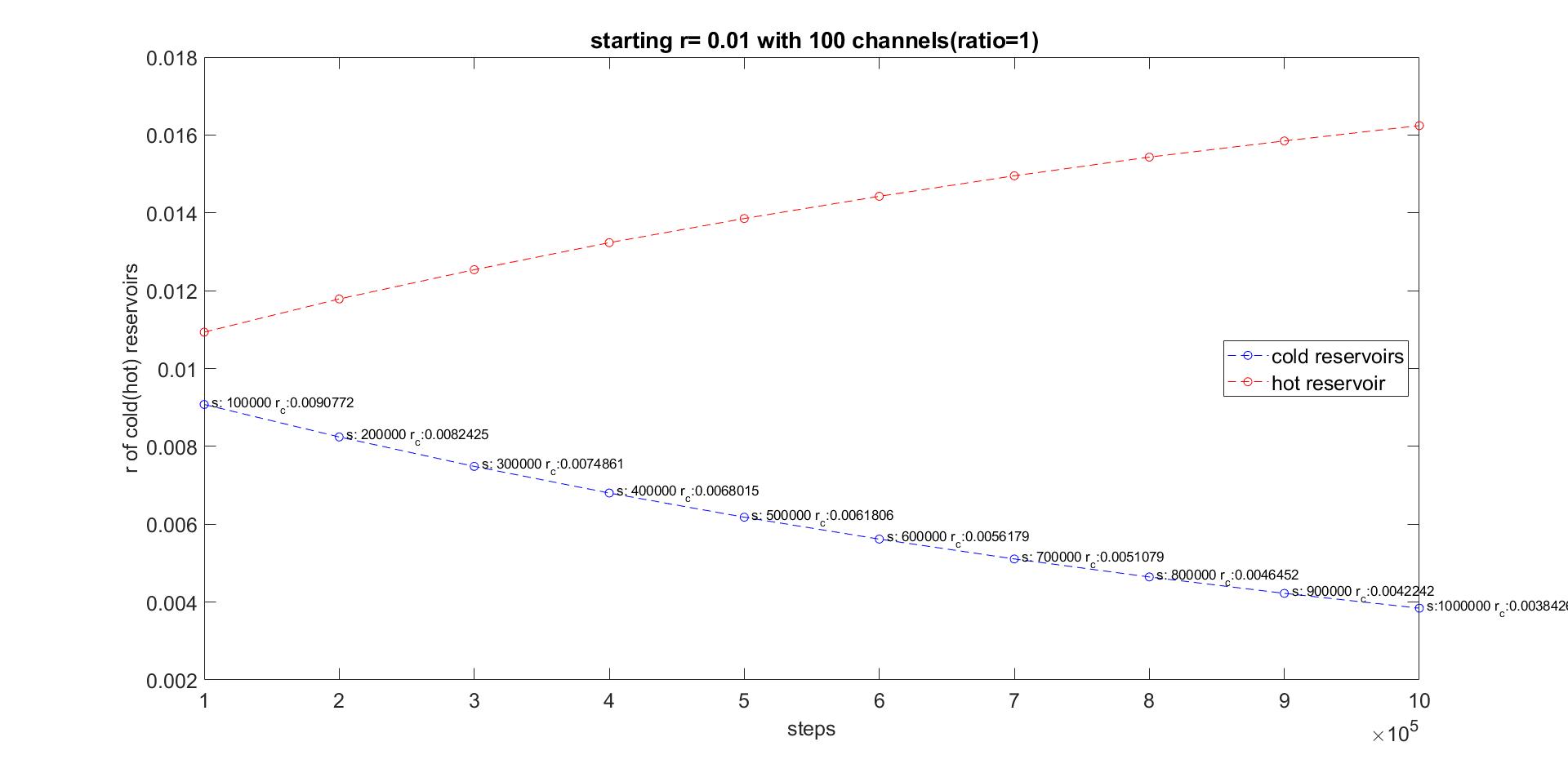}
	\caption{\textbf{100 reservoirs starting at $r_c=0.01$.} }
	\label{100r0.01}
\end{figure}

\begin{figure}
    \centering
	\includegraphics[width=1\textwidth]{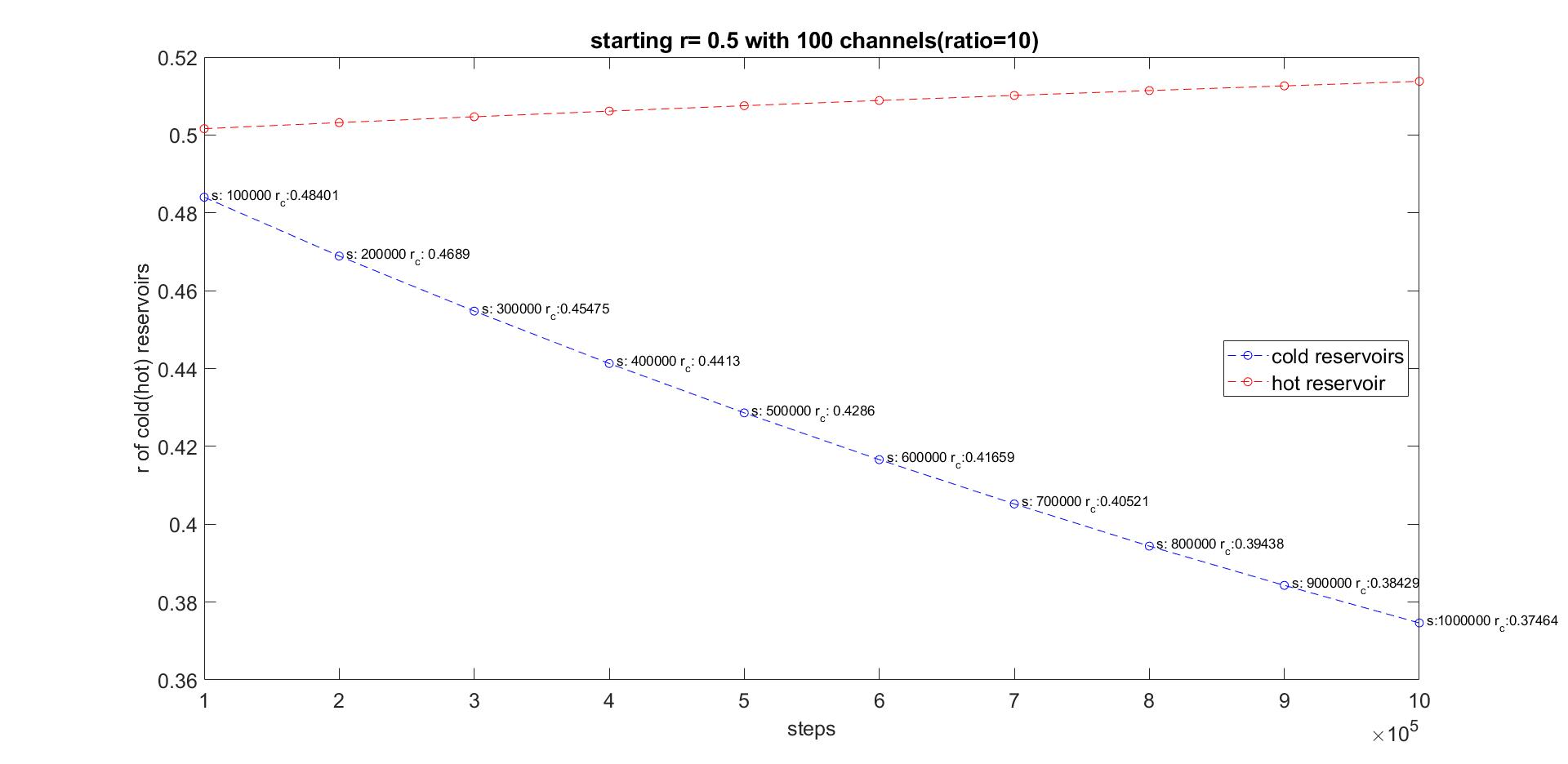}
	\caption{\textbf{100 reservoirs starting at $r_c=0.5$.}}
	\label{100r0.5}
\end{figure}

\begin{figure}
    \centering
	\includegraphics[width=1\textwidth]{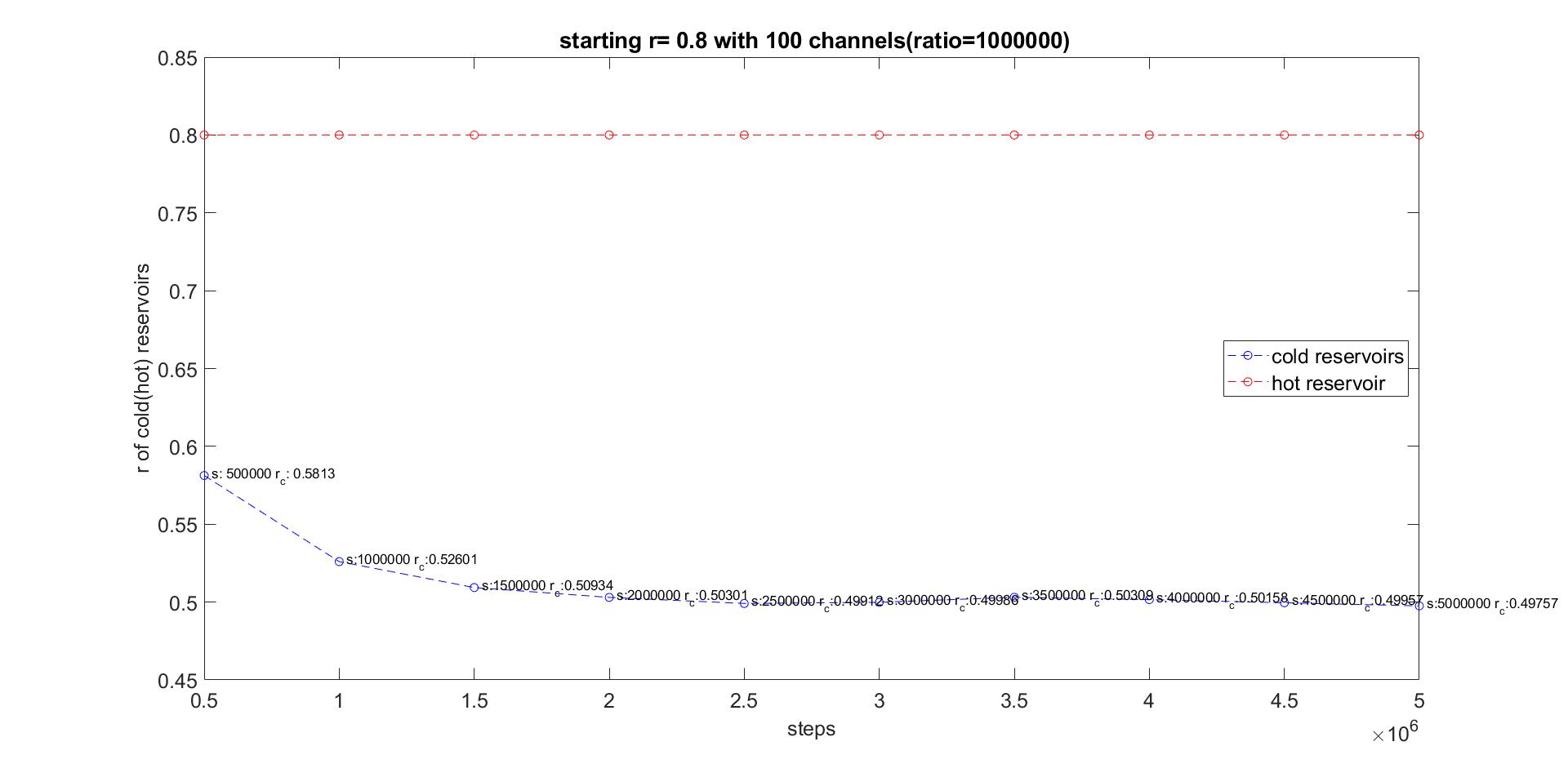}
	\caption{\textbf{The smallest attainable $r_c$ for the cold reservoirs starting at $r_{c}=0.8$. No matter how large the size of the hot reservoir is compared to the cold ones, the cold reservoirs can't be further cooled down.}}
	\label{0.8lowest}
\end{figure}

\begin{figure}
    \centering
	\includegraphics[width=1\textwidth]{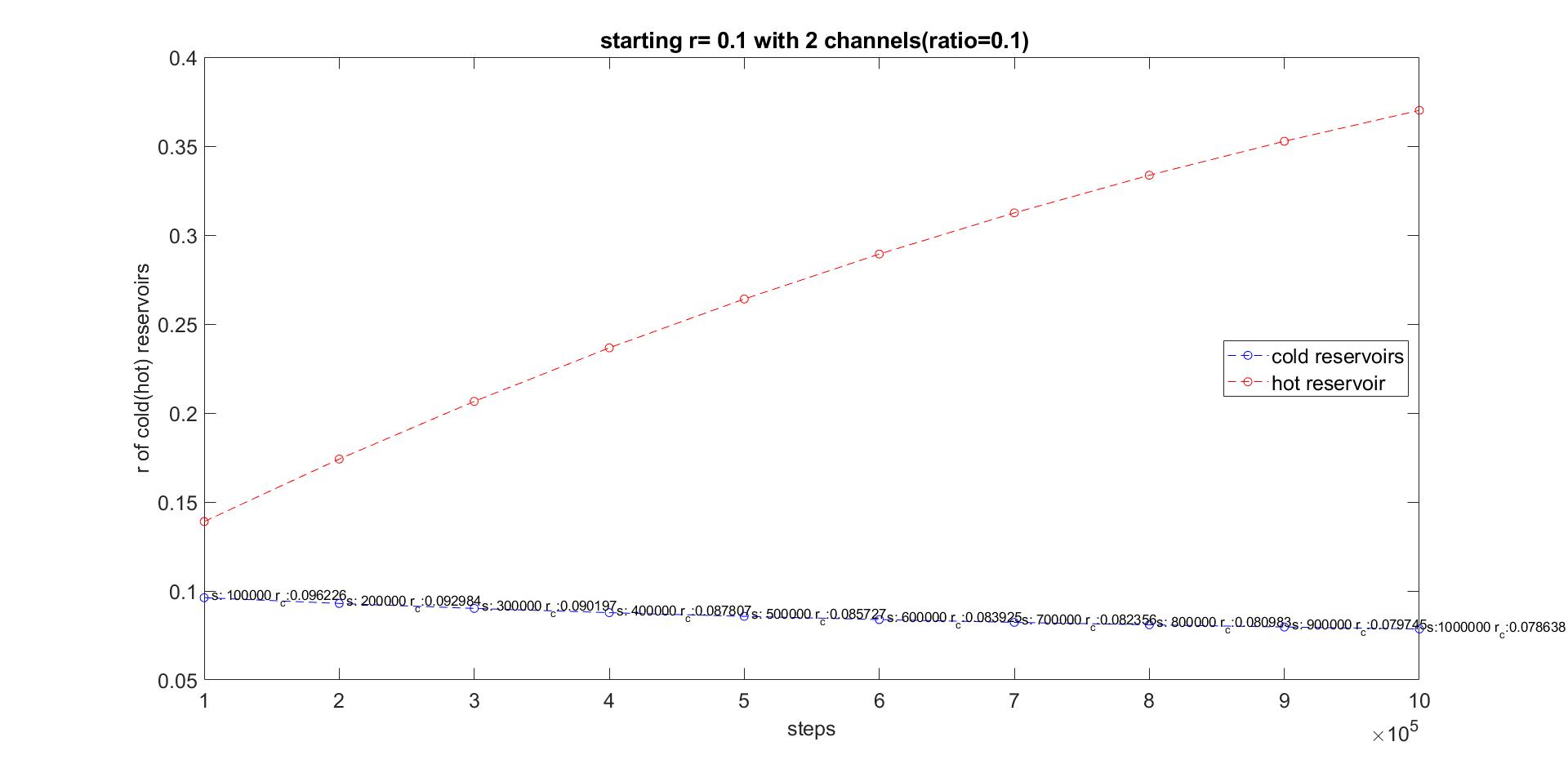}
	\caption{\textbf{When $n_H/n_c$ is relatively small, the ICO fridge can't effectively cool down the cold reservoirs.}}
	\label{2chancant}
\end{figure}

\newpage
\subsection{For fixed number of reservoirs with a D-dimensional target system} \label{NfixedD}

\begin{figure}
    \centering
	\includegraphics[width=1\textwidth]{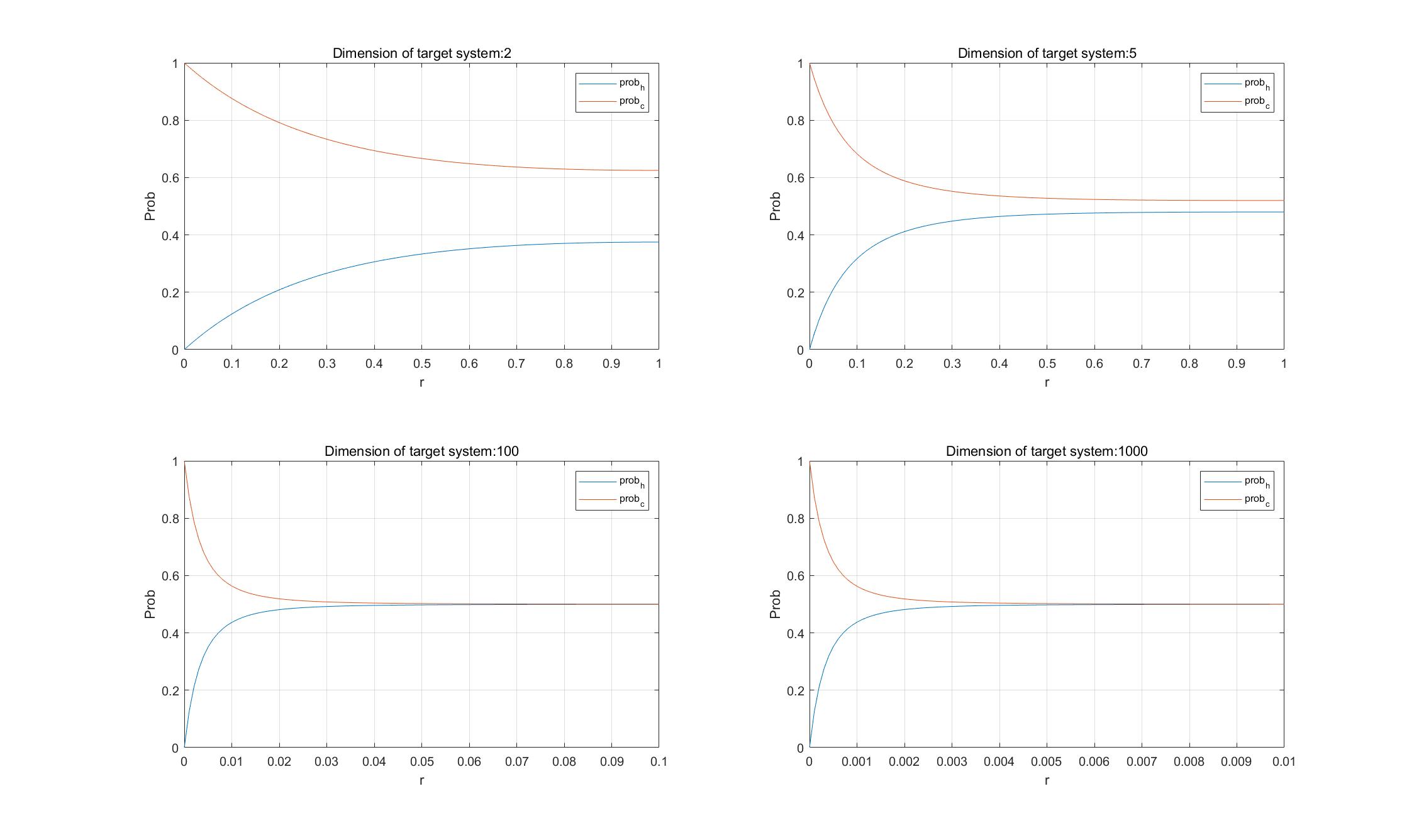}
	\caption{\textbf{How the probability of getting cooling branch varies with the dimension of the target system (2 reservoirs)}. }
	\label{Probico2D}
\end{figure}

\begin{figure}
	\includegraphics[width=1\textwidth]{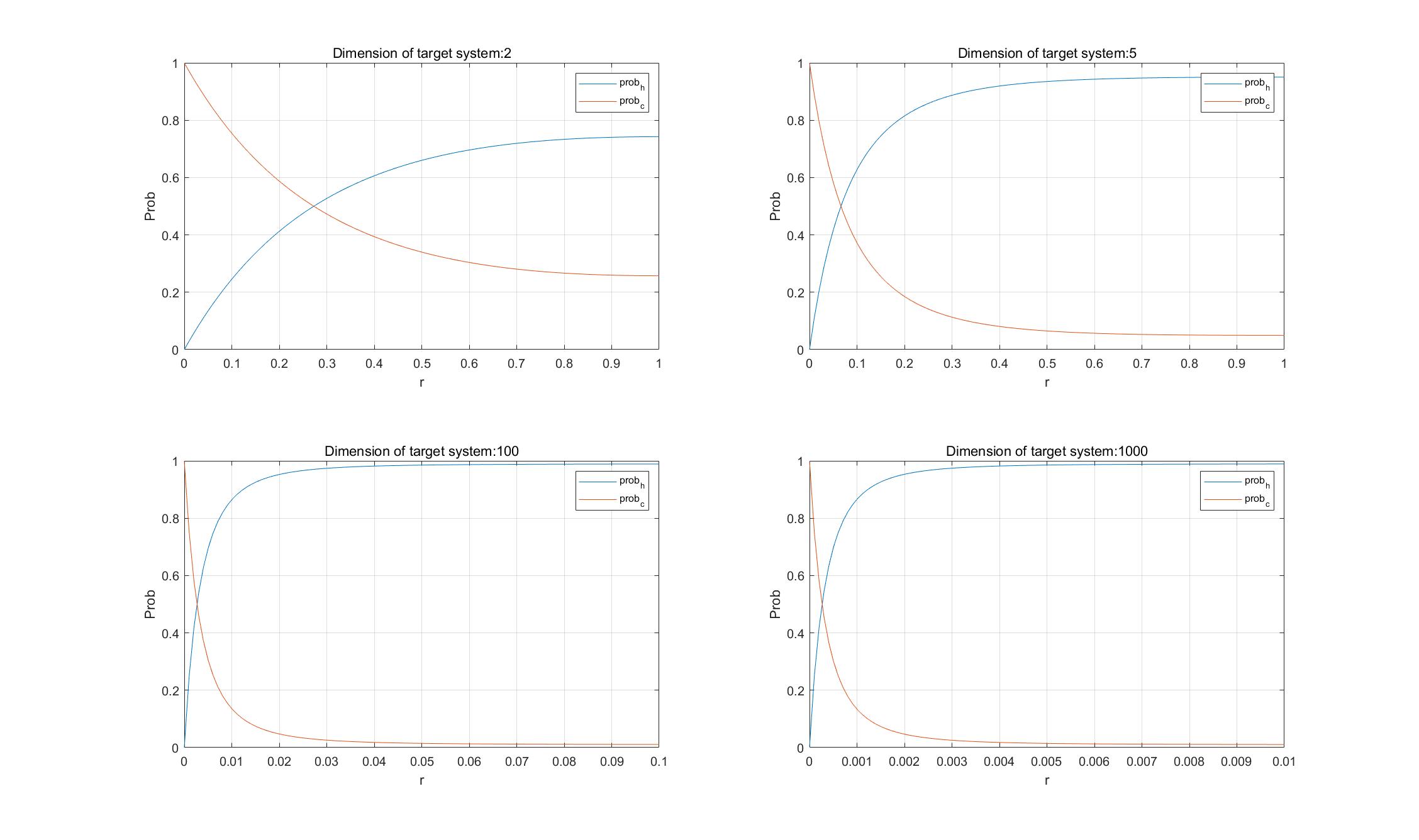}
	\caption{\textbf{How the probability of getting cooling branch varies with the dimension of the target system (100 reservoirs).}}
	\label{Probico100D}
\end{figure}

\begin{figure}
	\includegraphics[width=1\textwidth]{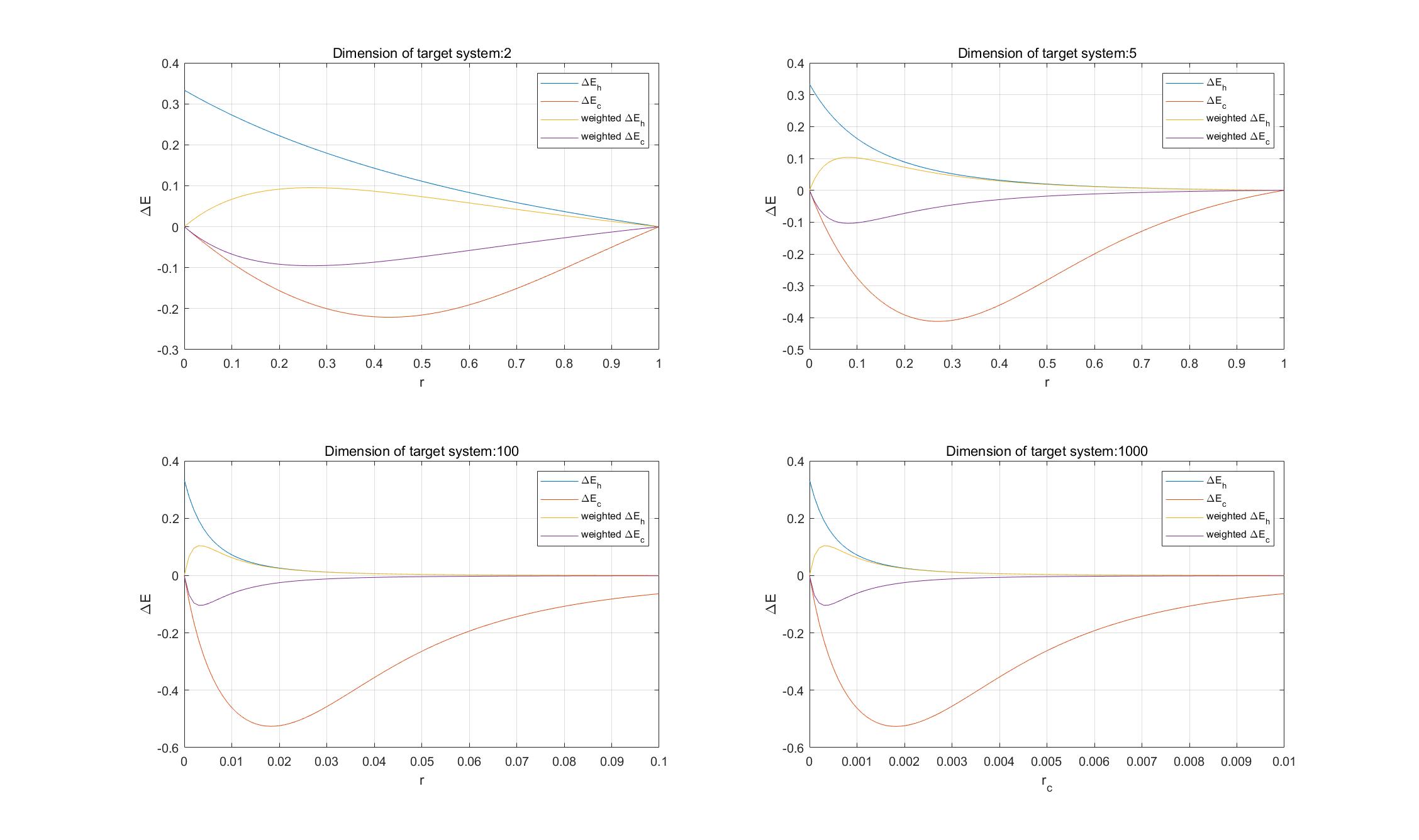}
	\caption{\textbf{How the weighted energy changes varies with the dimension of the target system (100 reservoirs).} }
	\label{Wico100D}
\end{figure}

\begin{figure*}[htbp]
    \centering
	\begin{minipage}[t]{0.6\textwidth}
	\centering
		\includegraphics[width=3.5in]{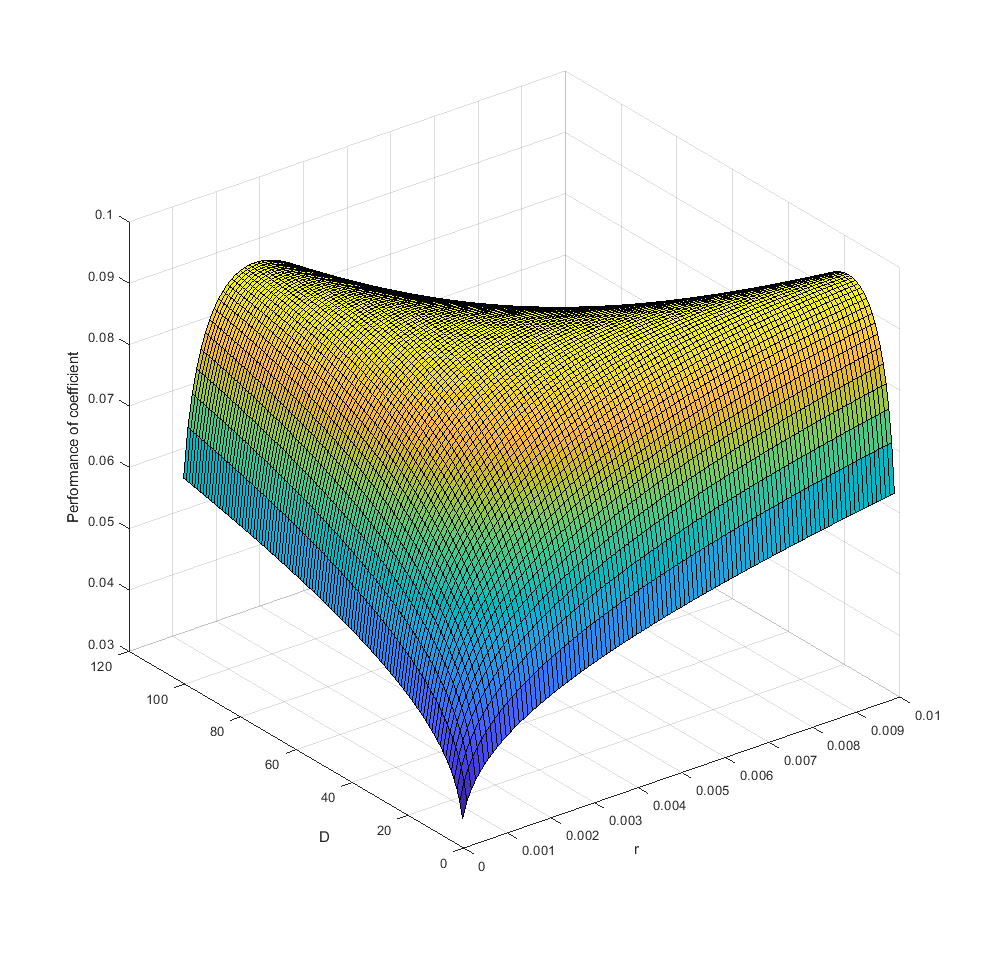}
		\caption{\textbf{$r\in[0,10^{-2}]$}}
		\label{fig:side:a}
	\end{minipage}
	\begin{minipage}[t]{0.6\textwidth}
	\centering
		\includegraphics[width=3.5in]{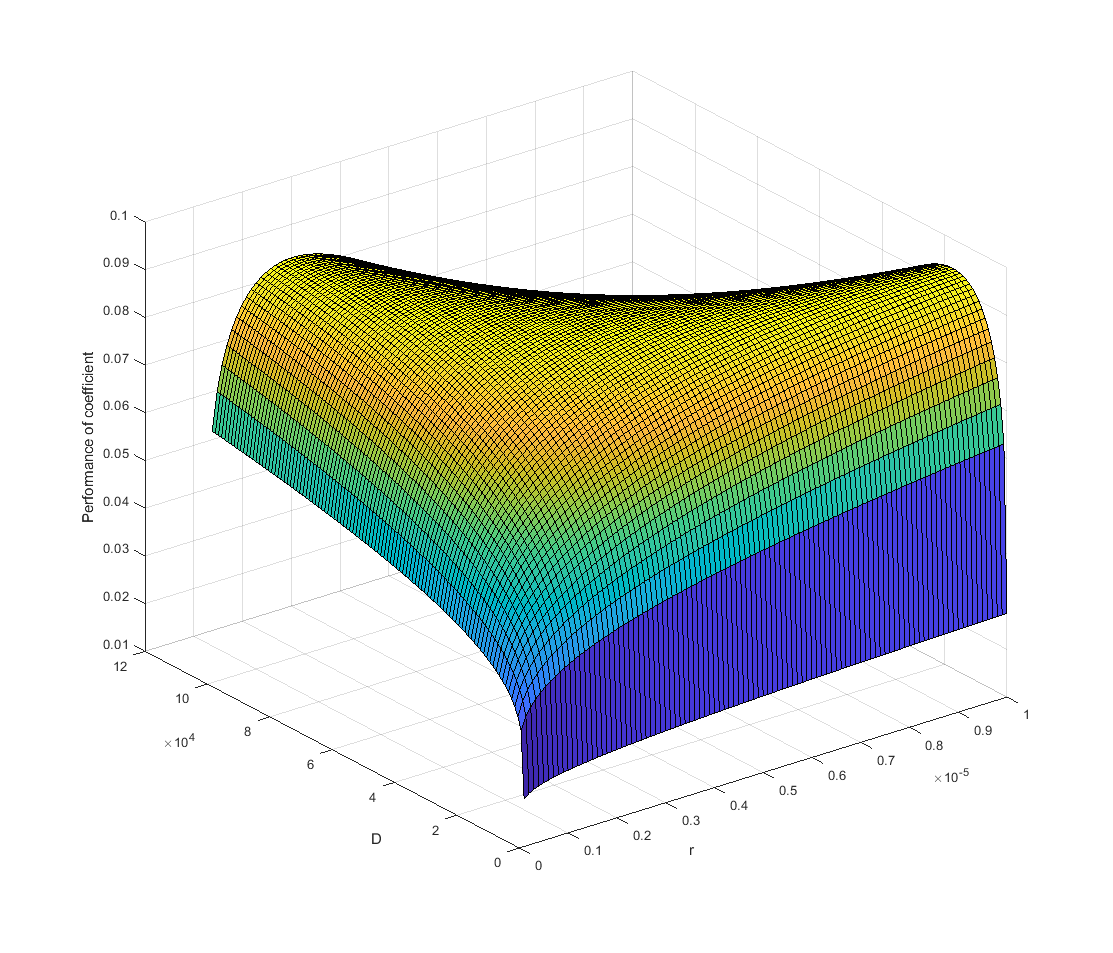}
		\caption{\textbf{$r\in[0,10^{-5}]$}}
		\label{fig:side:b}
	\end{minipage}
\end{figure*}

\newpage
\section{Local Gibbs states of target system and reservoirs qubits when attaining cooling branch in controlled-SWAPs scheme}\label{mgs}
Following Eqn.(\ref{36}), we can get the final quantum correlated target-reservoirs (qubits) state after tracing out all the ancillary systems used to purify the initial thermal states when attaining cooling branch
\begin{equation}\label{marsswap}
    \bra{e_0}Tr_{anc}[\ket{T^{\beta}_{f}}\bra{T^{\beta}_{f}}]\ket{e_0}=\frac{1}{N}\underbrace{T\otimes T\cdots\otimes T}_{N+1  terms}+\frac{1}{N^2}\sum_{a\cdots a_{N}}p_{a}\cdots p_{a_{N}}\sum_{k}\sum_{k'\ne k}\ket{a_{k}\cdots aa_{k+1}\cdots}\bra{a_{k'}\cdots aa_{k'+1}\cdots}.
\end{equation}To determine the local Gibbs state of the target system or each reservoir qubit, the key is to evaluate the terms come from off-diagonal terms of $ Tr_{anc}[\ket{T^{\beta}_{f}}\bra{T^{\beta}_{f}}]$. Take $N=5$ as an example
	\begin{equation}\label{offswap}
\begin{split}%
    &\sum_{aa_{1}a_{2}a_{3}a_{4}a_{5}}p_{a}p_{a_{1}}p_{a_{2}}p_{a_{3}}p_{a_{4}}p_{a_{5}}\\
    &(\ket{a_{1}aa_{2}\highlight{a_{3}a_{4}a_{5}}}\bra{a_{2}a_{1}a\highlight{a_{3}a_{4}a_{5}}}+\ket{a_{1}a\highlight{a_{2}}a_{3}\highlight{a_{4}a_{5}}}\bra{a_{3}a_{1}\highlight{a_{2}}a\highlight{a_{4}a_{5}}}+\\
    &\ket{a_{1}a\highlight{a_{2}a_{3}}a_{4}\highlight{a_{5}}}\bra{a_{4}a_{1}\highlight{a_{2}a_{3}}a\highlight{a_{5}}}+\ket{a_{1}a\highlight{a_{2}a_{3}a_{4}}a_{5}}\bra{a_{5}a_{1}\highlight{a_{2}a_{3}a_{4}}a}\\
    &+\ket{a_{2}\highlight{a_{1}}aa_{3}\highlight{a_{4}a_{5}}}\bra{a_{3}\highlight{a_{1}}a_{2}a\highlight{a_{4}a_{5}}}+\ket{a_{2}\highlight{a_{1}}a\highlight{a_{3}}a_{4}\highlight{a_{5}}}\bra{a_{4}\highlight{a_{1}}a_{2}\highlight{a_{3}}a\highlight{a_{5}}}+\ket{a_{2}\highlight{a_{1}}a\highlight{a_{3}a_{4}}a_{5}}\bra{a_{5}\highlight{a_{1}}a_{2}\highlight{a_{3}a_{4}}a}\\
    &+\ket{a_{3}\highlight{a_{1}a_{2}}aa_{4}\highlight{a_{5}}}\bra{a_{4}\highlight{a_{1}a_{2}}a_{3}a\highlight{a_{5}}}+\ket{a_{3}\highlight{a_{1}a_{2}}a\highlight{a_{4}}a_{5}}\bra{a_{5}\highlight{a_{1}a_{2}}a_{3}\highlight{a_{4}}a}\\
    &+\ket{a_{4}\highlight{a_{1}a_{2}a_{3}}aa_{5}}\bra{a_{5}\highlight{a_{1}a_{2}a_{3}}a_{4}a}+h.c),\\ % 
\end{split}    
	\end{equation}
the terms come for $\dyad{0}{1}$, $\dyad{0}{2}$,$\dyad{0}{3}$, $\dyad{0}{4}$, $\dyad{1}{2}$, $\dyad{1}{3}$, $\dyad{1}{4}$, $\dyad{2}{3}$, $\dyad{2}{4}$, $\dyad{3}{4}$ terms in  $ Tr_{anc}[\ket{T^{\beta}_{f}}\bra{T^{\beta}_{f}}]$ respectively. The subsystems denoted by the highlighted terms can be discarded from the sum, for example 
\begin{equation*}
    \sum_{aa_{1}a_{2}a_{3}a_{4}a_{5}}p_{a}p_{a_{1}}p_{a_{2}}p_{a_{3}}p_{a_{4}}p_{a_{5}}\ket{a_{1}aa_{2}\highlight{a_{3}a_{4}a_{5}}}\bra{a_{2}a_{1}a\highlight{a_{3}a_{4}a_{5}}}=(\sum_{aa_{1}a_{2}}p_{a}p_{a_{1}}p_{a_{2}}\dyad{a_{1}aa_{2}}{a_{2}a_{1}a})\otimes T\otimes T\otimes T,
\end{equation*}So for the local Gibbs state of target system, the contribution of the term 
\begin{equation*}
    \sum_{aa_{1}a_{2}a_{3}a_{4}a_{5}}p_{a}p_{a_{1}}p_{a_{2}}p_{a_{3}}p_{a_{4}}p_{a_{5}}\ket{a_{1}aa_{2}\highlight{a_{3}a_{4}a_{5}}}\bra{a_{2}a_{1}a\highlight{a_{3}a_{4}a_{5}}},   
\end{equation*}  and one can obtain 
\begin{equation*}
\begin{split}
    &Tr_{R_{1}R_{2}R_{3}R_{4}R_{5}}[(\sum_{aa_{1}a_{2}}p_{a}p_{a_{1}}p_{a_{2}}\dyad{a_{1}aa_{2}}{a_{2}a_{1}a})\otimes T\otimes T\otimes T],\\
    &=\sum_{aa_{1}a_{2}}p_{a}p_{a_{1}}p_{a_{2}}\braket{a}{a_{1}}\braket{a_{2}}{a}\dyad{a_{1}}{a_{2}}Tr(T)Tr(T)Tr(T),\\
    &=\sum_{aa_{1}a_{2}}p_{a}p_{a_{1}}p_{a_{2}}\delta_{a,a_{1}}\delta_{a,a_{2}}\dyad{a_{1}}{a_{2}},\\
    &=\sum_{a}p^3_{a}\dyad{a}{a}=T^3.
\end{split}
\end{equation*}and all the terms in Eqn.(\ref{offswap}) give rise to the same term $T^3$ in local Gibbs state of target system. But when we calculate the local Gibbs state of the reservoir qubit, we see something different than $T^3$. Take the reservoir qubit 1 as an example, the contribution of term $\sum_{aa_{1}a_{2}a_{3}a_{4}a_{5}}p_{a}p_{a_{1}}p_{a_{2}}p_{a_{3}}p_{a_{4}}p_{a_{5}}\ket{a_{2}\highlight{a_{1}}aa_{3}\highlight{a_{4}a_{5}}}\bra{a_{3}\highlight{a_{1}}a_{2}a\highlight{a_{4}a_{5}}}$ in local Gibbs state of reservoir qubit 1 is

\begin{equation*}
\begin{split}
    &\sum_{aa_{1}a_{2}a_{3}a_{4}a_{5}}p_{a}p_{a_{1}}p_{a_{2}}p_{a_{3}}p_{a_{4}}p_{a_{5}}Tr_{WR_{2}R_{3}R_{4}R_{5}}[\ket{a_{2}\highlight{a_{1}}aa_{3}\highlight{a_{4}a_{5}}}\bra{a_{3}\highlight{a_{1}}a_{2}a\highlight{a_{4}a_{5}}}],\\
    &=\sum_{aa_{1}a_{2}a_{3}}p_{a}p_{a_{1}}p_{a_{2}}p_{a_{3}}\braket{a_{2}}{a_{3}}\braket{a_{2}}{a}\braket{a_{3}}{a}\dyad{a_{1}}{a_{1}}Tr(T)Tr(T),\\
    &=\sum_{aa_{1}a_{2}a_{3}}\delta_{a_{2},a_{3}}\delta_{a_{2},a}\delta_{a_{3},a}p_{a}p_{a_{1}}p_{a_{2}}p_{a_{3}}\dyad{a_{1}}{a_{1}},\\
    &= (\sum_{a}p^3_{a})\sum_{a_{i}}p_{a_{1}}\dyad{a_{1}}{a_{1}}=\frac{1+r^3}{(1+r)^3}T.
\end{split}
\end{equation*}this term occurs when the system you don't want to trace over is denoted by one of the highlighted symbols in Eqn.(\ref{offswap}). And there are $5(5-1)-2(5-1)$ terms in Eqn.(\ref{offswap}) produce $\frac{1+r^3}{(1+r)^3}T$ for the local Gibbs state of reservoir qubit 1. It is not hard to generalise the idea to arbitrary $N$, then there will be $N-2$ highlighted symbols in those terms from off-diagonal terms in $ Tr_{anc}[\ket{T^{\beta}_{f}}\bra{T^{\beta}_{f}}]$. And actually only $\dyad{0}{k}$, $\dyad{k}{0}$ ($k=1,2,\cdots,N-1$) terms from $ Tr_{anc}[\ket{T^{\beta}_{f}}\bra{T^{\beta}_{f}}]$ can produce $T^3$. There are $2(N-1)$ those terms and the reason is when the control state is in state other than $\ket{0}$, the controlled-SWAPs exchange a and $a_{i}$ ($i\ne 1)$ and leaves everything unchanged, so when you calculate the local Gibbs state of reservoir qubit 1, these $\dyad{k}{k'}$ ($k,k'\ne 0, k\ne k')$ terms from $ Tr_{anc}[\ket{T^{\beta}_{f}}\bra{T^{\beta}_{f}}]$ contribute the same as $\sum_{aa_{1}a_{2}a_{3}a_{4}a_{5}}p_{a}p_{a_{1}}p_{a_{2}}p_{a_{3}}p_{a_{4}}p_{a_{5}}\ket{a_{2}\highlight{a_{1}}aa_{3}\highlight{a_{4}a_{5}}}\bra{a_{3}\highlight{a_{1}}a_{2}a\highlight{a_{4}a_{5}}}$. Similar logic can be applied for the calculation of local Gibbs states of other reservoir qubits.

So when we attain cooling branch, the local Gibbs state of the working system is $\frac{1}{N}[T+(N-1)T^3]$, but for the reservoir qubit, for the terms from off-diagonal terms in  $ Tr_{anc}[\ket{T^{\beta}_{f}}\bra{T^{\beta}_{f}}]$,  $\frac{N-1}{\frac{N(N-1)}{2}}=\frac{2}{N}$ of the terms are $T^3$ while the reset of them are $\frac{1+r^3}{(1+r)^3}T$. So we get the expression of the local Gibbs state of the reservoir qubit when attaining cooling branch
\begin{equation}
     \rho_{R_{i}}= \frac{1}{N}T+\frac{N(N-1)}{N^2}[\frac{2}{N}T^3+\frac{N-2}{N}\frac{r^3+1}{(1+r)^3}T].
\end{equation}where $N(N-1)$ is the total number of off-diagonal terms in $ Tr_{anc}[\ket{T^{\beta}_{f}}\bra{T^{\beta}_{f}}]$.

\subsection{Thermalisation of 1 of N+1 subsystems in the final quantum correlated system is local-Gibbs-state-preserving for the remaining subsystems}\label{msp}

To see whether the local Gibbs states of the qubits in remaining quantum correlated system remain unchanged when we discard one qubit each time by thermalising with one of the cold reservoirs (with thermal state $T$), it is enough to focus on the second term in Eqn.(\ref{marsswap}). Take $N=3$ case as an example, we have 

\begin{equation*}\label{offswap3}
\begin{split}
    &\sum_{aa_{1}a_{2}a_{3}}p_{a}p_{a_{1}}p_{a_{2}}p_{a_{3}}(\ket{a_{1}aa_{2}\highlight{a_{3}}}\bra{a_{2}a_{1}a\highlight{a_{3}}}\\
    &+\ket{a_{1}a\highlight{a_{2}}a_{3}}\bra{a_{3}a_{1}\highlight{a_{2}}a}+\ket{a_{2}\highlight{a_{1}}aa_{3}}\bra{a_{3}\highlight{a_{1}}a_{2}a}+h.c),
\end{split}    
\end{equation*}we can rewrite it as
\begin{equation}\label{offswap31}
\begin{split}
    &\sum_{aa_{1}a_{2}}p_{a}p_{a_{1}}p_{a_{2}}(\dyad{a_{1}}{a_{2}}\otimes\dyad{a}{a_{1}}\otimes\dyad{a_{2}}{a})\otimes T\\
    &+\sum_{aa_{1}a_{3}}p_{a}p_{a_{1}}p_{a_{3}}(\dyad{a_{1}}{a_{3}}\otimes\dyad{a}{a_{1}}\otimes T\otimes\dyad{a_{3}}{a})\\
    &+\sum_{aa_{2}a_{3}}p_{a}p_{a_{2}}p_{a_{3}}(\dyad{a_{2}}{a_{3}}\otimes T\otimes\dyad{a}{a_{2}}\otimes\dyad{a_{3}}{a})+h.c.
\end{split}    
\end{equation}recall that the action of thermalising channel is 
\begin{equation*}
    \aleph^{T}(\rho) = \text{Tr}[\rho]T = \frac{1}{d}\sum^{d^2}_{i}A(U_{i}\rho U^{\dagger}_{i})A^{\dagger}.
\end{equation*}so after thermalising the target system by one of the cold reservoirs, the terms in (\ref{offswap31}) evolve to

\begin{equation}\label{dis1}
    \begin{split}
    &T\otimes\sum_{aa_{1}a_{2}}p_{a}p_{a_{1}}p_{a_{2}}\delta_{a_{1},a_{2}}(\dyad{a}{a_{1}}\otimes\dyad{a_{2}}{a})\otimes T\\
    &+T\otimes\sum_{aa_{1}a_{3}}p_{a}p_{a_{1}}p_{a_{3}}\delta_{a_{1},a_{3}}(\dyad{a}{a_{1}}\otimes T\otimes\dyad{a_{3}}{a})\\
    &+T\otimes\sum_{aa_{2}a_{3}}p_{a}p_{a_{2}}p_{a_{3}}\delta_{a_{2},a_{3}}(T\otimes \dyad{a}{a_{2}}\otimes\dyad{a_{3}}{a})+h.c\\
    &=T\otimes\sum_{aa_{1}}p_{a}p^2_{a_{1}}(\dyad{a}{a_{1}}\otimes\dyad{a_{1}}{a})\otimes T+T\otimes\sum_{aa_{1}}p_{a}p^2_{a_{1}}(\dyad{a}{a_{1}}\otimes T\otimes\dyad{a_{1}}{a})\\
    &+T\otimes\sum_{aa_{2}}p_{a}p^2_{a_{2}}( T\otimes \dyad{a}{a_{2}}\otimes\dyad{a_{2}}{a})+h.c.
    \end{split}
\end{equation}when we calculate the local Gibbs state of the reservoir qubit 1 by tracing out the target system and reservoir qubits 2, 3, we see that the terms in (\ref{dis1}) give rise to the same terms as those in (\ref{offswap31}), where the first 2 terms in (\ref{dis1}) give $T^3$ while the third one gives $\frac{1+r^3}{(1+r)^3}T$.

Then we discard reservoir qubit 1 by thermalising with one of the cold reservoirs with thermal state $T$, we get 
\begin{equation}\label{dis2}
    \begin{split}
    T\otimes T\otimes\sum_{a}p^3_{a}(\dyad{a}{a})\otimes T+T\otimes T\otimes T\otimes\sum_{a}p^3_{a}(\dyad{a}{a})+T\otimes T\otimes\sum_{aa_{2}}p_{a}p^2_{a_{2}}(\dyad{a}{a_{2}}\otimes\dyad{a_{2}}{a})+h.c.
    \end{split}
\end{equation}Similarly, when we determine local Gibbs state of reservoir qubit 2, we should see that the terms in (\ref{dis2}) give rise to the same terms as those in (\ref{offswap31}) and (\ref{dis1}) , where the first 2 terms in (\ref{dis2}) give $T^3$ while the third one gives $\frac{1+r^3}{(1+r)^3}T$.

Finally, we discard reservoir qubit  by thermalising with one of the cold reservoirs with thermal state $T$
\begin{equation}\label{dis3}
    \begin{split}
    T\otimes T\otimes\frac{1+r^3}{(1+r)^3}T\otimes T+T\otimes T\otimes T\otimes T^3+T\otimes T\otimes T\otimes\sum_{a}p^3_{a}(\dyad{a}{a})+h.c.
    \end{split}
\end{equation}and we can easily see that, when we trace out the target system and reservoir qubits 1, 2, the terms in (\ref{dis3}) give rise to the same terms as those in (\ref{offswap31}), (\ref{dis1}) and (\ref{dis2}).

And the whole procedure can be generalised to shceme with arbitrary number of reservoirs, the conclusion will still hold. The key is that discarding a subsystem from the $N+1$-qubit quantum correlated sytsem by thermalising it with a reservoir with thermal state $T$ in the quantum-controlled-SWAPs of thermal qubits scheme won't change the ratio of $T^3$ (so does $\frac{1+r^3}{(1+r)^3}T$) terms among all terms in the local Gibbs states of target system and reservoir qubits.

\section{Maxwell-demon-like scenario with thermalisation in superposition of quantum trajectories} \label{mlt}

\begin{figure}
    \centering
	\includegraphics[width=1\textwidth]{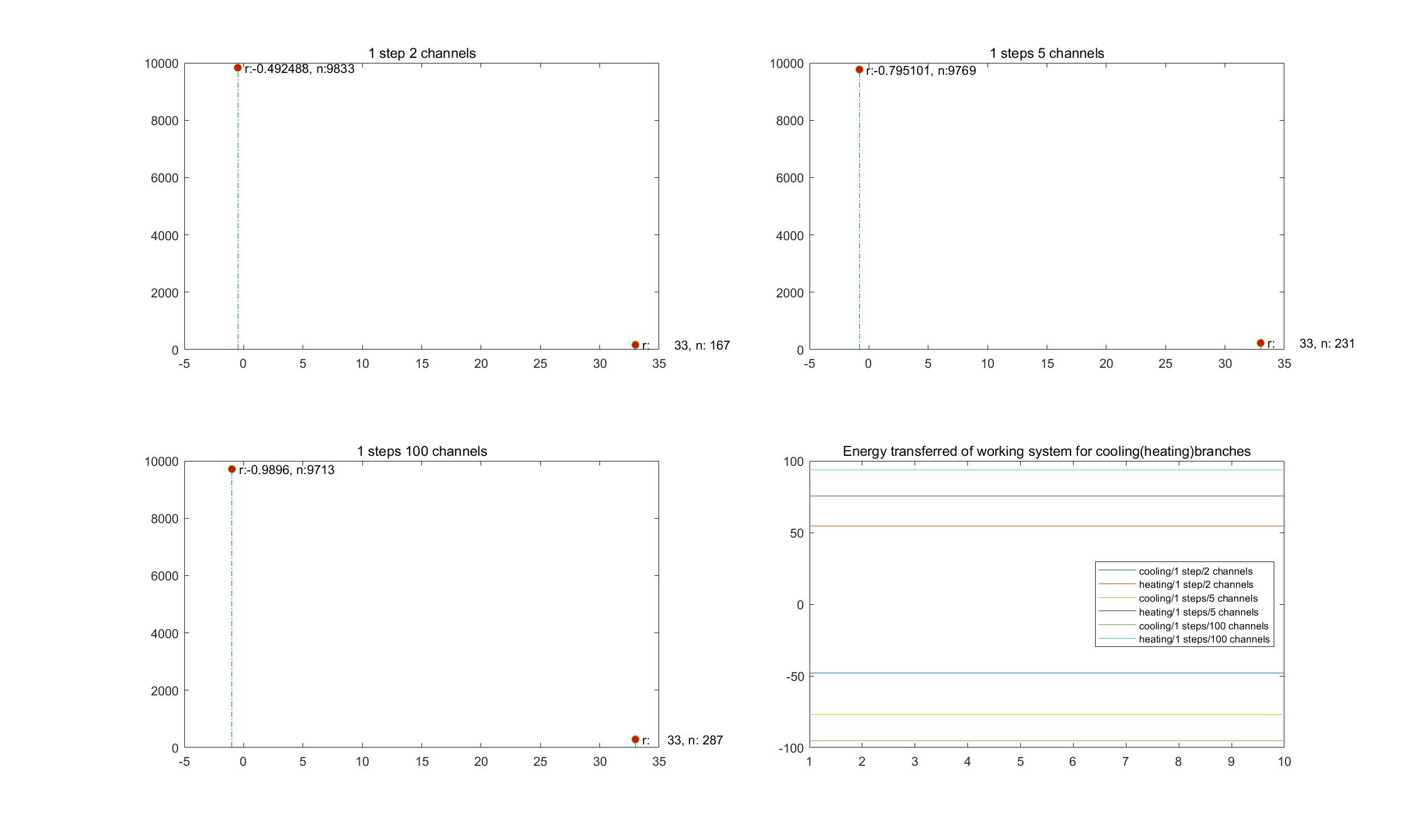}
	\caption{\textbf{Number of particles in sample C or D and their corresponding energy changes compared to the initial local Gibbs state with $r=0.01$.}}
	\label{multi0.01}
\end{figure}
\begin{figure}
    \centering
	\includegraphics[width=1\textwidth]{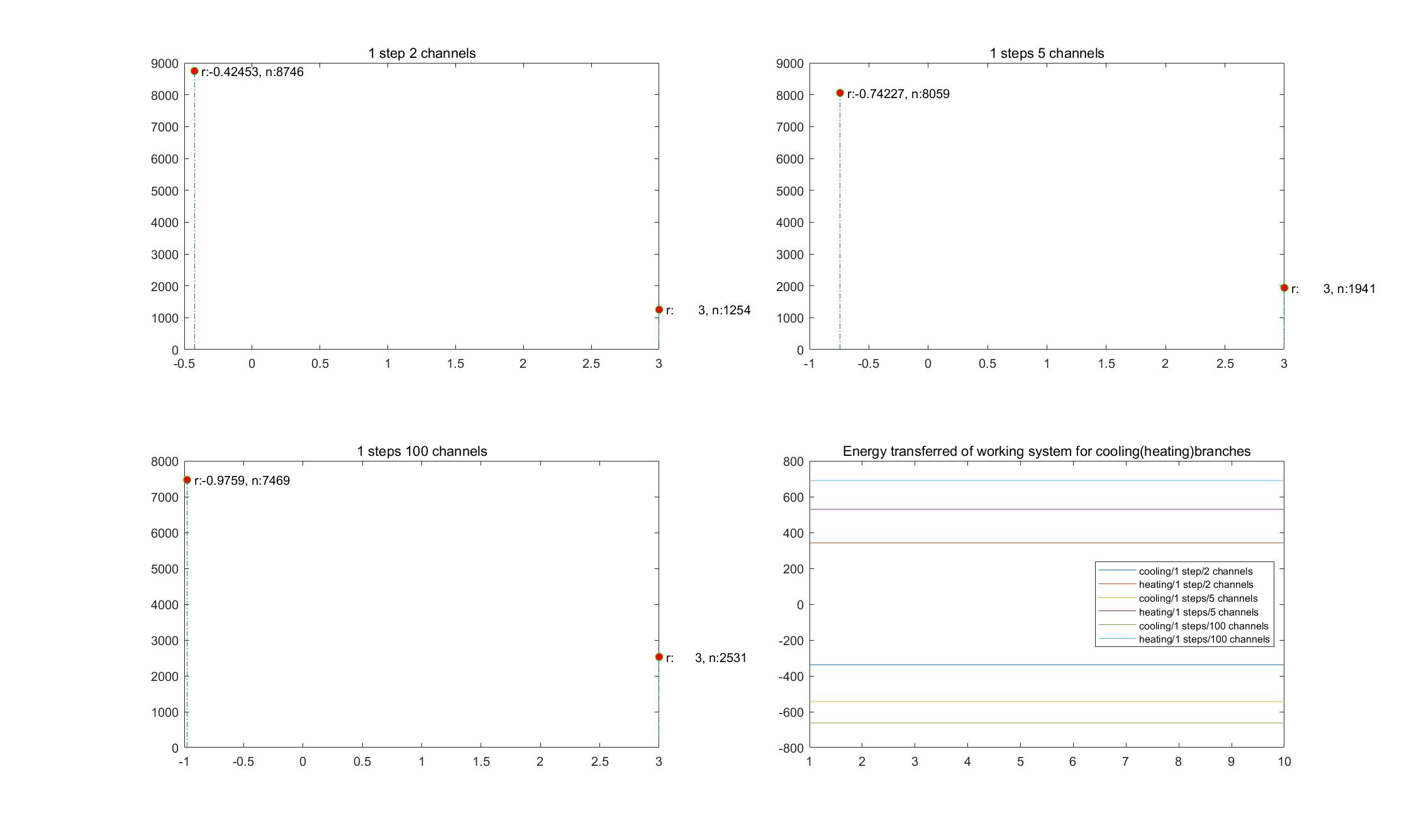}
	\caption{\textbf{Number of particles in sample C or D and their corresponding energy changes compared to the initial local Gibbs state with $r=0.1$.}}
	\label{multi0.1}
\end{figure}

\begin{figure}
    \centering
	\includegraphics[width=1\textwidth]{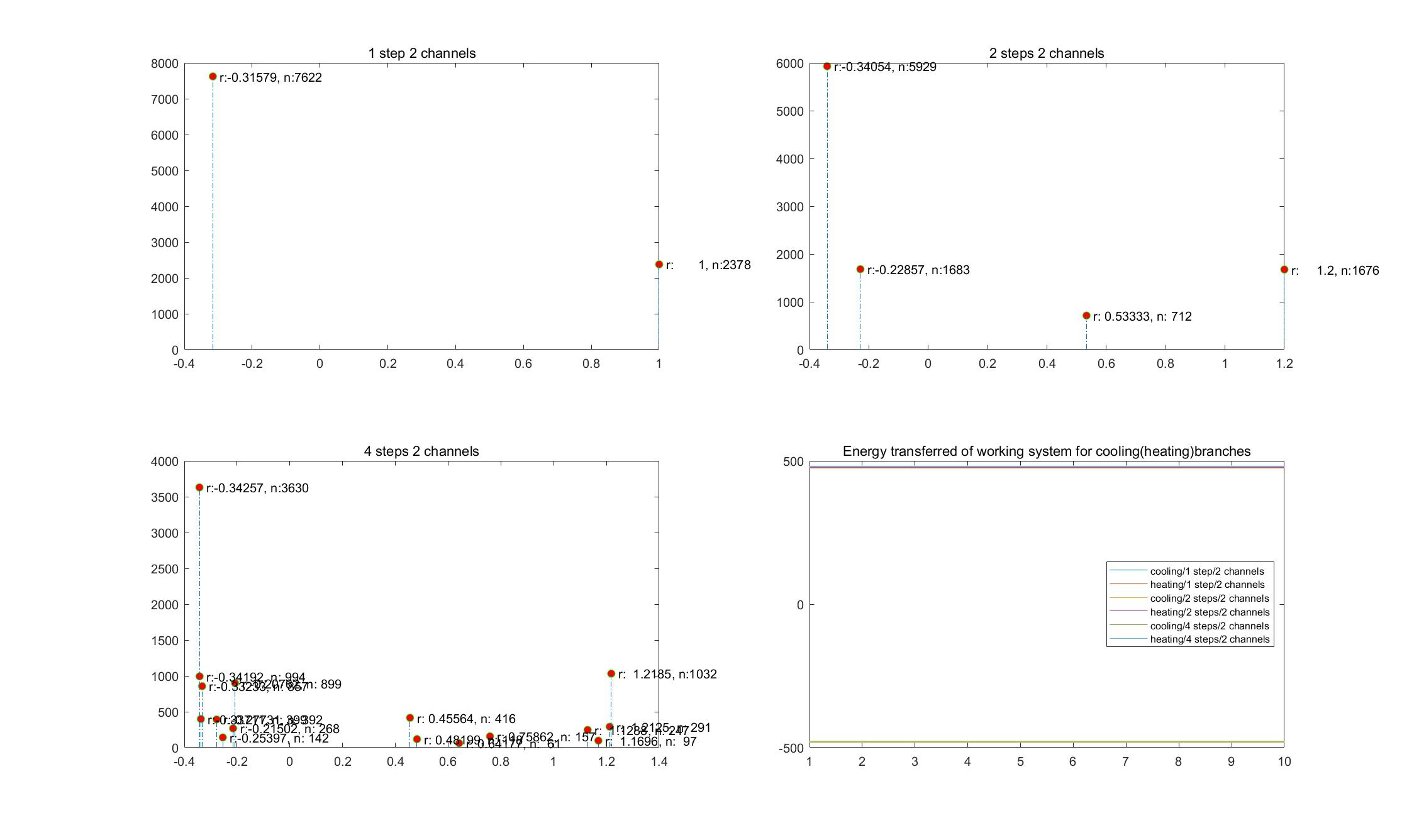}
	\caption{\textbf{Multiple runs of the process with sample initialised in $r=0.1$.}}
	\label{multi2N}
\end{figure}

\begin{figure}
    \centering
	\includegraphics[width=1\textwidth]{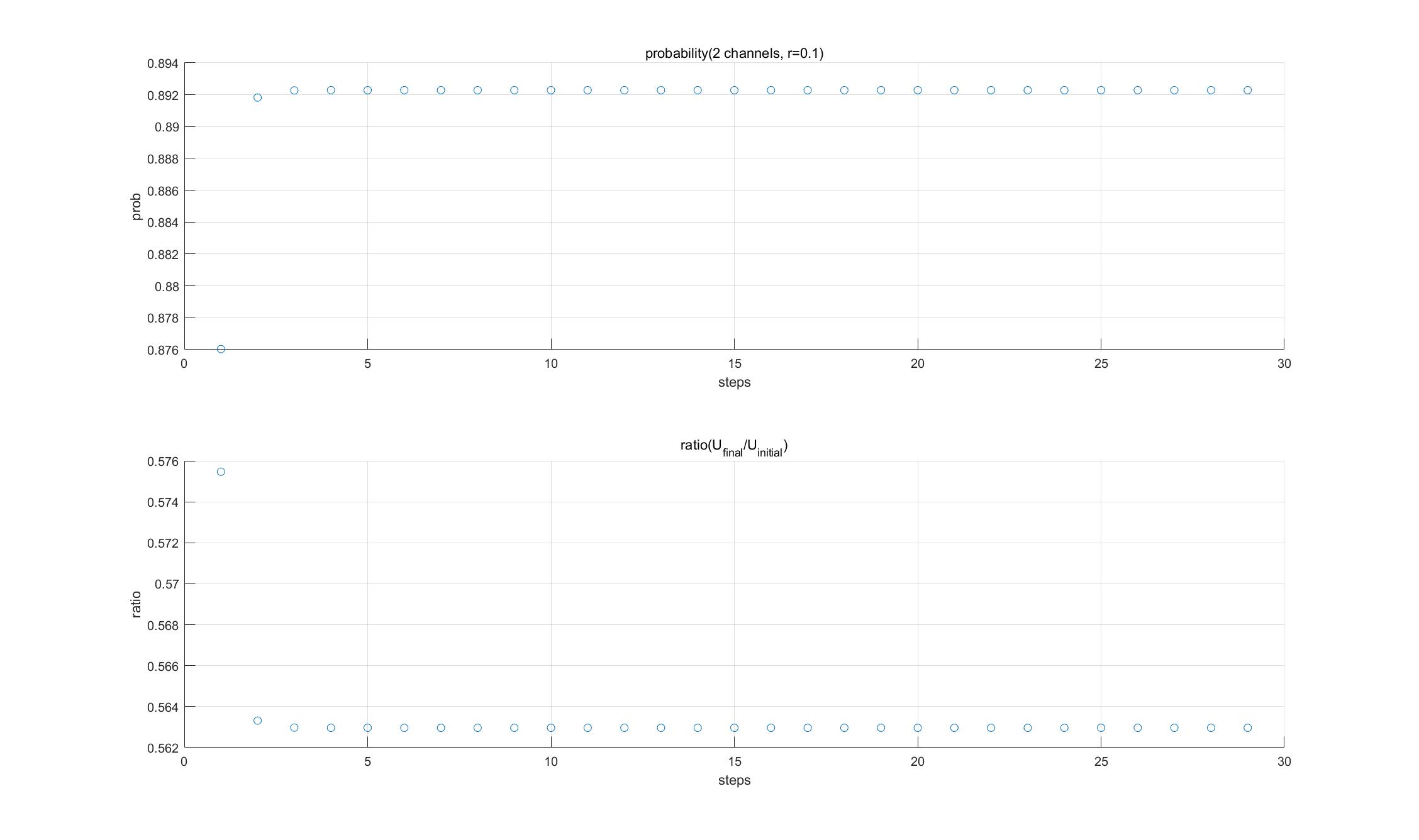}
	\caption{\textbf{Multiple runs of the process with sample initialised in $r=0.1$ for 2 reservoirs.}}
	\label{2chan0.1}
\end{figure}

\begin{figure}
    \centering
	\includegraphics[width=1\textwidth]{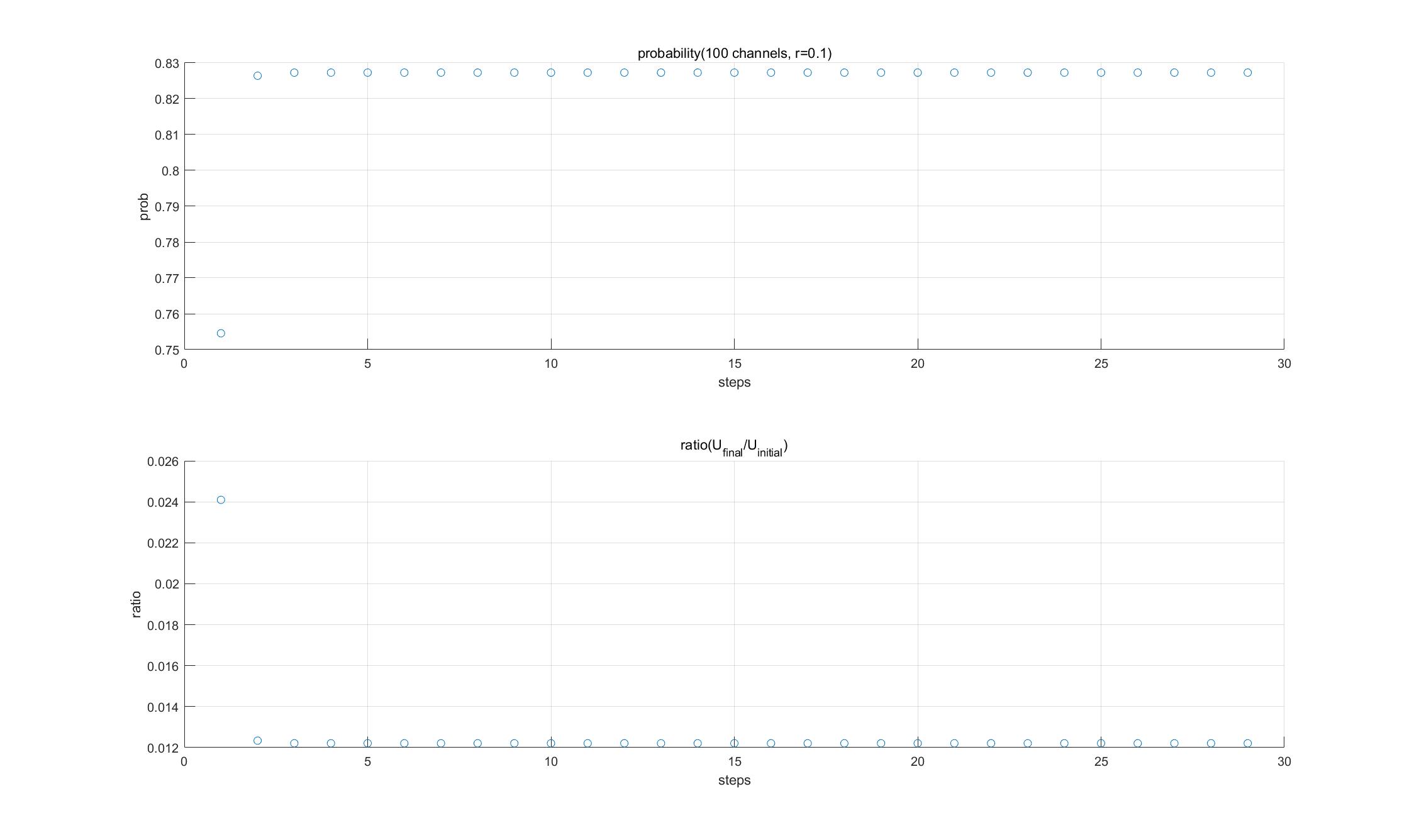}
	\caption{\textbf{Multiple runs of the process with sample initialised in $r=0.1$ for 100 reservoirs.}}
	\label{100chan0.1}
\end{figure}

\begin{figure}
    \centering
	\includegraphics[width=1\textwidth]{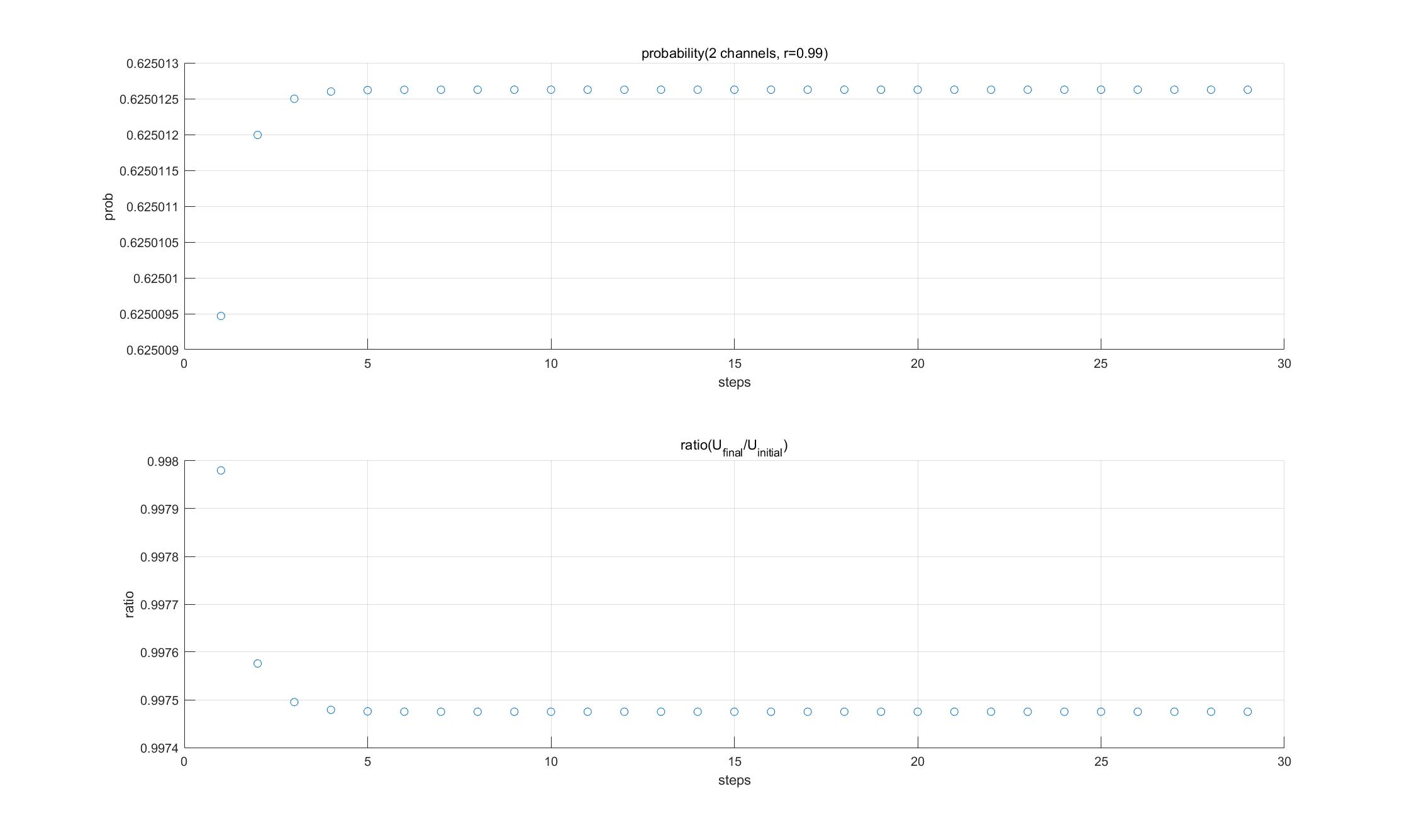}
	\caption{\textbf{Multiple runs of the process with sample initialised in $r=0.99$ for 2 reservoirs.}}
	\label{2chan0.99}
\end{figure}

\begin{figure}
    \centering
	\includegraphics[width=1\textwidth]{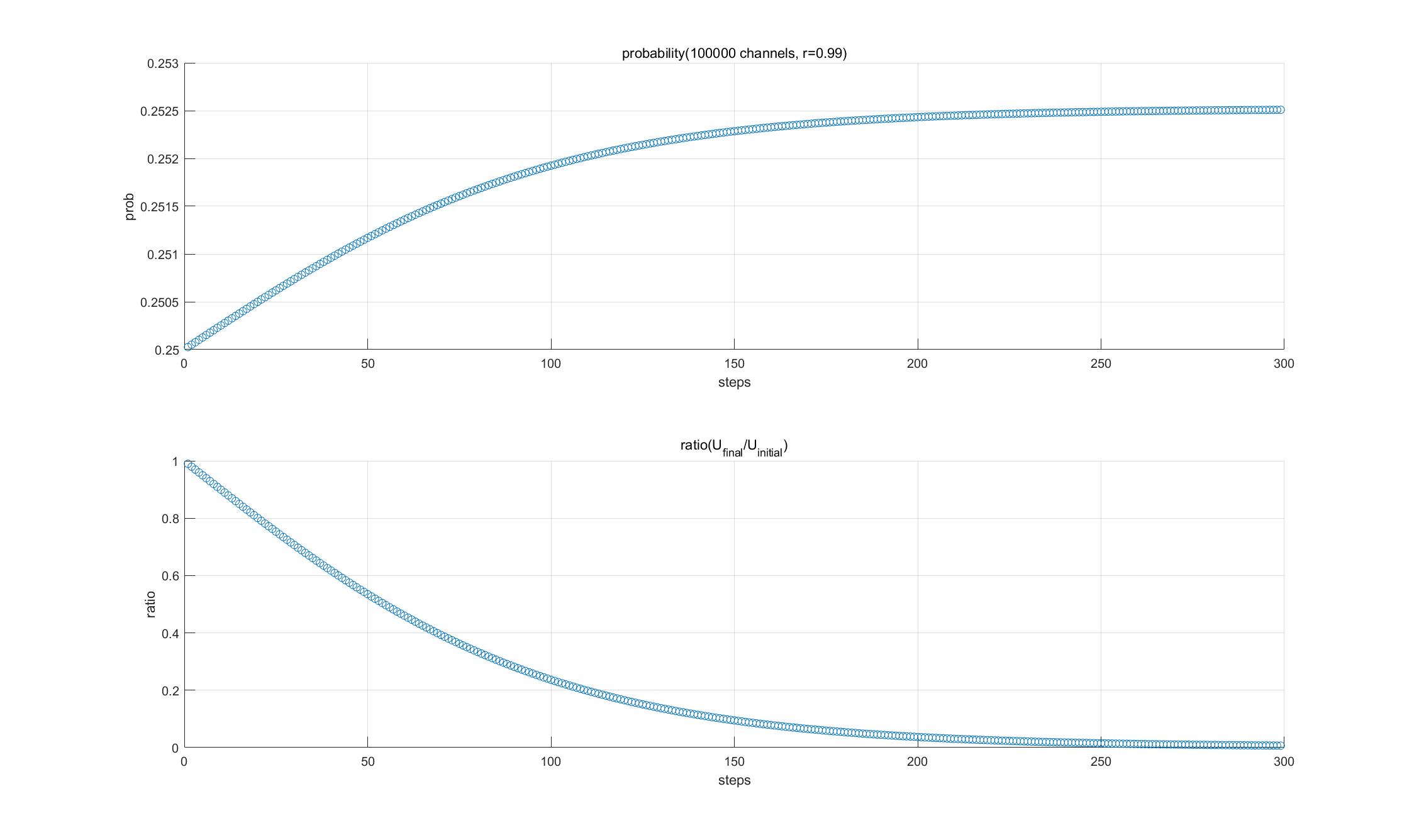}
	\caption{\textbf{Multiple runs of the process with sample initialised in $r=0.99$ for 100000 reservoirs.}}
	\label{100000chan0.99}
\end{figure}

\begin{figure}
    \centering
	\includegraphics[width=1\textwidth]{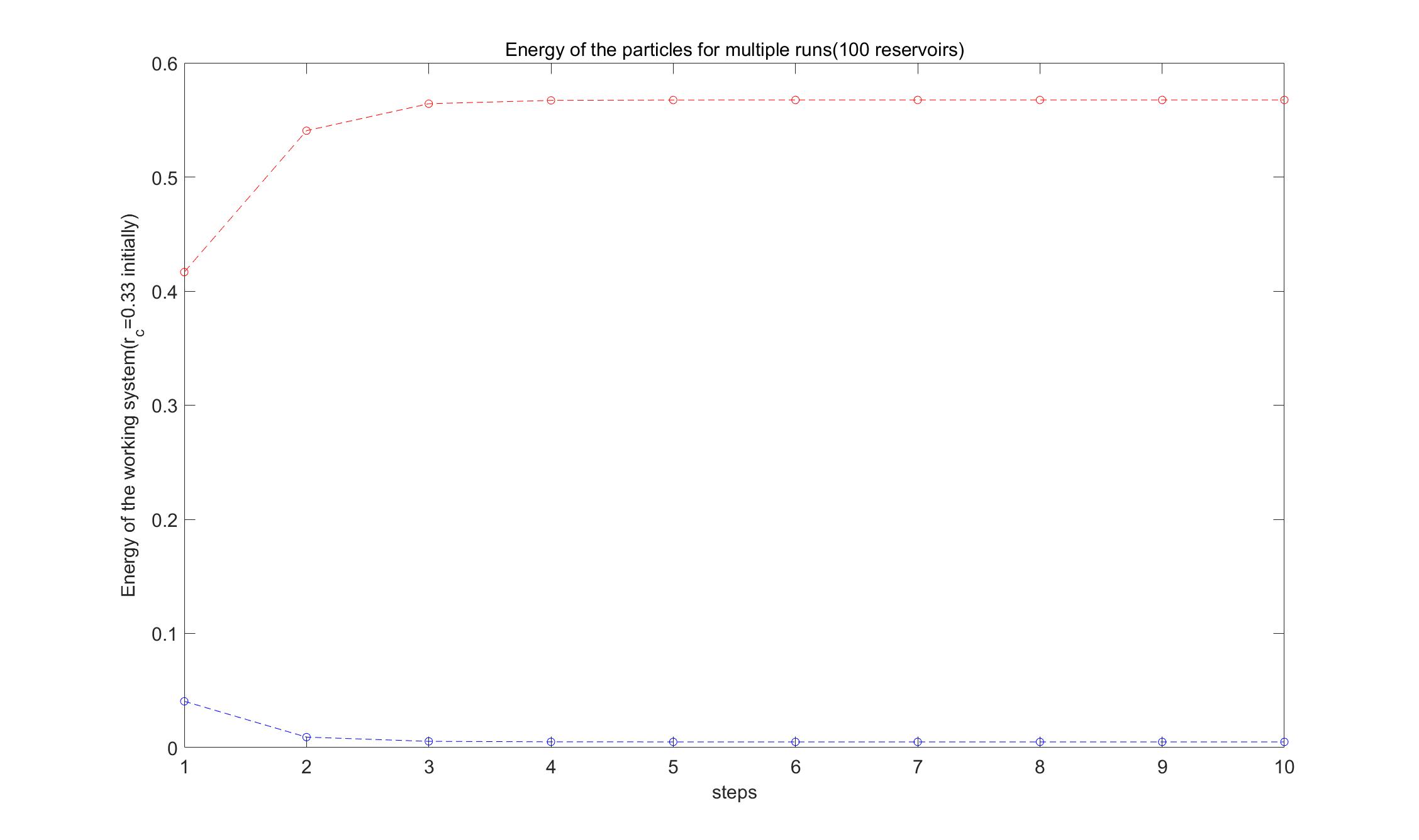}
	\caption{\textbf{Heat jump for the multiple runs scheme for $N=100$ and start at $r=0.33$. }}
	\label{heatj}
\end{figure}

\end{widetext}

\end{document}